\documentclass[a4paper,12pt,oneside,openany]{book}
\usepackage[utf8]{inputenc}
\usepackage[nottoc]{tocbibind}
\usepackage{amsmath,amssymb,amsfonts,latexsym}
\usepackage{dsfont}
\usepackage{amsthm}
\usepackage{cancel}
\usepackage[left=3cm, right=2cm, top=3cm, bottom=2cm]{geometry}
\usepackage{graphicx}
\usepackage{setspace}
\usepackage{anyfontsize}
\usepackage{float}
\usepackage{slashed}
\DeclareMathOperator{\Tr}{Tr}
\newtheorem*{Gol}{Goldstone’s Theorem}

\usepackage{ragged2e}
\usepackage[usenames]{color}
\usepackage[hyperindex=true,breaklinks=true,colorlinks=true,citecolor={blue},linkcolor=blue]{hyperref}
\usepackage{fancyhdr} 
\usepackage{setspace}
\begin{document}
\pagestyle{empty}
\pagenumbering{roman} 
\begin{center}
{
\fontsize{16}{0}{\selectfont University of São Paulo}\\[0.1cm]
\fontsize{16}{0}{\selectfont Physics Institute}\\[3.3cm]
\fontsize{22}{0}{\selectfont \textbf{Composite Higgs models}}\\[2cm]
\fontsize{18}{0}{\selectfont Juan Pablo Hoyos Daza}\\[5cm]
}
\leftskip 6cm
\noindent {Dissertation submitted to the Physics Institute of the University of São Paulo in partial fulfillment of the requirements 
for the degree of Master of Science.}\\[6cm]
\begin{center}
São Paulo \\  2018
\end{center}
\end{center}
\newpage

\chapter*{Acknowledgments}
I would like to thank the working people of Brazil, to share me his progress and wealth. I would also like to extend my gratitude to:
\begin{itemize}
 \item My family for all the collaboration provided;
 \item The USP and IFUSP for hospitality, the amazing research environment and resources necessary for my academic education;
 \item My research advisor, Dr. Enrico Bertuzzo for his great disposition and patience;
 \item The brothers Gina Ordoñez and Gustavo Ordoñez for their valuable collaboration;
 \item The CNPq for a Master Scholarship.
\end{itemize}
\chapter*{Abstract} 
One of the solutions to the hierarchy problem of the Standard Model is the composite Higgs scenario, where the Higgs emerges as a composite 
pseudo-Nambu-Goldstone boson.\\
In this work we present and study the basic characteristics of the composite Higgs scenario, based on the 
$\textup{SO}\left(5\right)/\textup{SO}\left(4\right)$ and $\textup{SO}\left(6\right)/\textup{SO}\left(5\right)$ cosets. We construct their 
effective Lagrangians through the Callan-Coleman-Wess-Zumino construction. The first coset does not differ much from the Standard Model and the 
second contains a singlet scalar in addition to the Higgs doublet. In these models we study the gauge sector, the fermion sector and we 
estimate the composite Higgs potential.\\[0.5 cm]
\textit{Keywords}: Composite Higgs; Higgs boson; Callan-Coleman-Wess-Zumino construction; Standard Model.
\chapter*{Notation} 
\addcontentsline{toc}{chapter}{Notation}
In this work we will use the natural units $\hbar=c=1$ where $\hbar=h/2\pi$ with $h$ the Planck constant and $c$ the velocity of light.\\[0.5cm]
Greek indices $\mu$ and $\nu$ run over the four spacetime coordinate, usually taken as 0, 1, 2, 3.\\[0.5cm]
The Einstein’s summation convention is used, i.e. indices that are repeated are summed over.\\[0.5cm]
$\mathcal{L}$ Lagrangian density, frequently called Lagrangian.\\[0.5cm]
$\varepsilon_{\alpha\beta\gamma}$ Levi -Civita symbol.\\[0.5cm]
The transpose of a matrix $A$ is $A^{T}$.\\[0.5cm]
The complex conjugate of a matrix $A$ is $A^{\ast}$.\\[0.5cm]
The Hermitian adjoint of a matrix A is $A^{\dagger}=A^{\ast T}$.\\[0.5cm]
$\left(A^{a}_{Ad}\right)_{c}\hspace{0.01 cm}^{b}=if^{abc}$ is the adjoint representation.\\[0.5cm]
$\mathds{1}$ is the identity matrix, or sometimes called a unit matrix.\\[0.5 cm]
Dirac conjugation is expressed by $\overline{\psi}=\psi^{\dagger}\gamma^{0}$.\\[0.5 cm]
$+h.c$ is the addition of the Hermitian adjoint or complex conjugate.\\[0.5 cm]
The Pauli matrix are given by 
$$\sigma_{1}=\begin{pmatrix}0&&1\\1&&0\end{pmatrix},\hspace{0.5cm}\sigma_{2}=\begin{pmatrix}0&&-i\\i&&0\end{pmatrix},\hspace{0.5cm}\sigma_{3}=\begin{pmatrix}1&&0\\0&&-1\end{pmatrix}.$$
\pagestyle{fancy} 
\spacing{1.5}
\tableofcontents
\chapter{Introduction}
\pagenumbering{arabic}
Experimental tests such as the discovery of the weak neutral interaction, the production of the $W$ and $Z$ bosons, the recent discovery of a 
Higgs-like boson \cite{ATLAS,CMS}, among another data sets \cite{PDG}, confirm the extraordinary success of the Standard Model (SM) to explain 
the interactions of the fundamental particles.\\
However some drawbacks observed experimentally (neutrino masses, baryon asymmetry in the universe, dark matter), and other of theoretical type 
(strong Charge-Parity violation, electric charge quantization and the hierarchy or ``naturalness" problem) require an extension of the theory.
We will focus on the Hierarchy problem \footnote{The first references to the problem are in \cite{Hooft} and \cite{Dimopoulos,Susskind}.}, 
since in this thesis we will study a scenario which addresses this problem.
\section{Hierarchy problem}
The problem stems out when we assume that the SM is an Effective Field Theory (EFT), where the Lagrangian of the SM is generated at a smaller 
scale than that of the fundamental theory. If there is just the SM, then there is no naturalness problem, but as there is no 
description of gravity in the SM and we expect gravity to kick in at a certain energy (around the Planck scale $M_{P}\approx 10^{19}$ GeV), 
we can consider the SM as an EFT.\\
Almost everything we see in Nature is described by $d=4$ operators of the SM Lagrangian (see Appendix \ref{SM}). In the SM however 
there is one operator of dimension $d=2$, the Higgs mass term. The coefficients in front of operators of dimension $d$ are proportional to 
$1/\Lambda_{SM}^{d-4}$ ($\Lambda$ is the cut-off scale \cite{Barbieri}). This means that the Higgs mass is enhanced by $\Lambda_{SM}^{2}$, 
so we can write the Higgs mass term as
\begin{equation}
k\Lambda_{SM}^{2}H^{\dagger}H,
\end{equation}
where $k$ is a numerical coefficient. Currently we know that the coefficient in front of $H^{\dagger}H$ in the SM is 
$\mu^{2}=m_{H}^{2}/2=\left(89\hspace{0.1 cm}\textup{GeV}\right)^{2}$, but if we take the SM cutoff as reference\footnote{This cutoff of reference 
is not necessarily the cutoff of the new physics.} at $\Lambda_{SM}\sim M_{GUT}\sim 10^{16}\hspace{0.1 cm}\textup{GeV}$ then
\begin{equation}
k\sim 10^{-28}\lll 1. 
\end{equation}
This enormous hierarchy is basically the Naturalness problem. This problem is sometimes presented also as a problem of quadratic 
divergences, due to the fact that if we use the cut-off regularization, the radiative corrections to the Higgs mass are quadratically 
divergent, a feature not present when applying the dimensional regularization method. This does not mean that the problem 
disappears (physics does not depend on the regularization method), rather that it is a problem of quadratic divergences in the sense that 
if we calculate the Renormalization Group Equation (RGE) for the running scalar mass\footnote{We obtain this result from 
Lagrangian $\mathcal{L}=\frac{1}{2}\partial_{\mu}\phi\partial^{\mu}\phi-\frac{m^{2}}{2}\phi^{2}+
\lambda\phi\overline{\psi}\psi+\overline{\psi}\left(i\slashed{\partial}-M\right)\psi$, with $M\gg m$, but it is totally valid in the SM if this is
considered as EFT.} we obtain
\begin{equation}\label{running}
\mu\frac{dm^{2}}{d\mu}=-\frac{3\lambda^{2}}{\pi^{2}}M^{2}+\dots 
\end{equation}
It is apparent that a quadratic sensitivity to a heavy mass scale exists even in dimensional regularization. For comparison, for a running fermion
mass we have 
\begin{equation}
\mu\frac{dm_{f}}{d\mu}\propto m_{f}, 
\end{equation}
that is to say, there is no dependence on the details of the high energy theory at low energy. 
On the other hand, effects analogous to Eq. (\ref{running}) are obtained from the interactions with vectors and scalar particles, so that the 
running scalar mass receives contributions from the mass of all kinds of particle it couples to. In the Standard Model the problem is present 
because we have an elementary scalar in the theory (the Higgs boson). With $\lambda$ and $M$ constants we can write down a solution 
to (\ref{running}), given by
\begin{equation}
m^{2}\left(\Lambda\right)-m^{2}\left(\Lambda_{EW}\right)=-\frac{3\lambda^{2}}{\pi^{2}}M^{2}\log\frac{\Lambda}{\Lambda_{EW}}.
\end{equation}
Let us see the consequences arising from the term $M^{2}$. In order to do that we must change a little bit the details pertaining to the 
high energy limit: $M\rightarrow M+\delta M$. Then we obtain
\begin{equation}
\begin{split}
m^{2}\left(\Lambda_{EW}\right)\rightarrow&m^{2}\left(\Lambda\right)+\frac{3\lambda^{2}}{\pi^{2}}M^{2}\log\frac{\Lambda}{\Lambda_{EW}}+\frac{6\lambda^{2}}{\pi^{2}}M\delta M\log\frac{\Lambda}{\Lambda_{EW}}\\
&=\left[m^{2}\left(\Lambda_{EW}\right)\right]_{old}+\frac{6\lambda^{2}}{\pi^{2}}M\delta M\log\frac{\Lambda}{\Lambda_{EW}}
\end{split}
\end{equation}
from which
\begin{equation}
\delta m^{2}\left(\Lambda_{EW}\right)=m^{2}\left(\Lambda_{EW}\right)-\left[m^{2}\left(\Lambda_{EW}\right)\right]_{old}=\frac{6\lambda^{2}}{\pi^{2}}M^{2}\log\left(\frac{\Lambda}{\Lambda_{EW}}\right)\frac{\delta M}{M}.
\end{equation}
In this way, we can define the sensitivity of the observable $\delta m^{2}\left(m_{pole}\right)$ with respect to the changes of the high energy, 
that is to say
\begin{equation}
\frac{\delta m^{2}\left(m_{pole}\right)}{m^{2}_{pole}}=\Delta\frac{\delta M}{M},
\end{equation}
where the sensitivity factor is 
\begin{equation}
\Delta=\frac{6\lambda^{2}}{\pi^{2}}\frac{M^{2}}{m^{2}_{pole}}\log\frac{\Lambda}{m_{pole}}.
\end{equation}
We see that if $\Delta\gg 1\Rightarrow\delta m^{2}\left(m_{pole}\right)/m^{2}_{pole}\gg 1$, which means that to small changes in the high 
energy scale correspond large changes in the observables. This goes against the conception of the effective field theory, which should not 
depend on the details of the high energy theory.\\
The way to keep the large changes under control is by tuning $m^{2}\left(\Lambda\right)$ 
against $\frac{6\lambda^{2}}{\pi^{2}}M^{2}\log\frac{\Lambda}{\Lambda_{EW}}$ (which are two unrelated quantities, in principle). This is the 
hierarchy problem in terms of fine tuning.\\
In the case of SM, if we think of it as an effective field theory where we are integrating out some states with mass $M_{P}$, 
the Higgs field interacts with these states and we will have a sensitivity factor given by
\begin{equation}
\Delta=\frac{const}{16\pi^{2}}\frac{M^{2}_{P}}{m^{2}_{h,Pole}}\log\frac{M_{P}}{m_{h,Pole}}\simeq 10^{31},
\end{equation}
so we have a huge tuning. This is the so-called fine tuning problem of the Standard Model.\\[0.5 cm]
The composite Higgs scenario \cite{Kaplan, Georgi} addresses the problem by considering the Higgs as a low energy bound state of some more 
fundamental degree of freedom, 
so the RGE is valid only up to the energy at which the more fundamental theory kicks in. From the sensitivity factor, we can estimate the scale 
at which the New Physics  should appear without fine tuning, $\Delta\simeq 1$, with which $\Lambda\simeq 1$ TeV. The Large Hadron Collider 
can test the TeV region and determine if scenarios like the composite Higgs are viable. In addition to being a bound state, we will also require
the Higgs boson to be a pseudo Nambu-Goldstone boson to allow for its mass to be lower than the scale $\Lambda\approx 1$ TeV at which the bound
state is formed. This is analog to what happens in low energy QCD: typical bound states like protons and neutrons have masses
$m_{p,n}\sim\Lambda_{\textup{cond}}\sim 1$ GeV, where $\Lambda_{\textup{cond}}$ is the scale at which the QCD interaction is becoming strong. To
allow for states with masses $m\ll\Lambda_{\textup{cond}}$ (as is the case for the pseudoscalar mesons), we need to interpret them as pseudo
Nambu-Goldstone bosons.\\[0.5 cm]
In this work, we will present two composite Higgs models and determine the modifications with respect to the Standard Model. The work is organized as 
follows: in Chapter 2, we express the idea of the composite Higgs scenario and we present the CCWZ formalism necessary to construct
effective theories Lagrangians. In Chapter 3, we will present the minimal composite Higgs model in
the coset $\textup{SO}\left(5\right)/\textup{SO}\left(4\right)$ and we will show some of the modifications respect to the SM. In Chapter 4,
we will explore a non-minimal composite Higgs scenario for the coset $\textup{SO}\left(6\right)/\textup{SO}\left(5\right)$ and will see the 
consequences and modifications when inserting one additional scalar field. In Chapter 5 we estimate the Higgs potential to the two studied 
cosets, through the method of spurions. Finally, in Chapter 6, we offer the conclusions of the study carried out. 
\chapter{Composite Higgs models}
\begin{figure}[t]
\centering
\includegraphics[scale=0.55]{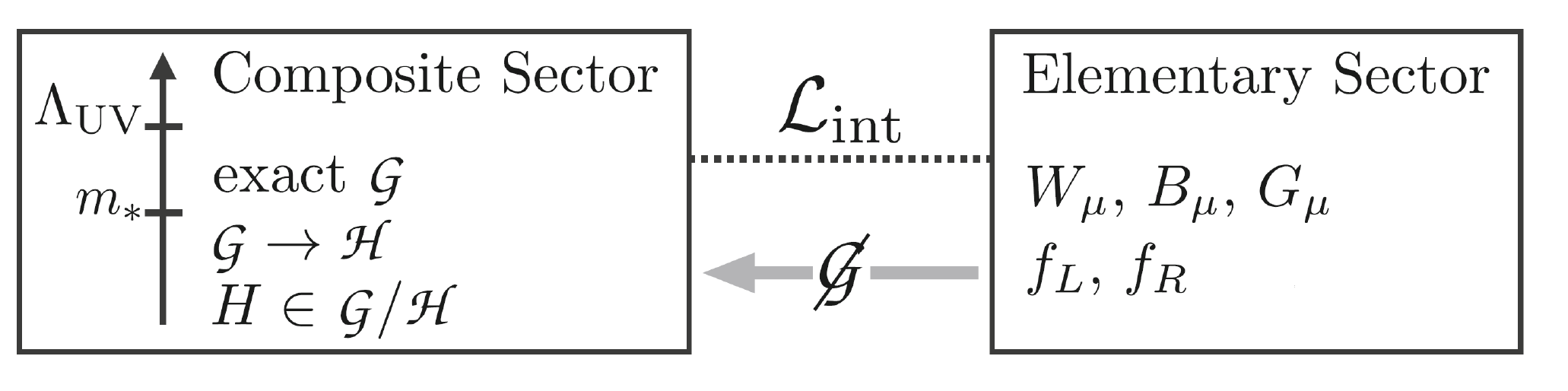}
\caption{Composite Higgs scenario, with its main ingredients. Taken from \cite{Panico}.}\label{res}
\end{figure}

In this chapter the composite Higgs scenario is presented. We mainly follow the presentation given by \cite{Panico}.\\
The idea of a composite Higgs boson, which can be naturally lighter than other resonances, was first pointed out by Georgi and Kaplan \cite{Kaplan, Georgi}. The Higgs emerges as pseudo Nambu-Goldstone boson (pNGB) of an enlarged global symmetry of the strong dynamics.
In the composite Higgs framework, there are three main ingredients.\\
The first ingredient is a ``composite sector" which has a global Lie group of symmetries $\mathcal{G}$. This sector will deliver the Higgs as 
a bound state. The second ingredient is an ``elementary" sector, $\mathcal{H}$, which is contained in $\mathcal{G}$. This sector must contain all 
the known particles. In addition, to produce NGB, $\mathcal{G}$ must be spontaneously broken to a subgroup $\mathcal{H}$. 
It is important to remember the Goldstone’s Theorem, which says 
\begin{Gol}
When a continuous global symmetry group $\mathcal{G}$ is broken down to a subgroup $\mathcal{H}\subset\mathcal{G}$ in which the broken generators do not 
leave the vacuum invariant, then there will be a massless scalar for every broken generator, called Nambu-Goldstone Boson (NGB).
\end{Gol}
\justify
The last ingredient is a Lagrangian of interaction, $\mathcal{L}_{int}$, between the composite and the elementary sector. 
$\mathcal{L}_{int}$ explicitly breaks $\mathcal{G}$, with the breaking allowing the NGB's to acquire mass. Figure \ref{res} summarizes the idea of the composite Higgs scenario. 
\section{Vacuum misalignment}
\begin{figure}[b]
\centering
\includegraphics[scale=1.0]{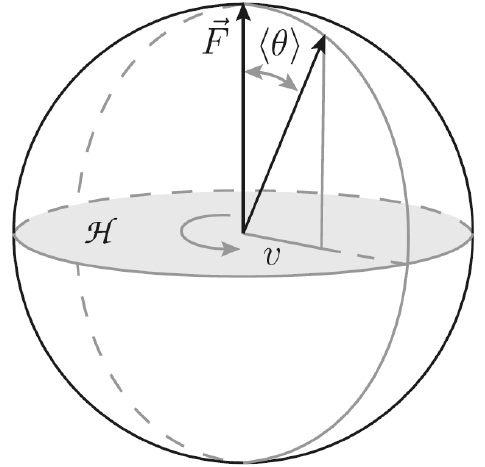}  
\caption{EWSB by means of vacuum misalignment, in the case $\textup{SO}\left(3\right)\rightarrow\textup{SO}\left(2\right)$. Taken from \cite{Panico}.}\label{geo}
\end{figure}

The vacuum misalignment is a mechanism by which the Higgs boson can behave as if it were elementary.
To see this, let us consider the composite sector in isolation and let us assume that it's vacuum state is only invariant under $\mathcal{H}\subset\mathcal{G}$, so there will be a spontaneous symmetry breaking $\mathcal{G}\rightarrow\mathcal{H}$, and according 
to the Goldstone's theorem dim$\mathcal{G}-$dim$\mathcal{H}$ NGB's will appear in the coset $\mathcal{G}/\mathcal{H}$.\\
Let us keep in mind that the group $\mathcal{G}$ must be large enough so that it at least contains one Higgs doublet in the coset, and the 
electro-weak group (EW) which is given by a Lie symmetry group, $G_{EW}=\textup{SU}\left(2\right)_{L}\times\textup{U}\left(1\right)_{Y}$, is contained in $\mathcal{H}$.\\
A basis for the Lie group is chosen with the generators $T^{A}\in\mathcal{G}$. The generators are divided into two sets given by
\begin{equation}
\left\lbrace T^{A}\right\rbrace=\left\lbrace T^{a},\hat{T}^{\hat{a}}\right\rbrace,
\end{equation}
where $T^{a}$ are the ``unbroken" generators $\left(a=1,...,dim\left[\mathcal{H}\right]\right)$, and $\hat{T}^{\hat{a}}$ 
are the ``broken" generators $\left(\hat{a}=\hat{1},...,dim\left[\mathcal{G}/\mathcal{H}\right]\right)$ .\\
Taking one of the degenerate vacuums of $\mathcal{G}$, a vacuum field configuration $\boldsymbol{F}$ is such that
\begin{equation}
T^{a}\boldsymbol{F}=0,\hspace{0.5 cm}\hat{T}^{\hat{a}}\boldsymbol{F}\neq 0. 
\end{equation}
As we will see in section \ref{CCWZ}, the $\mathcal{G}/\mathcal{H}$ coset can be parametrized by
\begin{equation}
\boldsymbol{\Phi}=e^{i\theta^{\hat{a}}\left(x\right)\hat{T}^{\hat{a}}}\boldsymbol{F},
\end{equation}
in which the NGB's are fluctuations along the broken generators $\hat{T}^{\hat{a}}$.
However the $\theta$ fields are massless and their Vacuum Expectation Values, $\langle\theta\rangle$, are not observable. This is because by 
redefining the $\theta$ fields through the transformation
$\boldsymbol{\Phi}\rightarrow\exp\left[-i\langle\theta^{\hat{a}}\rangle\hat{T}^{\hat{a}}\right]\boldsymbol{\Phi}$, we can set up
$\langle\theta^{\hat{a}}\rangle=0$. To change this, $\mathcal{G}$ must be explicitly broken so that $\theta$ becomes a PNGB, with which
\begin{itemize}
 \item $\theta$ develops a potential and then a VEV;
 \item $\langle\theta\rangle$ becomes observable, with the physical effect of breaking $G_{EW}$, thus causing the breaking of the Electro-Weak 
 symmetry (EWSB).
\end{itemize}
Geometrically $\langle\theta\rangle$ represents the measurement of the angle by which the vacuum is misaligned with respect to $\boldsymbol{F}$, 
as depicted in Fig. \ref{geo} (case $\mathcal{G}=\textup{SO}(3)$ and $\mathcal{H}=\textup{SO}(2)$).
\\From the figure, $\boldsymbol{F}$ is perpendicular to $\mathcal{H}$, so it is orthogonal to $G_{EW}$. So the projection of $\boldsymbol{F}$ 
on the $G_{EW}$ plane control EWSB (for example the mass of the particles of the SM). In other words, the scale of the EWSB 
is set by $v=f\sin\langle\theta\rangle$, 
being $f=\arrowvert\boldsymbol{F}\arrowvert$ the scale of spontaneous symmetry breaking $\mathcal{G}\rightarrow\mathcal{H}$.
\justify
The striking issues of the composite Higgs scenario will only occur when the misalignment angle is small $\langle\theta\rangle\ll 1$
which generates a gap between the $f$  and $v$ scale. The aforementioned can be expressed as follows
\begin{equation}\label{gap}
\xi\equiv\frac{v^{2}}{f^{2}}=\sin^{2}\langle\theta\rangle\ll 1,
\end{equation}
and parameter $\xi$ will always be present in the composite Higgs scenario. When $v\sim f$ (which is to say $\xi\rightarrow 1$), 
we have the case of a technicolor model \cite{technicolor}. Notice that for at fixed $v$ and taking the limit $\xi\rightarrow 0$ 
(which would imply to send the scale $f$ to infinity), the composite sector is decoupled from the low energy degrees of freedom, leaving only 
the Goldstone boson Higgs in the spectrum. Also, in this limit the theory is reduced to the SM and the Higgs becomes elementary. Finally we can 
say that the composite Higgs theories have an adjustable parameter $\xi$ which controls the deviations from the standard Higgs model.
\section{The CCWZ construction}\label{CCWZ}
Whenever a theory is characterized by a symmetry breaking pattern $\mathcal{G}\rightarrow\mathcal{H}$, such that $\mathcal{H}\subset\mathcal{G}$, the Callan-Coleman-Wess-Zumino (CCWZ) construction \cite{F_CCWZ1,F_CCWZ2} provides a systematic way to write effective Lagrangians that 
allow to manifest the symmetries of the theory.\\
By the definition of coset, any element $g\left[\alpha_{A}\right]\in\mathcal{G}$ in a neighborhood of the identity can be decomposed as
\begin{equation}
g\left[\alpha_{A}\right]=e^{i\alpha_{A}T^{A}}=e^{if_{\hat{a}}\left[\alpha\right]\hat{T}^{\hat{a}}}\cdot e^{if_{a}\left[\alpha\right]T^{a}},
\end{equation}
where $T^{a}$ and $\hat{T}^{\hat{a}}$ are the unbroken and broken generators, respectively.
This can be easily seen by expanding the group elements up to first order in the $\{\alpha\}$ parameters
\begin{equation}
\begin{split}
&e^{i\alpha_{A}T^{A}}=e^{i\left(\alpha_{\hat{a}}T^{\hat{a}}+\alpha_{a}T^{a}\right)}=\mathds{1}+i\alpha_{\hat{a}}T^{\hat{a}}+i\alpha_{a}T^{a}+\mathcal{O}\left(\alpha^{2}\right)\\
&e^{if_{\hat{a}}\left[\alpha\right]\hat{T}^{\hat{a}}}\cdot e^{if_{a}\left[\alpha\right]T^{a}}=\mathds{1}+if_{\hat{a}}T^{\hat{a}}+if_{a}T^{a}+\mathcal{O}\left(f_{\hat{a}}f_{a},f_{\hat{a}}^{2},f_{a}^{2}\right),                       
\end{split}
\end{equation}
which allows us to conclude that
\begin{equation}
\begin{split} 
&f_{\hat{a}}\left[\alpha\right]=\alpha_{\hat{a}}+\mathcal{O}\left(\alpha^{2}\right),\\
&f_{\hat{a}}\left[\alpha\right]=\alpha_{a}+\mathcal{O}\left(\alpha^{2}\right).\qedhere
\end{split}
\end{equation}
Since the directions in the coset are in a one to one correspondence with the NGB's, we promote the corresponding parameters to dynamical fields. 
The NGB's matrix is thus given by
\begin{equation}\label{MGoldstone}
U\left[\Pi\right]=e^{i\frac{\sqrt{2}}{f}\Pi_{\hat{a}}\left(x\right)\hat{T}^{\hat{a}}}. 
\end{equation}
As the CCWZ construction manifest the symmetries of a theory, let us see the properties of the Goldstone fields under $\mathcal{G}$, $\mathcal{H}$ 
and $\mathcal{G}/\mathcal{H}$.\\
First, let us determine how $U\left[\Pi\right]$ transforms under a generic element $g\in\mathcal{G}$. To this end, we use the decomposition of
a generic group element, finding
\begin{equation}\label{undergini}
gU\left[\Pi\right]=U\left[\Pi^{g}\right]h\left[\Pi;g\right];\hspace{0.5 cm}h\in\mathcal{H},
\end{equation}
or, equivalently,
\begin{equation}\label{underg}
U\left[\Pi\right]\rightarrow U\left[\Pi^{g}\right]=gU\left[\Pi\right]h^{-1}\left[\Pi;g\right].
\end{equation}
Because $\Pi^{(g)}$ has a non-linear dependency on $\Pi$, it is called a non-linear realization of $\mathcal{G}$.\\[0.5 cm]
Now let us see how $U\left[\Pi\right]$ transforms under $\mathcal{H}$. For this we need the commutation relations between generators $T^{a},T^{\hat{a}}$. First
\begin{equation}\label{comnobroken}
\left[T^{a},T^{b}\right]=if^{ab}\hspace{0.01 cm}_{c}T^{c}+i\cancel{f^{ab}\hspace{0.01 cm}_{\hat{c}}}\hat{T}^{\hat{c}}\equiv T^{c}\left(t^{a}_{Ad}\right)_{c}\hspace{0.01 cm}^{b},
\end{equation}
with the previous relation following from $f^{ab}\hspace{0.01 cm}_{\hat{c}}=0$, since $\mathcal{H}$ is a subgroup. The next relation is
\begin{equation}
\left[T^{a},\hat{T}^{\hat{b}}\right]=if^{a\hat{b}}\hspace{0.01 cm}_{\hat{c}}\hat{T}^{\hat{c}}+i\cancel{f^{a\hat{b}}\hspace{0.01 cm}_{c}}T^{c}\equiv\hat{T}^{\hat{c}}\left(t_{\pi}\hspace{0.01 cm}^{a}\right)_{\hat{c}}\hspace{0.01 cm}^{\hat{b}},
\end{equation}
due to the fact that the structure constants $f$ are completely antisymmetric and, from (\ref{comnobroken}), 
$f^{a\hat{b}}\hspace{0.01 cm}_{c}=0$. The matrices $t_{\pi}\hspace{0.01 cm}^{a}$ form the representation in which the NGB's transform under $\mathcal{H}$ . Finally
\begin{equation}
\left[\hat{T}^{\hat{a}},\hat{T}^{\hat{b}}\right]=if^{\hat{a}\hat{b}}\hspace{0.01 cm}_{c}T^{c}+if^{\hat{a}\hat{b}}\hspace{0.01 cm}_{\hat{c}}T^{\hat{c}}.
\end{equation}
In the last relation both terms are present, where we have terms for the broken and unbroken generators. If $f^{\hat{a}\hat{b}}_{\hspace{0.3 cm}\hat{c}}=0$ the coset is called symmetric.\\[0.5 cm]
Continuing with the transformations of $U\left[\Pi\right]$ under $g\in\mathcal{H}$, namely for $g=g_{\mathcal{H}}=e^{i\alpha_{a}T^{a}}$, we have
\begin{equation}
\begin{split}
g_{\mathcal{H}}U\left[\Pi\right]&=g_{\mathcal{H}}e^{b\Pi_{\hat{a}}\hat{T}^{\hat{a}}},\hspace{1 cm}b=i\frac{\sqrt{2}}{f}\\
&=g_{\mathcal{H}}\left(\mathds{1}+b\Pi_{\hat{a}}\hat{T}^{\hat{a}}+\frac{b^{2}}{2}\Pi_{\hat{a}}\Pi_{\hat{b}}\hat{T}^{\hat{a}}\hat{T}^{\hat{b}}+\cdots\right)\\
&=\left(\mathds{1}+b\Pi_{\hat{a}}g_{\mathcal{H}}\hat{T}^{\hat{a}}g_{\mathcal{H}}^{-1}+\frac{b^{2}}{2}\Pi_{\hat{a}}\Pi_{\hat{b}}g_{\mathcal{H}}\hat{T}^{\hat{a}}g^{-1}_{\mathcal{H}}g_{\mathcal{H}}\hat{T}^{\hat{b}}g_{\mathcal{H}}^{-1}+\cdots\right)g_{\mathcal{H}}\\
&=\exp\left[i\frac{\sqrt{2}}{f}\Pi_{\hat{a}}g_{\mathcal{H}}\hat{T}^{\hat{a}}g_{\mathcal{H}}^{-1}\right]g_{\mathcal{H}}.
\end{split}
\end{equation}
Developing the expression $g_{\mathcal{H}}\hat{T}^{\hat{a}}g_{\mathcal{H}}^{-1}$ using the Baker–Campbell–Hausdorff formula
\begin{equation}\label{BCH}
e^{i\lambda A}B e^{-i\lambda A}=B+i\lambda\left[A,B\right]+\frac{i^2\lambda^{2}}{2!}\left[A,\left[A,B\right]\right]+\cdots,
\end{equation}
we have
\begin{equation}
\begin{split}
g_{\mathcal{H}}\hat{T}^{\hat{a}}g_{\mathcal{H}}^{-1}&=e^{i\alpha_{b}T^{b}}\hat{T}^{\hat{a}}e^{-i\alpha_{b}T^{b}}\\
&=\hat{T}^{\hat{a}}+i\alpha_{b}\left[T^{b},\hat{T}^{\hat{a}}\right]+\mathcal{O}\left(\alpha^{2}\right)\\
&=\hat{T}^{\hat{a}}+i\alpha_{b}\hat{T}^{\hat{c}}\left(t_{\pi}\hspace{0.01 cm}^{b}\right)_{\hat{c}}^{\hspace{0.1 cm}\hat{a}}+\mathcal{O}\left(\alpha^{2}\right)\\
&=\hat{T}^{\hat{c}}\left(\delta^{\hat{a}}_{\hat{c}}+i\alpha_{b}\left(t_{\pi}\hspace{0.01 cm}^{b}\right)_{\hat{c}}^{\hspace{0.1 cm}\hat{a}}\right)+\mathcal{O}\left(\alpha^{2}\right)\\
&=\hat{T}^{\hat{c}}\left[\exp\left(i\alpha_{b}t^{\hspace{0.1 cm}b}_{\pi}\right)\right]_{\hat{c}}^{\hspace{0.1 cm}\hat{a}}.
\end{split}
\end{equation}
Using the previous result and comparing with Eq. (\ref{undergini}), we have
\begin{equation}
\begin{split}
g_{\mathcal{H}}U\left[\Pi\right]&=\exp\left[i\frac{\sqrt{2}}{f}\hat{T}^{\hat{c}}\left[\exp\left(i\alpha_{b}t^{\hspace{0.1 cm}b}_{\pi}\right)\right]_{\hat{c}}^{\hspace{0.1 cm}\hat{a}}\Pi_{\hat{a}}\right]g_{\mathcal{H}}\\
&=U\left[e^{i\alpha_{a}t_{\pi}^{\hspace{0.1 cm}a}}\boldsymbol{\Pi}\right]g_{\mathcal{H}},
\end{split}
\end{equation}
from which we see that $\Pi$ transforms under $g\in\mathcal{H}$  as
\begin{equation}
\Pi_{\hat{a}}\rightarrow\Pi_{\hat{a}}^{\hspace{0.1 cm}\left(g_{\mathcal{H}}\right)}=\left(e^{i\alpha_{a}t_{\pi}^{\hspace{0.1 cm}a}}\right)_{\hat{a}}^{\hspace{0.1 cm}\hat{b}}\Pi_{\hat{b}}.
\end{equation}
So under $\mathcal{H}$ we have linear transformations of the Goldstones, unlike what happens for generic $\mathcal{G}$ transformations.\\[0.5 cm]
Finally, it is not possible to obtain an exact form for the transformation under $\mathcal{G}/\mathcal{H}$, but an infinitesimal expression can be determined instead.
Let us assume an infinitesimal element of $\mathcal{G}/\mathcal{H}$, which is defined by
$g_{\mathcal{G}/\mathcal{H}}\simeq\mathds{1}+i\alpha_{\hat{a}}\hat{T}^{\hat{a}}$, $U\left[\Pi\right]$ transforms as
\begin{equation}
\begin{split}
g_{\mathcal{G}/\mathcal{H}}U\left[\Pi\right]&=\left(\mathds{1}+i\alpha_{\hat{a}}\hat{T}^{\hat{a}}+\mathcal{O}\left(\alpha^{2}\right)\right)\left(\mathds{1}+i\frac{\sqrt{2}}{f}\Pi_{\hat{a}}\hat{T}^{\hat{a}}+\mathcal{O}\left(\frac{\Pi^{2}}{f}\right)\right)\\
&=\mathds{1}+i\frac{\sqrt{2}}{f}\hat{T}^{\hat{a}}\left(\Pi_{\hat{a}}+\frac{f}{\sqrt{2}}\alpha_{\hat{a}}+\mathcal{O}\left(\alpha\frac{\Pi^{2}}{f}+\alpha\frac{\Pi^{3}}{f^{2}}+\cdots\right)\right),
\end{split}
\end{equation}
from which we see that $\Pi$ transforms as 
\begin{equation}
\Pi_{\hat{a}}\rightarrow\Pi^{\left(g_{\mathcal{G}/\mathcal{H}}\right)}_{\hspace{1 cm}\hat{a}}=\Pi_{\hat{a}}+\frac{f}{\sqrt{2}}\alpha_{\hat{a}}+\mathcal{O}\left(\alpha\frac{\Pi^{2}}{f}+\alpha\frac{\Pi^{3}}{f^{2}}+\cdots\right).
\end{equation}
This ``shift symmetry" is responsible for the absence of non-derivative terms in the Goldstone Lagrangian.\\
With the above we see that in the CCWZ construction, $U\left[\Pi\right]$ is the main element to construct a Lagrangian 
invariant under $\mathcal{G}$. To introduce derivatives in the Lagrangian we introduce the $d$ and $e$ symbols through
the Maurer-Cartan one form, which is built with $U\left[\Pi\right]$,
that is to say
\begin{equation}\label{MaurerCartan}
iU\left[\Pi\right]^{-1}\partial_{\mu}U\left[\Pi\right]=d_{\mu,\hat{a}}\left[\Pi\right]\hat{T}^{\hat{a}}+e_{\mu,a}\left[\Pi\right]T^{a}\equiv d_{\mu}+e_{\mu}.
\end{equation}
Using (\ref{underg}) the one-form transforms as
\begin{equation}
\begin{split} 
iU\left[\Pi\right]^{-1}\partial_{\mu}U\left[\Pi\right]&\rightarrow ih\left[\Pi;g\right]U\left[\Pi\right]^{-1}g^{-1}\partial_{\mu}\left(gU\left[\Pi\right]h^{-1}\left[\Pi;g\right]\right)\\
&=h\left[\Pi;g\right]\left(iU\left[\Pi\right]^{-1}\partial_{\mu}U\left[\Pi\right]\right)h\left[\Pi;g\right]^{-1}+ih\left[\Pi;g\right]\partial_{\mu}h\left[\Pi;g\right]^{-1}.     
\end{split}
\end{equation}
In terms of $d$ and $e$ symbols, we have
\begin{equation}
\begin{split} 
iU\left[\Pi\right]^{-1}\partial_{\mu}U\left[\Pi\right]&\rightarrow h\left[\Pi;g\right]\left(d_{\mu}+e_{\mu}\right)h\left[\Pi;g\right]^{-1}+ih\left[\Pi;g\right]\partial_{\mu}h\left[\Pi;g\right]^{-1}\\
&=h\left[\Pi;g\right]d_{\mu}h\left[\Pi;g\right]^{-1}+h\left[\Pi;g\right]\left(e_{\mu}+i\partial_{\mu}\right)h\left[\Pi;g\right]^{-1}.
\end{split}
\end{equation}
The association of $e_{\mu}$ with the derivative is due to the fact that the term $ih\partial_{\mu}h^{-1}$ is a one form associated with $h$
and is decomposed on the $\mathcal{H}$ Lie algebra that has no components in the broken generators.
Thus, $d_{\mu}$ and $e_{\mu}$ transform under any group element $g\in\mathcal{G}$ as
\begin{equation}
\begin{split}
&d_{\mu}\rightarrow h\left[\Pi;g\right]d_{\mu}h\left[\Pi;g\right]^{-1}\\
&e_{\mu}\rightarrow h\left[\Pi;g\right]\left(e_{\mu}+i\partial_{\mu}\right)h\left[\Pi;g\right]^{-1}.
\end{split}
\end{equation}
Using (\ref{BCH}) the transformation of the $d$ symbol is written in components
\begin{equation}
d_{\mu,\hat{a}}\rightarrow d^{\left(g\right)}_{\hspace{0.4 cm}\mu,\hat{a}}=\left(e^{i\zeta_{a}\left[\Pi;g\right]t_{\pi}^{\hspace{0.1 cm}a}}\right)_{\hat{a}}^{\hspace{0.1 cm}\hat{b}}d_{\mu,\hat{b}}.
\end{equation}
Given the transformation of $d_{\mu}$ and using the cyclic property of the trace, we see that 
\begin{equation}
\begin{split} 
\Tr\left(d_{\mu}d^{\mu}\right)&\rightarrow\Tr\left(h\left[\Pi;g\right]d_{\mu}h\left[\Pi;g\right]^{-1}h\left[\Pi;g\right]d^{\mu}h\left[\Pi;g\right]^{-1}\right)\\
&=\Tr\left(h\left[\Pi;g\right]d_{\mu}d^{\mu}h\left[\Pi;g\right]^{-1}\right)\\
&=\Tr\left(h\left[\Pi;g\right]^{-1}h\left[\Pi;g\right]d_{\mu}d^{\mu}\right)\\
&=\Tr\left(d_{\mu}d^{\mu}\right).
\end{split}
\end{equation}
Since $d_{\mu,\hat{a}}$ transforms as the Goldstone bosons and $e_{\mu,a}$ transforms as a gauge field (with $\mathcal{H}$ as the local group gauge), 
we can construct invariant operators under $\mathcal{G}$ combining the symbols $d_{\mu,\hat{a}}$, $e_{\mu,a}$ and its derivatives. For example the  
lowest dimensional term without no other fields than the Goldstone bosons is \footnote{The Lagrangian of the dimension four involves the terms:
$\left[\Tr\left(d_{\mu}d^{\mu}\right)\right]^{2}$, $\Tr\left(d_{\mu}d_{\nu}d^{\mu}d^{\nu}\right)$, $\Tr\left(d_{\mu}d^{\mu}d_{\nu}d^{\nu}\right)$,
 $\Tr\left(d_{\mu}d^{\nu}\right)\cdot\Tr\left(d^{\mu}d_{\nu}\right)$.}
\begin{equation}\label{LowLagrangian}
\mathcal{L}^{\left(2\right)}=\frac{f^{2}}{4}d_{\mu}^{\hat{a}}d^{\mu}_{\hat{a}},
\end{equation}
which is of dimension two in energy because the $d_{\mu}$ symbol is of dimension one in energy, and the $ 1/4 $ factor allows us to obtain 
the canonical Goldstone kinetic term.
\chapter{The minimal composite Higgs model SO(5)/SO(4)}\label{ChapMCHM}
In this chapter we make use of the ideas of the composite Higgs scenario and the CCWZ construction. To have a model of EWSB we must take into 
account that we need four NGB in order to form a complex Higgs doublet, and that $G_{EW}$ must be contained in the subgroup $\mathcal{H}$. 
The pattern $\textup{SO}(N)\rightarrow\textup{SO}(N-1)$ provides $N-1$ NGB according to the Goldstone's theorem, so for 
$\textup{SO}(5)\rightarrow\textup{SO}(4)$ we get the four NGB needed to form the Higgs doublet. 
Since $\textup{SO}(4)$ is isomorphic to $\textup{SU}(2)_{L}\times\textup{SU}(2)_{R}$ \footnote{For a proof see \cite{Contino,Panico}}(both groups 
have the same number of generators and it can be shown that they have the same algebra at a local level), we can embed the $G_{EW}$ group 
in $\textup{SO}(4)$ by identifying the global $\textup{SU}(2)_{L}$ with the gauged one and the third generator of $\textup{SU}(2)_{R}$ 
with the hypercharge.
\section{Group generators and \texorpdfstring{$\boldsymbol{d_{\mu}}$}{1} symbol}
In the fundamental representation \textbf{5} of the $\textup{SO}(5)$ algebra the 10 generators are given by
\begin{align}
&T^{a}=\left\lbrace T_{L}^{\alpha}=\begin{pmatrix}
t_{L}^{\alpha} & 0 \\
0 & 0\end{pmatrix}, T_{R}^{\alpha}=\begin{pmatrix}
t_{R}^{\alpha} & 0 \\
0 & 0\end{pmatrix}\right\rbrace,\\[0.5 cm]
&\left(\hat{T}^{k}\right)_{IJ}=-\frac{i}{\sqrt{2}}\left(\delta^{k}_{I}\delta^{5}_{J}-\delta^{k}_{J}\delta^{5}_{I}\right),\label{generatorsbroken}
\end{align}
with the generators $t^{\alpha}_{L,R}$ defined by
\begin{equation}\label{subgenerators}
\begin{split}
&\left(t^{\alpha}_{L}\right)_{kj}=-\frac{i}{2}\left[\varepsilon_{\alpha\beta\gamma}\delta^{\beta}_{k}\delta^{\gamma}_{j}+\left(\delta^{\alpha}_{k}\delta^{4}_{j}-\delta^{\alpha}_{j}\delta^{4}_{k}\right)\right],\\
&\left(t^{\alpha}_{R}\right)_{kj}=-\frac{i}{2}\left[\varepsilon_{\alpha\beta\gamma}\delta^{\beta}_{k}\delta^{\gamma}_{j}-\left(\delta^{\alpha}_{k}\delta^{4}_{j}-\delta^{\alpha}_{j}\delta^{4}_{k}\right)\right],
\end{split}
\end{equation}
where the indices take values $\alpha,\beta,\gamma=1,2,3$ and $k,j=1,2,3,4$. With the broken generators Eq. (\ref{generatorsbroken}), we can calculate the Goldstone matrix, and using the CCWZ construction we compute 
the non-linear Lagrangian of order 2 Eq. (\ref{LowLagrangian}). With this purpose the Maurer-Cartan form (\ref{MaurerCartan}) is multiplied 
by $\hat{T}^{\hat{b}}$ and the trace is taken, that is to say
\begin{equation}
\begin{split} 
\Tr\left(iU\left[\Pi\right]^{-1}\partial_{\mu}U\left[\Pi\right]\hat{T}^{\hat{b}}\right)=d_{\mu,\hat{a}}\Tr\left[\hat{T}^{\hat{a}}\hat{T}^{\hat{b}}\right]+e_{\mu,a}\Tr\left[T^{a}\hat{T}^{\hat{b}}\right].  
\end{split}
\end{equation}
Using the fact that $\Tr\left[\hat{T}^{\hat{a}}\hat{T}^{\hat{b}}\right]=\delta^{\hat{a},\hat{b}}$ and $\Tr\left[T^{a}\hat{T}^{\hat{b}}\right]=0$,
we obtain
\begin{equation}\label{symbold}
d_{\mu,\hat{a}}=\Tr\left(iU\left[\Pi\right]^{-1}\partial_{\mu}U\left[\Pi\right]\hat{T}^{\hat{a}}\right).
\end{equation}
The computation of the NGB's matrix, its inverse and of $\partial_{\mu}U\left[\Pi\right]$ is presented in Appendix \ref{CalGoldstone}. After some algebra, we find
\begin{equation}\label{symboldfinal}
d_{\mu}^{i} =\sqrt{2}\left[\frac{1}{2}\left(\frac{1}{\Pi}\sin\frac{\Pi}{f}-\frac{1}{f}\right)\frac{\Pi^{i}}{\Pi^{2}}\partial_{\mu}\Pi^{2}-\frac{1}{\Pi}\sin\frac{\Pi}{f}\partial_{\mu}\Pi^{i}\right].            
\end{equation}
From (\ref{symboldfinal}), we can calculate the invariant term in the Lagrangian
\begin{equation}
d_{\mu}^{i}d^{\mu}_{i}=\frac{1}{2\Pi^{4}}\left(\frac{\Pi^{2}}{f^{2}}-\sin^{2}\frac{\Pi}{f}\right)\partial_{\mu}\Pi^{2}\partial^{\mu}\Pi^{2}+\frac{2}{\Pi^{2}}\sin^{2}\frac{\Pi}{f}\partial_{\mu}\boldsymbol{\Pi}^{T}\partial^{\mu}\boldsymbol{\Pi}.
\end{equation}
Finally, the 2-derivative non-linear Lagrangian is
\begin{equation}\label{FLagrangian}
\mathcal{L}^{\left(2\right)}=\frac{f^{2}}{4}d_{\mu,i}d^{\mu,i}=\frac{f^{2}}{2\Pi^{2}}\sin^{2}\frac{\Pi}{f}\partial_{\mu}\boldsymbol{\Pi}^{T}\partial^{\mu}\boldsymbol{\Pi}+\frac{f^{2}}{8\Pi^{4}}\left(\frac{\Pi^{2}}{f^{2}}-\sin^{2}\frac{\Pi}{f}\right)\partial_{\mu}\Pi^{2}\partial^{\mu}\Pi^{2}.
\end{equation}
This is a general equation which holds for any $\textup{SO}(N)\rightarrow\textup{SO}(N-1)$ breaking pattern.
\section{Gauge Lagrangian}
In the construction of the Gauge Lagrangian the ordinary derivative is promoted to covariant by gauging four generators of $\textup{SO}(4)$ (three for $t^{\alpha}_{L}$ and one for $t^{3}_{R}$). To such generators correspond 
the couplings $g$ and $g'$ respectively, and we can write
\begin{equation}
D_{\mu}\boldsymbol{\Pi}=\left(\partial_{\mu}-igW_{\mu,a}t_{L}^{a}-ig'B_{\mu}t^{3}_{R}\right)\boldsymbol{\Pi}.
\end{equation}
The fact that only 4 generators out of 6 are gauged causes the explicit breaking of the symmetry. We can identify the Higgs with the 
$\boldsymbol{\Pi}$ components by writing explicitly the \textbf{4} of $\textup{SO}(4)$ as a $\left(\boldsymbol{2},\boldsymbol{2}\right)$ 
of $\textup{SU}(2)_{L}\times\textup{SU}(2)_{R}$. This is done writing
\begin{equation}\label{correspondence}
\begin{split}
\Sigma&=\frac{1}{\sqrt{2}}\left(i\sigma_{\alpha}\Pi^{\alpha}+\mathds{1}_{2}\Pi^{4}\right)=\frac{1}{\sqrt{2}}\bar{\sigma}_{j}\Pi^{j};\hspace{1 cm}\bar{\sigma}_{j}=\{i\sigma_{\alpha},\mathds{1}_{2}\}.\\
&=\frac{1}{\sqrt{2}}\begin{pmatrix}\Pi^{4}+i\Pi^{3} & \Pi^{2}+i\Pi^{1}\\
-\Pi^{2}+i\Pi^{1} & \Pi^{4}-i\Pi^{3}\end{pmatrix},
\end{split}
\end{equation}
where $\Sigma$ is pseudo-real matrix and $\sigma_{\alpha}$ are the Pauli's matrix. In this $\left(\boldsymbol{2},\boldsymbol{2}\right)$ representation of the 4 components of $\boldsymbol{\Pi}$, we have a Higgs doublet $H$ with 1/2 of hypercharge, its components being
\begin{equation}\label{DHiggs}
H=\begin{pmatrix}h_{u}\\h_{d}\end{pmatrix}
=\frac{1}{\sqrt{2}}\begin{pmatrix}\Pi^{2}+i\Pi^{1}\\\Pi^{4}-i\Pi^{3}\end{pmatrix}.
\end{equation}
Notice that
\begin{equation}
i\sigma_{2}H^{*}=H^{c}=\frac{1}{\sqrt{2}}\begin{pmatrix}\Pi^{4}+i\Pi^{3}\\-\Pi^{2}+i\Pi^{1}\end{pmatrix},
\end{equation}
then
\begin{equation}
\Sigma=\left(H^{c},H\right).
\end{equation}
With the correspondence between the four components of the real vector $\boldsymbol{\Pi}$ and the complex Higgs doublet, from the expression (\ref{DHiggs}) we have
\begin{equation}
\begin{split}
&D_{\mu}H^{\dagger}D^{\mu}H=\frac{1}{2}D_{\mu}\Pi_{j}D^{\mu}\Pi^{j}=\frac{1}{2}D_{\mu}\boldsymbol{\Pi}^{T}\cdot D^{\mu}\boldsymbol{\Pi},\\
&H^{\dagger}H=|H|^{2}=\frac{1}{2}\Pi_{j}\Pi^{j}=\frac{1}{2}\Pi^{2},
\end{split}
\end{equation}
in such a way that the Lagrangian (\ref{FLagrangian}) becomes
\begin{equation}\label{FFLagrangian}
\mathcal{L}^{\left(2\right)}=\frac{f^{2}}{2|H|^{2}}\sin^{2}\frac{\sqrt{2}|H|}{f}D_{\mu}H^{\dagger}D^{\mu}H+\frac{f^{2}}{8|H|^{4}}\left(2\frac{|H|^{2}}{f^{2}}-\sin^{2}\frac{\sqrt{2}|H|}{f}\right)\left(\partial_{\mu}|H|^{2}\right)^{2},
\end{equation}
where the covariant derivative of the $H$ doublet is given in Eq. (\ref{DcovHiggs}) in the Appendix. Yang-Mills kinetic terms are also introduced
in the theory by the following Lagrangian
\begin{equation}
\mathcal{L}_{G}=-\frac{1}{4}W^{a}_{\mu\nu}W_{a}^{\mu\nu}-\frac{1}{4}B_{\mu\nu}B^{\mu\nu}.
\end{equation}
From now on, we will assume the Higgs doublet to develop a VEV; a sketch of the properties of the Higgs potential will be presented in Chapter 6.
Using the unitary gauge defined in Appendix \ref{SM} we can determine the implications of (\ref{FFLagrangian}). The unitary gauge for the Higgs
doublet is in this case 
\begin{equation}\label{unitary}
H=\frac{1}{\sqrt{2}}\begin{pmatrix}0\\V+h\left(x\right)\end{pmatrix},
\end{equation}
where $V$ is the vacuum expectation value (VEV) of the Higgs and $h(x)$ is the quantum fluctuation (physical Higgs) around the VEV.  
The unitary gauge implies
\begin{equation}
|H|^{2}=\frac{1}{2}\left(V+h\right)^{2},\hspace{0.5 cm}\partial_{\mu}|H|^{2}=\left(V+h\right)\partial_{\mu}h,
\end{equation}
and we obtain
\begin{equation}
\begin{split}
&D_{\mu}H^{\dagger}D^{\mu}H=\frac{1}{2}\left[\partial_{\mu}h\partial^{\mu}h+\frac{g^{2}}{2}\left(V+h\right)^{2}W_{\mu}^{-}W_{+}^{\mu}+\frac{g_{z}^{2}}{4}\left(V+h\right)^{2}Z_{\mu}Z^{\mu}\right],\\
&g_{z}^{2}=g^{2}+g'^{2}.     
\end{split}
\end{equation}
Writing Eq. (\ref{FFLagrangian}) in the unitary gauge, we obtain
\begin{align}
\notag
\mathcal{L}^{\left(2\right)}&=\frac{f^{2}}{2\left(V+h\right)^{2}}\sin^{2}\frac{V+h}{f}\left[\partial_{\mu}h\partial^{\mu}h+\frac{g^{2}}{2}\left(V+h\right)^{2}W_{\mu}^{-}W_{+}^{\mu}+\frac{g_{z}^{2}}{4}\left(V+h\right)^{2}Z_{\mu}Z^{\mu}\right]\\
&+\frac{f^{2}}{2\left(V+h\right)^{2}}\left[\frac{\left(V+h\right)^{2}}{f^{2}}-\sin^{2}\frac{V+h}{f}\right]\partial_{\mu}h\partial^{\mu}h\\\notag
&=\frac{1}{2}\partial_{\mu}h\partial^{\mu}h+\frac{g^{2}}{4}f^{2}\sin^{2}\frac{V+h}{f}\left[W_{\mu}^{-}W^{\mu}_{+}+\frac{1}{2c_{w}^{2}}Z_{\mu}Z^{\mu}\right],
\end{align}
where $W$ and $Z$ are the SM gauge bosons, $c_{w}$ is the cosine of the Weinberg angle, which relates the couplings $g$ and $g'$ by
$\cos\theta_{w}=g/\sqrt{g^{2}+g'^{2}}$. We can immediately read the gauge bosons masses, which result 
\begin{equation}\label{masa}
m_{W}=c_{w}m_{Z}=\frac{1}{2}gf\sin\frac{V}{f}\equiv\frac{1}{2}gv.
\end{equation}
From low energy experiments, in particular the muon decay rate, we can measure the Fermi coupling constant with great accuracy, 
$G_{F}=1.1663787(6)\times 10^{-5}\textup{GeV}^{-2}$ \cite{PDG}. Since we know that in the Fermi theory 
$G_{F}=\sqrt{2}g^{2}/8m_{W}^{2}$ we have, $v=\left(\sqrt{2}G_{F}\right)^{-1/2}\simeq 246\hspace{0.1cm}\textup{GeV}$.\\
Recalling that $\rho=m^{2}_{W}/(c_{w}m_{z})^{2}$, from Eq. (\ref{masa}), we see that in this model the tree-level relation $\rho=1$ is respected.
The reason for that is that the custodial symmetry is still present when EWSB occurs, since the VEV causes 
$\textup{SO}(4)\rightarrow\textup{SO}(3)\simeq\textup{SU}(2)$ (see Appendix \ref{Custodial}).\\
Given the non-linear form of the Lagrangian, there are infinite interactions between the gauge fields and the Higgs field. 
We can see this by expanding in Taylor series around $h=0$
\begin{equation}
\begin{split}
f^{2}\sin^{2}\frac{V+h}{f}\approx &f^{2}\sin^{2}\frac{V}{f}+\left(2fh-\frac{4h^{3}}{3f}\right)\sin\frac{V}{f}\cos\frac{V}{f}\\
&+\left(h^{2}-\frac{h^{4}}{3f^{2}}\right)\left(1-2\sin^{2}\frac{V}{f}\right)+\cdots.\\
\end{split}
\end{equation}
Using the definition Eq. (\ref{gap}), the expansion is written as
\begin{equation}
\begin{split}
f^{2}\sin^{2}\frac{V+h}{f}\approx&v^{2}\bigg\{1+2\sqrt{1-\xi}\frac{h}{v}+\left(1-2\xi\right)\frac{h^{2}}{v^{2}}\\
&-\frac{4}{3}\xi\sqrt{1-\xi}\frac{h^{3}}{v^{3}}-\left(1-2\xi\right)\frac{\xi}{3}\frac{h^{4}}{v^{4}}+\cdots\bigg\}.
\end{split}
\end{equation}
Finally we have
\begin{equation}\label{finalLagrangian}
\begin{split}
\mathcal{L}^{\left(2\right)}=&\frac{1}{2}\partial_{\mu}h\partial^{\mu}h+\frac{g^{2}}{4}v^{2}\left[W_{\mu}^{-}W^{\mu}_{+}+\frac{1}{2c_{w}^{2}}Z_{\mu}Z^{\mu}\right]\\
&+\frac{g^{2}}{4}v^{2}\left[W_{\mu}^{-}W^{\mu}_{+}+\frac{1}{2c_{w}^{2}}Z_{\mu}Z^{\mu}\right]\bigg\{2\sqrt{1-\xi}\frac{h}{v}+\left(1-2\xi\right)\frac{h^{2}}{v^{2}}\\
&-\frac{4}{3}\xi\sqrt{1-\xi}\frac{h^{3}}{v^{3}}-\left(1-2\xi\right)\frac{\xi}{3}\frac{h^{4}}{v^{4}}+\cdots\bigg\}.
\end{split}
\end{equation}
We see from Eq. (\ref{finalLagrangian}) that the couplings of the composite Higgs to the gauge boson are modified with respect to the SM by
\begin{equation}\label{compa}
\frac{g^{CH}_{hVV}}{g^{SM}_{hVV}}=\sqrt{1-\xi},\hspace{0.5 cm}\frac{g^{CH}_{hhVV}}{g^{SM}_{hhVV}}=1-2\xi.
\end{equation}
When $\xi\rightarrow 0$, we recover the couplings of the SM and the interactions of the Higgs with the gauge bosons with powers of 
$h\geq 3$ are eliminated. In this limit the composite Higgs is considered as elementary. Given that the current measurements of the Higgs 
interactions at the LHC allow for deviations from the SM coupling up to about 20$\%$, we get from Eq. (\ref{compa}) the bound $\xi\leq0.4$.
\section{Fermions in the model}\label{FerMCHM}
To implement the fermions in the model and their interactions with the Higgs, the so-called ``partial compositeness'' ansatz is used \cite{KaplanParcial}, where the fermions are incorporated as elementary fields external to the composite sector.
The terms of interaction are linear in the fermions and the composite sector operators $\mathcal{O}^{L,R}_{F}$ are fermionic. 
The interactions are given by
\begin{equation}
\mathcal{L}_{\textup{Int}}=\lambda_{t_{L}}\overline{q}_{L}\mathcal{O}^{L}_{F}+\lambda_{t_{R}}\overline{t}_{R}\mathcal{O}^{R}_{F}+\cdots,
\end{equation}
plus similar terms for the other quarks. However, there is one detail to be considered; the representations of the fermionic operators must be 
specified under the global group $\mathcal{G}$ since the elementary/composite interactions occur at a large scale, far from where the spontaneous 
breaking $\mathcal{G}\rightarrow\mathcal{H}$ occurs. At that scale the operators are classified in multiplets of the unbroken group. This reflects 
on the ambiguity of choosing the $\textup{SO}(5)$ representations of $\mathcal{O}_{F}^{L,R}$ in the construction of the model. From now on during 
this discussion, the fundamental representation of the global group will be used\footnote{The embedding in the spinorial \textbf{4} is presented 
in \cite{Agashe}. Some comments on the representations \textbf{10} and \textbf{14} are given in \cite{Panico}}.\\
Under $G_{EW}$ the multiplet must contain the representations of the quarks of the standard model, $\textbf{2}_{1/6}$, $\textbf{1}_{2/3}$ 
and $\textbf{1}_{-1/3}$ (see Appendix \ref{SM}). But if $G_{EW}$ is completely incorporated in $\textup{SO}(5)$, the representations of the quarks 
do not exist because the hypercharge of the SM fermions is not reproduced. Therefore an extension of the symmetry group is required, and it is
achieved by adding a new factor $\textup{U}(1)_{X}$ which is unbroken, so the breaking pattern\footnote{The unbroken color $\textup{SU}(3)_{c}$ 
group must also be added. Then the complete group is\\
$\textup{SO}(5)\times\textup{U}(1)_{X}\times\textup{SU}(3)_{c}$.} will be
\begin{equation}
\textup{SO}\left(5\right)\times\textup{U}\left(1\right)_{X}\rightarrow\textup{SO}\left(4\right)\times\textup{U}\left(1\right)_{X}.
\end{equation}
The hypercharge is now defined as
\begin{equation}\label{hypercharge}
Y=T^{3}_{R}+X.
\end{equation}
Choosing $X=2/3$ as $\textup{U}(1)_{X}$ charge, the fundamental representation \textbf{5} of $\textup{SO}(5)$ decomposes as follows under $\textup{SO}(4)$ and $\textup{SU}(2)_{L}\times\textup{U}(1)_{Y}$
\begin{equation}
\textbf{5}_{2/3}\rightarrow\textbf{4}_{2/3}\oplus\textbf{1}_{2/3}\rightarrow\textbf{2}_{7/6}\oplus\textbf{2}_{1/6}\oplus\textbf{1}_{2/3},
\end{equation}
where the subscripts in the second term is the $X$ charge, while it is the hypercharge in the last term. As it can be seen, the last two terms can 
be identified with $q_{L}$ and $t_{R}$. The embedding of $t_{R}$ is
\begin{equation}
T_{R}=\{0,0,0,0,t_{R}\}^{T}=\{\textbf{0},t_{R}\}^{T},
\end{equation}
so the interaction is conveniently expressed as
\begin{equation}\label{tR}
\mathcal{L}^{t_{R}}_{\textup{Int}}=\lambda_{t_{R}}\overline{t}_{R}\left(\mathcal{O}_{F}^{R}\right)_{5}+h.c.=\lambda_{t_{R}}\left(\overline{T}_{R}\right)^{I}\left(\mathcal{O}_{F}^{R}\right)_{I}+h.c.
\end{equation}
For such interaction to be formally invariant, $T_{R}$ must transform like $\mathcal{O}^{R}_{F}$ under $SO(5)$, that is to say
\begin{equation}
\left(T_{R}\right)_{I}\rightarrow g_{I}^{\hspace{0.1 cm}J}\left(T_{R}\right)_{J}.
\end{equation}
To convert the source $T_{R}$ to a multiplets of $\textup{SO}(4)$, we ``dress" the source with $U[\Pi]^{-1}$, defining
\begin{equation}
\{T_{R}^{\textbf{4}},T_{R}^{\textbf{1}}\}^{T}=U\left[\Pi\right]^{-1}\cdot T_{R}.
\end{equation}
Using equation (\ref{InvGoldstone}) of the Appendix, we obtain
\begin{equation}
U\left[\Pi\right]^{-1}\cdot T_{R}=t_{R}\left[-\frac{\boldsymbol{\Pi}}{\Pi}\sin\frac{\Pi}{f},\cos\frac{\Pi}{f}\right]^{T},
\end{equation}
from which
\begin{equation}
T_{R}^{\textbf{4}}=-t_{R}\frac{\boldsymbol{\Pi}}{\Pi}\sin\frac{\Pi}{f},\hspace{1 cm}T_{R}^{\textbf{1}}=t_{R}\cos\frac{\Pi}{f}.
\end{equation}
In this way we have two different representations that belong to $\textup{SO}(4)$, namely 
\begin{equation}
T_{R}^{\textbf{4}}\in\textbf{4}_{2/3},\hspace{0.5 cm}T_{R}^{\textbf{1}}\in\textbf{1}_{2/3}.
\end{equation}
In the same way we write the $q_{L}$ interaction as
\begin{equation}\label{qlint}
\mathcal{L}^{q_{L}}_{\textup{Int}}=\lambda_{t_{L}}\left(\overline{Q}_{t_{L}}\right)^{I}\left(\mathcal{O}_{F}^{L}\right)_{I}+h.c.
\end{equation}
Now to determine the form of $Q_{t_{L}}$, we can proceed as in Eq. (\ref{correspondence}), writing
\begin{equation}\label{psi}
\begin{split}
\Psi&=\frac{1}{\sqrt{2}}\left(\psi^{4}+i\sigma_{\alpha}\psi^{\alpha}\right)=\frac{1}{\sqrt{2}}\overline{\sigma}_{j}\psi^{j}\\
&=\frac{1}{\sqrt{2}}\begin{pmatrix}\psi^{4}+i\psi^{3} & i\psi^{1}+\psi^{2}\\
i\psi^{1}-\psi^{2} & \psi^{4}-i\psi^{3}\end{pmatrix}\equiv\begin{pmatrix}\Psi_{-}^{u}&\Psi_{+}^{u}\\\Psi_{-}^{d}&\Psi_{+}^{d}\end{pmatrix}.
\end{split}
\end{equation}
The $\Psi_{-}$ and $\Psi_{+}$ fields are $\textup{SU}(2)_{L}$  doublets with hypercharge $\mp 1/2$ respectively.
From Eq. (\ref{psi}) we get the up and down components of the two doublets in terms of the fourplet fields, which are
\begin{equation}
\begin{split} 
&\psi_{1}=-\frac{i}{\sqrt{2}}\left(\Psi_{+}^{u}+\Psi_{-}^{d}\right),\hspace{0.5 cm}\psi_{2}=\frac{1}{\sqrt{2}}\left(\Psi_{+}^{u}-\Psi_{-}^{d}\right)\\  
&\psi_{3}=\frac{i}{\sqrt{2}}\left(\Psi_{+}^{d}-\Psi_{-}^{u}\right),\hspace{0.8 cm}\psi_{4}=\frac{1}{\sqrt{2}}\left(\Psi_{+}^{d}+\Psi_{-}^{u}\right).
\end{split}
\end{equation}
Then the fourplet components are written in terms of $\Psi_{\pm}^{u,d}$ as
\begin{equation}\label{cuadrupleto}
\boldsymbol{\psi}=\frac{1}{\sqrt{2}}\{-i\Psi_{+}^{u}-i\Psi_{-}^{d},\Psi_{+}^{u}-\Psi_{-}^{d},i\Psi_{+}^{d}-i\Psi_{-}^{u},\Psi_{+}^{d}+\Psi_{-}^{u}\}^{T}.
\end{equation}
Since the $q_{L}$ doublet has hypercharge $1/6$, we project the $Q_{t_{L}}$ source on the $T^{3}_{R}=-1/2$ doublet (see Eq. (\ref{hypercharge})),
that is to say $q_{L}=\Psi_{-}$ ($\Psi_{+}^{u}=\Psi_{+}^{d}=0$). So the $q_{L}$ doublet is embedded in the fourplet as
\begin{equation}
\boldsymbol{q}_{L_{-}}=\frac{1}{\sqrt{2}}\{-ib_{L},-b_{L},-it_{L},t_{L}\}^{T},
\end{equation}
and the multiplet is given by
\begin{equation}
Q_{t_{L}}=\{\boldsymbol{q}_{L_{-}},0\}^{T}. 
\end{equation}
To convert the $\textup{SO}(5)$ representation to a $\textup{SO}(4)$ representation, we dress $Q_{t_{L}}$ with $U[\Pi]^{-1}$, obtaining
\begin{equation}
\{Q_{t_{L}}^{\textbf{4}},Q_{t_{L}}^{\textbf{1}}\}^{T}=U\left[\Pi\right]^{-1}\cdot Q_{t_{L}}=
\begin{pmatrix}\left[\mathds{1}-\left(1-\cos\frac{\Pi}{f}\right)\frac{\boldsymbol{\Pi}\cdot\boldsymbol{\Pi}^{T}}{\Pi^{2}}\right]\cdot\boldsymbol{q}_{L_{-}}\\
\left(\frac{\boldsymbol{\Pi}^{T}}{\Pi}\sin\frac{\Pi}{f}\right)\cdot\boldsymbol{q}_{L_{-}}\end{pmatrix},
\end{equation}
where
\begin{equation}
Q_{t_{L}}^{\textbf{4}}=\left[\mathds{1}-\left(1-\cos\frac{\Pi}{f}\right)\frac{\boldsymbol{\Pi}\cdot\boldsymbol{\Pi}^{T}}{\Pi^{2}}\right]\cdot\boldsymbol{q}_{L_{-}},\hspace{1 cm}Q_{t_{L}}^{\textbf{1}}=\left(\frac{\boldsymbol{\Pi}^{T}}{\Pi}\sin\frac{\Pi}{f}\right)\cdot\boldsymbol{q}_{L_{-}}.
\end{equation}
So we get two new objects of $\textup{SO}(4)$, namely 
\begin{equation}
Q_{t_{L}}^{\textbf{4}}\in\textbf{4}_{2/3},\hspace{0.5 cm}Q_{t_{L}}^{\textbf{1}}\in\textbf{1}_{2/3}.
\end{equation}
From the contractions of $Q_{t_{L}}^{\textbf{4}}$ with $T_{R}^{\textbf{4}}$ and $Q_{t_{L}}^{\textbf{1}}$ with $T_{R}^{\textbf{1}}$ 
two invariants can be formed. However, these invariants are not independent, as it can be easily seen from
\begin{equation}\label{relinv}
\left(\overline{Q}^{\textbf{4}}_{t_{L}}\right)^{j}\left(T^{\textbf{4}_{R}}\right)_{j}+\overline{Q}_{t_{L}}^{\textbf{1}}T_{R}^{\textbf{1}}=\left(\overline{Q}_{t_{L}}\right)^{I}\left(T_{R}\right)_{I}=0,
\end{equation}
since the embeddings are orthogonal. The simplest invariant can thus be obtained from the singlets only,
\begin{equation}
\begin{split}
\overline{Q}_{t_{L}}^{\textbf{1}}T_{R}^{\textbf{1}}&=\cos\frac{\Pi}{f}\sin\frac{\Pi}{f}\overline{\boldsymbol{q}}_{L_{-}}\cdot\frac{\boldsymbol{\Pi}}{\Pi}t_{R}=\frac{1}{2\Pi}\sin\frac{2\Pi}{f}\left(\overline{t}_{L},\overline{b}_{L}\right)\frac{1}{\sqrt{2}}\begin{pmatrix}\Pi_{4}+i\Pi_{3}\\i\Pi_{1}-\Pi_{2}\end{pmatrix}t_{R}\\
&=\frac{1}{2\Pi}\sin\frac{2\Pi}{f}\overline{q}_{L}H^{c}t_{R}=\frac{1}{2\sqrt{2}|H|}\sin\frac{2\sqrt{2}|H|}{f}\overline{q}_{L}H^{c}t_{R}.
\end{split}
\end{equation}
We thus obtain a generalized top Yukawa Lagrangian of the form
\begin{equation}\label{YukLagrangian}
\begin{split}
\mathcal{L}^{t}_{\textup{Yuk}}&=-c^{t}\frac{\lambda_{t_{L}}\lambda_{t_{R}}}{g^{2}_{\ast}}m_{\ast}\overline{Q}_{t_{L}}^{\textbf{1}}T_{R}^{\textbf{1}}+h.c.\\
&=-c^{t}\frac{\lambda_{t_{L}}\lambda_{t_{R}}}{g^{2}_{\ast}}m_{\ast}\frac{1}{2\sqrt{2}|H|}\sin\frac{2\sqrt{2}|H|}{f}\overline{q}_{L}H^{c}t_{R}+h.c.,
\end{split}
\end{equation}
where the couplings $\lambda_{t_{L}}$ and $\lambda_{t_{R}}$ represent the interaction of the sources $Q_{L}$ and $T_{R}$ with the composite sector.
The scale of the composite sector is given by $m_{\ast}$, which guarantees the correct dimensionality of the operator, and we add a free parameter
$c^{t}$, with $\lambda_{t_{L}}$, $\lambda_{t_{R}}$ and $c^{t}\in R$.\\Now, when the Higgs is set to its VEV we obtain
\begin{equation}
\begin{split} 
\mathcal{L}_{\textup{Mass}}&=-c^{t}\frac{\lambda_{t_{L}}\lambda_{t_{R}}}{g^{2}_{\ast}}m_{\ast}\frac{1}{2V}\sin\frac{2V}{f}\left(\overline{t}_{L},\overline{b}_{L}\right)
\begin{pmatrix}\frac{V}{\sqrt{2}} \\ 0\end{pmatrix}t_{R}+h.c\\
&=-c^{t}\frac{\lambda_{t_{L}}\lambda_{t_{R}}}{g^{2}_{\ast}}m_{\ast}\frac{1}{\sqrt{2}}\sin\frac{V}{f}\cos\frac{V}{f}\overline{t}_{L}t_{R}+h.c\\
&=-c^{t}\frac{\lambda_{t_{L}}\lambda_{t_{R}}}{g^{2}_{\ast}}m_{\ast}\sqrt{\frac{\xi\left(1-\xi\right)}{2}}\overline{t}t,
\end{split}
\end{equation}
from which we can read the top mass
\begin{equation}
m_{t}=c^{t}\frac{\lambda_{t_{L}}\lambda_{t_{R}}}{g^{2}_{\ast}}m_{\ast}\sqrt{\frac{\xi\left(1-\xi\right)}{2}}.
\end{equation}
Using this expression for $m_{t}$ and the unitary gauge, the Yukawa Lagrangian of Eq. (\ref{YukLagrangian}) becomes
\begin{equation}
\begin{split}
\mathcal{L}^{t}_{\textup{Yuk}}&=-c^{t}\frac{\lambda_{t_{L}}\lambda_{t_{R}}}{g^{2}_{\ast}}m_{\ast}\frac{1}{2\left(V+h\right)}\sin\frac{2\left(V+h\right)}{f}\left(\overline{t}_{L},\overline{b}_{L}\right)\begin{pmatrix}\frac{V+h}{\sqrt{2}} \\ 0\end{pmatrix}t_{R}+h.c\\
&=-c^{t}\frac{\lambda_{t_{L}}\lambda_{t_{R}}}{g^{2}_{\ast}}m_{\ast}\frac{1}{2\sqrt{2}}\sin\frac{2\left(V+h\right)}{f}\overline{t}_{L}t_{R}+h.c\\
&=-\frac{m_{t}}{2}\frac{1}{\sqrt{\xi\left(1-\xi\right)}}\sin\frac{2\left(V+h\right)}{f}\overline{t}t.
\end{split}
\end{equation}
Now by Taylor-expanding the function $\sin\frac{2\left(V+h\right)}{f}$ around $h=0$, that is to say
\begin{equation}
\begin{split}
\sin\frac{2\left(V+h\right)}{f}&\simeq 2\sin\frac{V}{f}\cos\frac{V}{f}+\frac{2h}{f}\left(1-2\sin^{2}\frac{V}{f}\right)-\frac{8}{2!f^{2}}\sin\frac{V}{f}\cos\frac{V}{f}+\cdots\\
&\simeq 2\sqrt{\xi\left(1-\xi\right)}+2\frac{h}{v}\left(1-2\xi\right)\sqrt{\xi}-4\frac{h^{2}}{v^{2}}\sqrt{\xi\left(1-\xi\right)}\xi+\cdots,
\end{split}
\end{equation}
we get
\begin{equation}
\mathcal{L}^{t}_{\textup{Yuk}}=-m_{t}\overline{t}t-k_{t}^{\textbf{5}}\frac{m_{t}}{v}h\overline{t}t-c_{2}^{\textbf{5}}\frac{m_{t}}{v^{2}}h^{2}\overline{t}t+\cdots
\end{equation}
In addition to the top mass term, there are different top interactions with the Higgs, being the first one similar to the SM one. The difference is the modified coupling strength $k_{t}^{\textbf{5}}$. The second term with the  coefficient $c_{2}^{\textbf{5}}$  
represents the interaction of the top with two Higgses. Producing a five dimension vertex, such interaction is not present in the SM. 
The superscript \textbf{5} accounts for the representation of embedding the operators. The coefficients are given by
\begin{equation}\label{coeff}
k_{t}^{\textbf{5}}=\frac{1-2\xi}{\sqrt{1-\xi}},\hspace{0.5 cm}c_{2}^{\textbf{5}}=-2\xi.
\end{equation}
In the limit $\xi\rightarrow 0$, the couplings reduce to the SM, namely $k_{t}^{\textbf{5}}\rightarrow 1$ and 
$c_{2}^{\textbf{5}}\rightarrow 0$.\\[0.5 cm]
For the bottom quark sector the procedure is completely analogous. Choosing $X=-1/3$, $\textbf{5}_{-1/3}$ decomposes under $\textup{SO}(4)\times\textup{U}(1)_{X}$ and $\textup{SU}(2)_{L}\times\textup{U}(1)_{Y}$ as
\begin{equation}
\textbf{5}_{-1/3}\rightarrow\textbf{4}_{-1/3}\oplus\textbf{1}_{-1/3}\rightarrow\textbf{2}_{1/6}\oplus\textbf{2}_{-5/6}\oplus\textbf{1}_{-1/3}.
\end{equation}
The interactions are written as
\begin{equation}
\mathcal{L}^{b}_{\textup{Int}}=\lambda_{b_{L}}\left(\overline{Q}_{b_{L}}\right)^{I}\left(\mathcal{O}_{F}^{b_{L}}\right)_{I}+\lambda_{b_{R}}\left(\overline{B}_{R}\right)^{I}\left(\mathcal{O}_{F}^{b_{R}}\right)_{I},
\end{equation}
where $\mathcal{O}_{F}^{b^{L}}$ and $\mathcal{O}_{F}^{b^{R}}$ are respectively in the $\boldsymbol{2}_{1/6}$ and $\boldsymbol{1}_{-1/3}$ 
of the SM group. An embedding of $b_{R}$ is made in the \textbf{5} of $\textup{SO}(5)$
\begin{equation}
B_{R}=\{0,0,0,0,b_{R}\}^{T},
\end{equation}
and unlike what happens in the top sector, the source $Q_{b_{L}}$ is chosen to project on the $T^{3}_{R}=1/2$ doublet, namely 
$q_{L}=\Psi_{+}$ ($\Psi_{-}^{u}=\Psi_{-}^{d}=0$), so that
\begin{equation}
\boldsymbol{q}_{L_{+}}=\frac{1}{\sqrt{2}}\{-it_{L},t_{L},ib_{L},b_{L}\}^{T}.
\end{equation}
Then the new source fields are given by
\begin{equation}
\begin{split}
&Q_{b_{L}}=\{\boldsymbol{q}_{L_{+}},0\}^{T}\\
&B_{R}=\{\textbf{0},b_{R}\}^{T}.
\end{split}
\end{equation}
Performing the dressing process with $U[\Pi]^{-1}$, we obtain the following multiples of $\textup{SO}(4)$
\begin{equation}
\begin{split} 
&Q^{\textbf{4}}_{b_{L}}=\left[\mathds{1}-\left(1-\cos\frac{\Pi}{f}\right)\frac{\boldsymbol{\Pi}\cdot\boldsymbol{\Pi}^{T}}{\Pi^{2}}\right]\cdot\boldsymbol{q}_{L_{+}},\hspace{0.5 cm}Q^{\textbf{1}}_{b_{L}}=\sin\frac{\Pi}{f}\frac{\boldsymbol{\Pi}^{T}}{\Pi}\cdot\boldsymbol{q}_{L_{+}},\\    
&B^{\textbf{4}}_{R}=-b_{R}\frac{\boldsymbol{\Pi}}{\Pi}\sin\frac{\Pi}{f},\hspace{4.45 cm}B^{\textbf{1}}_{R}=b_{R}\cos\frac{\Pi}{f}.
\end{split}
\end{equation}
With these multiplets we can form invariants. A generalized down-type Yukawa Lagrangian is given by
\begin{equation}\label{Lagrangianbottom}
\begin{split}
\mathcal{L}^{b}_{\textup{Yuk}}&=-c^{b}\frac{\lambda_{b_{L}}\lambda_{b_{R}}}{g_{\ast}^{2}}m_{\ast}Q^{\textbf{1}}_{b_{L}}B^{\textbf{1}}_{R}+h.c.\\
&=-c^{b}\frac{\lambda_{b_{L}}\lambda_{b_{R}}}{g_{\ast}^{2}}m_{\ast}\frac{1}{2\sqrt{2}|H|}\sin\frac{2\left(V+h\right)}{f}\overline{q}_{L}Hb_{R}+h.c.
\end{split}
\end{equation}
Setting the Higgs to its VEV, Eq. (\ref{Lagrangianbottom}) becomes
\begin{equation}
\begin{split}
\mathcal{L}_{\textup{Mass}}&=-c^{b}\frac{\lambda_{b_{L}}\lambda_{b_{R}}}{g_{\ast}^{2}}m_{\ast}\frac{1}{\sqrt{2}}\sin\frac{V}{f}\cos\frac{V}{f}\overline{b}_{L}b_{R}+h.c\\
&=-c^{b}\frac{\lambda_{b_{L}}\lambda_{b_{R}}}{g_{\ast}^{2}}m_{\ast}\sqrt{\frac{\xi\left(1-\xi\right)}{2}}\overline{b}b=-m_{b}\overline{b}b,
\end{split}
\end{equation}
where the bottom mass is
\begin{equation}
m_{b}=c^{t}\frac{\lambda_{b_{L}}\lambda_{b_{R}}}{g^{2}_{\ast}}m_{\ast}\sqrt{\frac{\xi\left(1-\xi\right)}{2}}.
\end{equation}
After going to the unitary gauge and doing a Taylor expansion around $h=0$, we have
\begin{equation}
\begin{split} 
\mathcal{L}^{b}_{\textup{Yuk}}=-m_{b}\overline{b}b-k_{b}^{\textbf{5}}\frac{m_{b}}{v}h\overline{b}b+\cdots
\end{split}
\end{equation}
Unlike in the SM case, there is a modified coupling, $k_{b}^{\textbf{5}}$, in the bottom-Higgs interaction given by
\begin{equation}
k_{b}^{\textbf{5}}=\frac{1-2\xi}{\sqrt{1-\xi}}.
\end{equation}
The vertices of dimensions greater than or equal to 5 are suppressed due to the small bottom mass.
The model discussed in this chapter is called the ``Minimal Composite Higgs Model'' in the representation \textbf{5} (MCHM$_{\textbf{5}}$), since 
it provides the minimum number of pNGB Higgs fields, besides obeying the custodial symmetry\footnote{Another minimal possibility would be
$\textup{SU}(3)\rightarrow\textup{SU}(2)\times\textup{U}(1)$, but it lacks the custodial symmetry so it must be discarded.}.
\chapter{The Non-Minimal Composite Higgs Model SO(6)/SO(5)}\label{CNMCHM}
In the last chapter, we applied the CCWZ formalism to the construction of the MCHM. In this chapter we will instead deal 
with the pattern of spontaneous symmetry breaking $\textup{SO(6)}\rightarrow\textup{SO(5)}$, which provides five NGB's in the coset 
$\textup{SO(6)}/\textup{SO(5)}$. In addition to the four NGB's needed to form the Higgs doublet, we have an additional gauge singlet scalar field, 
which under certain conditions has been used as Dark Matter candidate \cite{Frigerio}. These types of models are known as non-minimal and denoted 
by NMCHM \cite{Gripaios}.
\section{From NBG's to Higgs doublet}
As the $\mathcal{L}^{\left(2\right)}$ Lagrangian of Eq. (\ref{FLagrangian}) is valid for any coset $\textup{SO}(N)/\textup{SO}(N-1)$, we have in 
this case that the NGB's vector of Eq. (\ref{FLagrangian}), is given by
\begin{equation}
\begin{split}
&\boldsymbol{\Pi}^{T}=\left[\Pi_{1},\Pi_{2},\Pi_{3},\Pi_{4},\zeta\right]^{T}=\left[\boldsymbol{G},\zeta\right]^{T},\\
&\boldsymbol{G}^{T}=\left[\Pi_{1},\Pi_{2},\Pi_{3},\Pi_{4}\right]^{T},
\end{split}
\end{equation}
where $\boldsymbol{G}$ contains the NGB's that we associate with the Higgs doublet. To make such a correspondence, let us see how 
$\boldsymbol{\Pi}$ transforms under $\textup{SU}\left(2\right)_{L}\times\textup{SU}\left(2\right)_{R}$ transformations, that is to say
\begin{equation}
\begin{split}
&\boldsymbol{\Pi}\rightarrow e^{i\alpha_{a}^{L}t^{a}_{L}}\boldsymbol{\Pi}\simeq\left(\mathds{1}+i\alpha_{a}^{L}t^{a}_{L}+\mathcal{O}\left(\alpha^{2}\right)\right)\boldsymbol{\Pi}\Rightarrow\delta_{L}\boldsymbol{\Pi}=i\alpha_{a}^{L}t^{a}_{L}\boldsymbol{\Pi}\\
&\boldsymbol{\Pi}\rightarrow e^{i\alpha_{a}^{R}t^{a}_{R}}\boldsymbol{\Pi}\simeq\left(\mathds{1}+i\alpha_{a}^{R}t^{a}_{R}+\mathcal{O}\left(\alpha^{2}\right)\right)\boldsymbol{\Pi}\Rightarrow\delta_{R}\boldsymbol{\Pi}=i\alpha_{a}^{R}t^{a}_{R}\boldsymbol{\Pi},
\end{split}
\end{equation}
where $t^{a}_{L}$ and $t^{a}_{R}$ are $5\times 5$ matrices given by
\begin{equation}
t^{a}_{L}=\begin{pmatrix}X^{a}_{L} & 0 \\
0 & 0
\end{pmatrix},\hspace{0.5 cm} t^{a}_{R}=\begin{pmatrix}X^{a}_{R} & 0 \\
0 & 0
\end{pmatrix},
\end{equation}
and $X^{a}_{L}$, $X^{a}_{R}$ can be found in Eq.(\ref{subgenerators}). In what follows, it will prove convenient to use the following unitary matrix
\begin{equation}
P=\begin{pmatrix}
0 & 0 & i & 1 & 0\\
i & -1 & 0 & 0 & 0\\
i & 1 & 0 & 0 & 0\\
0 & 0 & -i & 1 & 0\\
0 & 0 & 0 & 0 & \sqrt{2}
\end{pmatrix},
\end{equation}
to transform to a basis in which the form of the generators is simpler \cite{Ponton}. The basis is obtained by the similarity transformation 
\begin{equation}\label{similaridad}
T_{L,R}^{a}=Pt^{a}_{L,R}P^{-1}.
\end{equation}
Applying a transformation to the $\boldsymbol{\Pi}$ vector under $P$ we obtain
\begin{equation}
\boldsymbol{\Pi}\stackrel{P}{\rightarrow}\boldsymbol{\Pi}'=\frac{1}{\sqrt{2}}\begin{pmatrix}
\Pi_{4}+i\Pi_{3}\\
-\Pi_{2}+i\Pi_{1}\\
\Pi_{2}+i\Pi_{1}\\
\Pi_{4}-i\Pi_{3}\\
\sqrt{2}\zeta\end{pmatrix},
\end{equation}
while Eq. (\ref{similaridad}) explicitly gives
\begin{equation}
\begin{split} 
&T_{L}^{1}=\frac{1}{2}\begin{pmatrix}\sigma_{1}&0&0\\0&\sigma_{1}&0\\0&0&0\end{pmatrix},\hspace{0.1 cm}T_{L}^{2}=\frac{1}{2}\begin{pmatrix}\sigma_{2}&0&0\\0&\sigma_{2}&0\\0&0&0\end{pmatrix},\\[0.5 cm]
&T_{L}^{3}=\frac{1}{2}\begin{pmatrix}\sigma_{3}&0&0\\0&\sigma_{3}&0\\0&0&0\end{pmatrix},\hspace{0.1 cm}T_{R}^{3}=\frac{1}{2}\begin{pmatrix}-\mathds{1}&0&0\\0&\mathds{1}&0\\0&0&0\end{pmatrix}
\end{split}
\end{equation}
We see that the similarity transformation is particularly useful because it makes explicit the correspondence 
$\textup{SO(4)}\simeq\textup{SU(2)}_{L}\times\textup{SU(2)}_{R}$. Therefore the left and right transformations are
\begin{align}
\notag
\delta_{L}\boldsymbol{\Pi}'=&i\left(\alpha^{L}_{1}T^{1}_{L}+\alpha^{L}_{2}T^{2}_{L}+\alpha^{L}_{3}T^{3}_{L}\right)\boldsymbol{\Pi}'\\[0.5 cm]
=&\frac{1}{2\sqrt{2}}\begin{pmatrix}-\alpha^{L}_{1}\Pi_{1}+\alpha^{L}_{2}\Pi_{2}-\alpha^{L}_{3}\Pi_{3}+i\left(-\alpha^{L}_{1}\Pi_{2}+\alpha^{L}_{2}\Pi_{1}+\alpha^{L}_{3}\Pi_{4}\right)\\
-\alpha^{L}_{1}\Pi_{3}-\alpha^{L}_{2}\Pi_{4}+\alpha^{L}_{3}\Pi_{1}+i\left(\alpha^{L}_{1}\Pi_{4}-\alpha^{L}_{2}\Pi_{3}+\alpha^{L}_{3}\Pi_{2}\right)\\
\alpha^{L}_{1}\Pi_{3}+\alpha^{L}_{2}\Pi_{4}-\alpha^{L}_{3}\Pi_{1}+i\left(\alpha^{L}_{1}\Pi_{4}-\alpha^{L}_{2}\Pi_{3}+\alpha^{L}_{3}\Pi_{2}\right)\\
-\alpha^{L}_{1}\Pi_{1}-\alpha^{L}_{2}\Pi_{2}-\alpha^{L}_{3}\Pi_{3}+i\left(\alpha^{L}_{1}\Pi_{2}-\alpha^{L}_{2}\Pi_{1}-\alpha^{L}_{3}\Pi_{4}\right)\\
0\end{pmatrix},\\[0.5 cm]
\delta_{R}\boldsymbol{\Pi}'=&i\alpha^{R}_{3}T^{3}_{R}\boldsymbol{\Pi}'=\frac{\alpha^{R}_{3}}{2\sqrt{2}}\begin{pmatrix}\Pi_{3}-i\Pi_{4}\\\Pi_{1}+i\Pi_{2}\\-\Pi_{1}+i\Pi_{2}\\\Pi_{3}+i\Pi_{4}\\0
\end{pmatrix}=\begin{pmatrix}\Pi'^{\dagger}_{0}\\\Pi'^{\dagger}_{+}\\\Pi'_{+}\\\Pi'_{0}\\0\end{pmatrix},
\end{align}
from which it is already clear that the new NGB $\zeta$ is a gauge singlet.\\
Taking into account that the Higgs doublet, given by Eq. (\ref{Higgs}) in the Appendix, transforms under $\textup{SU}(2)_{L}\times\textup{U}(1)_{Y}$ as
\begin{equation}
\begin{split}
&\delta_{L}H=\frac{i}{2}\alpha^{L}_{a}\sigma^{a}H=\frac{1}{2\sqrt{2}}\begin{pmatrix}\alpha^{L}_{2}H_{4}-\alpha^{L}_{3}H_{1}-\alpha^{L}_{1}H_{3}+i\left(\alpha^{L}_{1}H_{4}+\alpha^{L}_{2}H_{3}+\alpha^{L}_{3}H_{2}\right)\\\alpha^{L}_{3}H_{3}-\alpha^{L}_{1}H_{1}-\alpha^{L}_{2}H_{2}+i\left(\alpha^{L}_{1}H_{2}-\alpha^{L}_{2}H_{1}-\alpha^{L}_{3}H_{4}\right)\end{pmatrix},\\
&\delta_{Y}H=\frac{i}{2}\beta\mathds{1}H=\frac{\beta}{2\sqrt{2}}\begin{pmatrix}-H_{1}+iH_{2}\\-H_{3}+iH_{4}\end{pmatrix}=\begin{pmatrix}\delta_{Y}H_{+}\\\delta_{Y}H_{0}\end{pmatrix},
\end{split}
\end{equation}
and matching the components $\Pi'_{+}$ and $\Pi'_{0}$ to $\delta_{Y}H_{+}$ and $\delta_{Y}H_{0}$ respectively, we get
\begin{equation}
\begin{split}
-&\Pi_{1}+i\Pi_{2}=-H_{1}+iH_{2}\Rightarrow\Pi_{1}=H_{1},\hspace{0.4 cm}\Pi_{2}=H_{2},\\
&\Pi_{3}+i\Pi_{4}=-H_{3}+iH_{4}\Rightarrow\Pi_{3}=-H_{3},\hspace{0.1 cm}\Pi_{4}=H_{4}.
\end{split}
\end{equation}
We thus obtain
\begin{equation}
\begin{split}
\delta_{L}\boldsymbol{\Pi}'&=\begin{pmatrix}\delta_{L}H^{\dagger}_{0}\\\delta_{L}H^{\dagger}_{+}\\\delta_{L}H_{+}\\\delta_{L}H_{0}\\0\end{pmatrix},
\end{split}
\end{equation}
and the correspondence between the NGB's and the fields in the Higgs doublet is
\begin{equation}
H=\frac{1}{\sqrt{2}}\begin{pmatrix}\Pi_{2}+i\Pi_{1}\\\Pi_{4}-i\Pi_{3}\end{pmatrix},
\end{equation}
which is the same correspondence obtained for the case of the minimal model Eq. (\ref{DHiggs}), for which some of the expressions of Chapter 3 still hold true in this case. 
\section{Kinetic and Gauge Lagrangians}
Due to the presence of the field $\zeta$, the product of ordinary derivatives of the NGB's field vector is modified to
\begin{equation}
\partial_{\mu}\boldsymbol{\Pi}^{T}\partial^{\mu}\boldsymbol{\Pi}=\partial_{\mu}\boldsymbol{G}^{T}\partial^{\mu}\boldsymbol{G}+\partial_{\mu}\zeta\partial^{\mu}\zeta,
\end{equation}
and again, the explicit break of the Goldstone symmetry is performed by gauging, which transforms the ordinary derivatives into covariant derivatives
\begin{equation}
\partial^{\mu}\boldsymbol{G}\rightarrow D^{\mu}\boldsymbol{G}=\left(\partial^{\mu}-igW^{\mu}_{a}t^{a}_{L}-ig'B^{\mu}t^{3}_{R}\right)\boldsymbol{G}.
\end{equation}
Now, combining the latter along with the correspondence obtained between the NGB's and the Higgs doublet fields, we get
\begin{equation}
\partial_{\mu}\boldsymbol{\Pi}^{T}\partial^{\mu}\boldsymbol{\Pi}\Rightarrow 2D^{\dagger}_{\mu}HD^{\mu}H+\partial_{\mu}\zeta\partial^{\mu}\zeta,\hspace{0.5 cm}\Pi^{2}=2H^{\dagger}H+\zeta^{2}.
\end{equation}
With the above equations, we can write the non-linear Lagrangian for the $\textup{SO}(6)\rightarrow\textup{SO}(5)$ breaking pattern, which will be
\begin{equation}
\begin{split}
\mathcal{L}^{\left(2\right)}=&\frac{f^{2}}{2\left(2|H|^{2}+\zeta^{2}\right)}\sin^{2}\frac{\sqrt{2|H|^{2}+\zeta^{2}}}{f}\left(2D^{\dagger}_{\mu}HD^{\mu}H+\partial_{\mu}\zeta\partial^{\mu}\zeta\right)\\
&+\frac{f^{2}}{8\left(2|H|^{2}+\zeta^{2}\right)^{2}}\left(\frac{2|H|^{2}+\zeta^{2}}{f^{2}}-\sin^{2}\frac{\sqrt{2|H|^{2}+\zeta^{2}}}{f}\right)\left(\partial_{\mu}\left(2|H|^{2}+\zeta^{2}\right)\right)^{2}.
\end{split}
\end{equation}
Let us see the implications of $\mathcal{L}^{\left(2\right)}$ in this case. We go to the unitary gauge and we expand $\zeta$ around a possible VEV,
$\zeta=\eta(x)+N$ ($N$ being the VEV of the field $\zeta$) such that the Lagrangian $\mathcal{L}^{(2)}$ will be
\begin{equation}
\begin{split}
\mathcal{L}^{(2)}=&\textup{F}_{1}\left(h,\eta\right)\partial_{\mu}h\partial^{\mu}h+\textup{F}_{2}\left(h,\eta\right)\partial_{\mu}\eta\partial^{\mu}\eta+\textup{F}_{3}\left(h,\eta\right)\partial_{\mu}h\partial^{\mu}\eta\\[0.1 cm]
&+\frac{f^{2}g^{2}\left(V+h\right)^{2}}{4\left[\left(V+h\right)^{2}+\left(\eta+N\right)^{2}\right]}\sin^{2}\frac{\sqrt{\left(V+h\right)^{2}+\left(\eta+N\right)^{2}}}{f}\left[|W_{\mu}|^{2}+\frac{1}{2c_{w}^{2}}Z_{\mu}^{2}\right], 
\end{split}
\end{equation}
where
\begin{align} 
\notag
\textup{F}_{1}=&\frac{1}{2}\Bigg\{1-\frac{N^{2}}{\left(V+h\right)^{2}+\left(N+\eta\right)^{2}}\\
&+\frac{N^{2}f^{2}}{\left[\left(V+h\right)^{2}+\left(N+\eta\right)^{2}\right]^{2}}\sin^{2}\frac{\sqrt{\left(V+h\right)^{2}+\left(\eta+N\right)^{2}}}{f}\Bigg\},\\\notag
\textup{F}_{2}=&\frac{1}{2}\Bigg\{\frac{N^{2}}{\left(V+h\right)^{2}+\left(N+\eta\right)^{2}}-\frac{f^{2}N^{2}}{\left[\left(V+h\right)^{2}+\left(N+\eta\right)^{2}\right]^{2}}\sin^{2}\frac{\sqrt{\left(V+h\right)^{2}+\left(\eta+N\right)^{2}}}{f}\\[0.1 cm]
&+\frac{f^{2}}{\left(V+h\right)^{2}+\left(N+\eta\right)^{2}}\sin^{2}\frac{\sqrt{\left(V+h\right)^{2}+\left(\eta+N\right)^{2}}}{f}\Bigg\},\\
\textup{F}_{3}=&\frac{\left(V+h\right)\left(\eta+N\right)}{\left(V+h\right)^{2}+\left(\eta+N\right)^{2}}-\frac{f^{2}\left(V+h\right)\left(\eta+N\right)}{\left[\left(V+h\right)^{2}+\left(\eta+N\right)^{2}\right]^{2}}\sin^{2}\frac{\sqrt{\left(V+h\right)^{2}+\left(\eta+N\right)^{2}}}{f}. 
\end{align}
Thus we can read the vector boson masses as
\begin{equation}
m_{W}=c_{w}m_{Z}=\frac{1}{2}g\frac{V}{\sqrt{V^{2}+N^{2}}}f\sin\frac{\sqrt{V^{2}+N^{2}}}{f},
\end{equation}
where, to make contact with the usual notation, we can define the ElectroWeak VEV as
\begin{equation}\label{VEVNM}
v=\frac{V}{\sqrt{V^{2}+N^{2}}}f\sin\frac{\sqrt{V^{2}+N^{2}}}{f}.
\end{equation}
Notice that if $\zeta$ is to be a dark matter candidate we need to guarantee that $N=0$, and we go back to the same EW scale as in the MCHM.\\
From $\textup{F}_{3}\left(h,\eta\right)$ we see that $h$ and $\eta$ are not physical eigenstates, because there is a kinetic mixing.
To remove such terms, we note that the Lagrangian can be written as
\begin{equation}
\mathcal{L}^{\left(2\right)}=\begin{pmatrix}\partial_{\mu}h&\partial_{\mu}\eta\end{pmatrix}\underbrace{\begin{pmatrix}\textup{F}_{1}&\frac{\textup{F}_{3}}{2}\\\frac{\textup{F}_{3}}{2}&\textup{F}_{2}\end{pmatrix}}
_{\mathcal{O}\begin{pmatrix}\lambda_{+}&0\\0&\lambda_{-}\end{pmatrix}\mathcal{O}^{T}}
\begin{pmatrix}\partial^{\mu}h\\\partial^{\mu}\eta\end{pmatrix}+\cdots,
\end{equation}
with $\mathcal{O}$ the orthogonal matrix that diagonalize the kinetic mixing matrix, such that
\begin{equation}
\begin{pmatrix}\partial^{\mu}h'\\\partial^{\mu}\eta'\end{pmatrix}=\mathcal{O}^{T}\begin{pmatrix}\partial^{\mu}h\\\partial^{\mu}\eta\end{pmatrix}.
\end{equation}
The diagonalization is done by evaluating the functions $\textup{F}_{1}$, $\textup{F}_{2}$, and $\textup{F}_{3}$ at $(0,0)$ point. The
matrix $\mathcal{O}$, and the eigenvalues $\lambda_{+}$ and $\lambda_{-}$ are given by
\begin{align}
&\mathcal{O}=\frac{1}{\sqrt{V^{2}+N^{2}}}\begin{pmatrix}V&-N\\N&V\end{pmatrix},\\
&\lambda_{+}=\frac{1}{2},\hspace{0.5 cm}\lambda_{-}=\frac{v^{2}}{2V^{2}},
\end{align}
from which we have
\begin{equation}
\mathcal{L}^{\left(2\right)}=\lambda_{+}\partial_{\mu}h'\partial^{\mu}h'+\lambda_{-}\partial_{\mu}\eta'\partial^{\mu}\eta'+\cdots
\end{equation}
We can next redefine the $h'$ and $\eta'$ fields as
\begin{align}
&h'\rightarrow\frac{h_{p}}{\sqrt{2\lambda_{+}}},\\
&\eta'\rightarrow\frac{\eta_{p}}{\sqrt{2\lambda_{-}}},
\end{align}
where $h_{p}$ and $\eta_{p}$ are now the physical fields, with kinetic terms in their canonical form 
\begin{equation}
\mathcal{L}^{\left(2\right)}=\frac{1}{2}\partial_{\mu}h_{p}\partial^{\mu}h_{p}+\frac{1}{2}\partial_{\mu}\eta_{p}\partial^{\mu}\eta_{p}+\cdots
\end{equation}
The $h$ and $\eta$ fields in terms of the physical fields are given by
\begin{align} 
&h=\frac{V}{\sqrt{V^{2}+N^{2}}}\left(h_{p}-\frac{N}{v}\eta_{p}\right),\\
&\eta=\frac{1}{\sqrt{V^{2}+N^{2}}}\left(Nh_{p}+\frac{V^{2}}{v}\eta_{p}\right).
\end{align}
We calculate the interactions among two gauge field and the infinite number of fields $h_{p}$ and $\eta_{p}$ by Taylor-expanding the 
Eq. (\ref{gaugint}) (see Appendix \ref{series}) in terms of the physical fields. Up to second order, we obtain
\begin{equation}
\begin{split}
&\frac{g^{2}v^{2}}{4}\left(|W_{\mu}|^{2}+\frac{1}{2c^{2}_{w}}Z_{\mu}^{2}\right)
\Bigg\{1+2V\sqrt{\frac{1-\xi}{N^{2}+V^{2}}}\frac{h_{p}}{v}+\frac{V^{2}}{V^{2}+N^{2}}\left(1-2\xi\right)\frac{h_{p}^{2}}{v^{2}}\\
&-\frac{2N}{\sqrt{V^{2}+N^{2}}}\frac{\eta_{p}}{v}+\frac{\left(N^{2}-V^{2}\right)}{V^{2}+N^{2}}\frac{\eta_{p}^{2}}{v^{2}}+\frac{V^{3}}{V^{2}+N^{2}}\sqrt{1-\xi}\frac{\eta_{p}^{2}}{v^{3}}\\
&-\frac{4NV}{V^{2}+N^{2}}\sqrt{1-\xi}\frac{h_{p}\eta_{p}}{v^{2}}+\frac{2N}{V^{2}+N^{2}}\frac{h_{p}\eta_{p}}{v}+\cdots\Bigg\}. 
\end{split}
\end{equation}
Therefore in the unitary gauge the kinetic and gauge Lagrangian is given by
\begin{equation}
\begin{split} 
\mathcal{L}^{\left(2\right)}=&\frac{1}{2}\partial_{\mu}h_{p}\partial^{\mu}h_{p}+\frac{1}{2}\partial_{\mu}\eta_{p}\partial^{\mu}\eta_{p}+\frac{g^{2}v^{2}}{4}\left(|W_{\mu}|^{2}+\frac{1}{2c^{2}_{w}}Z_{\mu}^{2}\right)\\
&+\frac{g^{2}v^{2}}{4}\left(|W_{\mu}|^{2}+\frac{1}{2c^{2}_{w}}Z_{\mu}^{2}\right)
\Bigg\{2V\sqrt{\frac{1-\xi}{N^{2}+V^{2}}}\frac{h_{p}}{v}+\frac{V^{2}}{V^{2}+N^{2}}\left(1-2\xi\right)\frac{h_{p}^{2}}{v^{2}}\\
&-\frac{2N}{\sqrt{V^{2}+N^{2}}}\frac{\eta_{p}}{v}+\frac{\left(N^{2}-V^{2}\right)}{V^{2}+N^{2}}\frac{\eta_{p}^{2}}{v^{2}}+\frac{V^{3}}{V^{2}+N^{2}}\sqrt{1-\xi}\frac{\eta_{p}^{2}}{v^{3}}\\
&-\frac{4NV}{V^{2}+N^{2}}\sqrt{1-\xi}\frac{h_{p}\eta_{p}}{v^{2}}+\frac{2N}{V^{2}+N^{2}}\frac{h_{p}\eta_{p}}{v}+\cdots\Bigg\}+\cdots
\end{split}
\end{equation}
\section{Embedding Fermions}
As was the case in the MCHM, in order to reproduce the hypercharge of the SM fermions the global symmetry should be enlarged
to $\textup{SO}\left(6\right)\times\textup{U}\left(1\right)_{X}$. In the embedding of the fermions we decompose $\textup{SO}\left(6\right)$ 
under the maximal subgroup $\textup{SO}\left(4\right)\times\textup{SO}\left(2\right)_{\zeta}$ which is isomorphic 
to the subgroup $\textup{SU}\left(2\right)_{L}\times\textup{SU}\left(2\right)_{R}\times\textup{U}\left(1\right)_{\zeta}$.\\
Choosing $X=2/3$ as $\textup{U(1)}_{X}$ charge, the fundamental representation \textbf{6} of $\textup{SO}\left(6\right)$ decomposes under
$\textup{SO}\left(4\right)\times\textup{SO}\left(2\right)_{\zeta}$ and $\textup{SU}\left(2\right)_{L}\times\textup{SU}\left(2\right)_{R}\times\textup{U}\left(1\right)_{\zeta}$ as
\begin{equation}
\textbf{6}_{2/3}\rightarrow\textbf{4}_{2/3}\oplus\textbf{1}_{2/3}\oplus\textbf{1}_{2/3}\rightarrow\textbf{2}_{7/6}\oplus\textbf{2}_{1/6}\oplus\textbf{1}_{2/3}\oplus\textbf{1}_{2/3}, 
\end{equation}
where the subscripts in the last term denote the hypercharge given by the expression (\ref{hypercharge}). We see that $q_{L}$ and $t_{R}$ 
can be identified with the last three terms. The singlet $t_{R}$ can actually be embedded in two inequivalent ways  forms given by
\begin{equation}
T_{R1}=\{0,0,0,0,0,t_{R}\}^{T},\hspace{0.5 cm}T_{R2}=\{0,0,0,0,t_{R},0\}^{T},
\end{equation}
so the interaction Lagrangian is expressed as
\begin{equation}
\mathcal{L}^{t_{R}}_{\textup{Int}}=\lambda_{t_{R1}}\left(\overline{T}_{R1}\right)^{I}\left(\mathcal{O}_{F}^{t_{R1}}\right)_{I}+\lambda_{t_{R2}}\left(\overline{T}_{R2}\right)^{I}\left(\mathcal{O}_{F}^{t_{R2}}\right)_{I}+h.c.
\end{equation}
When we ``dress'' the source with the Goldstone matrix, we obtain the following objects
\begin{equation}
U\left[\Pi\right]^{-1}T_{R1}=\{T^{\textbf{5}}_{R1},T^{\textbf{1}}_{R1}\},\hspace{0.5 cm}U\left[\Pi\right]^{-1}T_{R2}=\{T^{\textbf{5}}_{R2},T^{\textbf{1}}_{R2}\}
\end{equation}
where
\begin{equation}
T^{\textbf{5}}_{R1}=-\frac{\boldsymbol{\Pi}}{\Pi}t_{R}\sin\frac{\Pi}{f},\hspace{0.5 cm}T^{\textbf{1}}_{R1}=t_{R}\cos\frac{\Pi}{f},
\end{equation}
and
\begin{equation}
T^{\textbf{5}}_{R2}=-\frac{t_{R}}{\Pi^{2}}\begin{pmatrix}\zeta\boldsymbol{G}\left(1-\cos\frac{\Pi}{f}\right)\\G^{2}+\zeta^{2}\cos\frac{\Pi}{f}
\end{pmatrix},\hspace{0.5 cm}T^{\textbf{1}}_{R2}=t_{R}\frac{\zeta}{\Pi}\sin\frac{\Pi}{f}.
\end{equation}
The $q_{L}$ interaction is instead again given by the expression (\ref{qlint}). The embedding of the doublet $q_{L}$ in the $Q_{t_{L}}$ source is
done as in Eqs. (\ref{psi}) and (\ref{cuadrupleto}), thus by using the projection of the $Q_{t_{L}}$ source on the doublet $T^{3}_{R}=-1/2$, we obtain
\begin{equation}
Q_{t_{L}}=\frac{1}{\sqrt{2}}\{-ib_{L},-b_{L},-it_{L},t_{L},0,0\}^{T}=\{\boldsymbol{q}_{L_{-}},0,0\}^{T}.
\end{equation}
By dressing $Q_{t_{L}}$ with $U\left[\Pi\right]^{-1}$, we obtain $\textup{SO}\left(5\right)$ multiplets 
\begin{equation}
U\left[\Pi\right]^{-1}Q_{t_{L}}=\{Q_{t_{L}}^{\textbf{5}},Q_{t_{L}}^{\textbf{1}}\},
\end{equation}
where
\begin{equation}
Q_{t_{L}}^{\textbf{5}}=\left[\mathds{1}-\left(1-\cos\frac{\Pi}{f}\right)\frac{\boldsymbol{\Pi}\cdot\boldsymbol{\Pi}^{T}}{\Pi^{2}}\right]\cdot\boldsymbol{q}_{L_{-}},\hspace{0.5 cm}Q_{t_{L}}^{\textbf{1}}=\left(\frac{\boldsymbol{\Pi}^{T}}{\Pi}\sin\frac{\Pi}{f}\right)\cdot\boldsymbol{q}_{L_{-}}.
\end{equation}
With the objects $T^{\textbf{5}}_{R1}$, $T^{\textbf{1}}_{R1}$, $T^{\textbf{5}}_{R2}$, $T^{\textbf{1}}_{R2}$, $Q^{\textbf{5}}_{t_{L}}$ 
and $Q^{\textbf{1}}_{t_{L}}$ we can form the invariants. Again, as in the case of (MCHM$_{\textbf{5}}$) we have an equation analogous to 
Eq. (\ref{relinv}) that is, the invariants are not independent. We form the simplest invariants from the singlets
\begin{equation}
\begin{split}
&\overline{Q}^{\textbf{1}}_{t_{L}}T^{\textbf{1}}_{R1}=\frac{1}{2\sqrt{2|H|^{2}+\zeta^{2}}}\sin\frac{2\sqrt{2|H|^{2}+\zeta^{2}}}{f}\overline{q}_{L}H^{c}t_{R},\\
&\overline{Q}^{\textbf{1}}_{t_{L}}T^{\textbf{1}}_{R2}=\frac{\zeta}{2|H|^{2}+\zeta^{2}}\sin^{2}\frac{\sqrt{2|H|^{2}+\zeta^{2}}}{f}\overline{q}_{L}H^{c}t_{R},
\end{split}
\end{equation}
from which the top Yukawa Lagrangian is
\begin{equation}
\begin{split} 
\mathcal{L}^{t}_{\textup{Yuk}}=&-c^{t}\frac{\lambda_{t_{L}}\lambda_{t_{R1}}}{g^{2}_{\ast}}m_{\ast}\frac{1}{2\sqrt{2|H|^{2}+\zeta^{2}}}\sin\frac{2\sqrt{2|H|^{2}+\zeta^{2}}}{f}\overline{q}_{L}H^{c}t_{R}\\
&-c^{t}\frac{\lambda_{t_{L}}\lambda_{t_{R2}}}{g^{2}_{\ast}}m_{\ast}\frac{\zeta}{2|H|^{2}+\zeta^{2}}\sin^{2}\frac{\sqrt{2|H|^{2}+\zeta^{2}}}{f}\overline{q}_{L}H^{c}t_{R}+h.c.
\end{split}
\end{equation}
In the unitary gauge and around the VEV of $\zeta$ the top Yukawa Lagrangian becomes
\begin{equation}
\begin{split} 
\mathcal{L}^{t}_{\textup{Yuk}}=&-c^{t}\frac{\lambda_{t_{L}}\lambda_{t_{R1}}}{g^{2}_{\ast}}\frac{m_{\ast}\left(V+h\right)}{2\sqrt{2\left[\left(V+h\right)^{2}+\left(N+\eta\right)^{2}\right]}}\sin\frac{2\sqrt{\left(V+h\right)^{2}+\left(N+\eta\right)^{2}}}{f}\overline{t}t\\
&-c^{t}\frac{\lambda_{t_{L}}\lambda_{t_{R2}}}{g^{2}_{\ast}}\frac{m_{\ast}\left(N+\eta\right)\left(V+h\right)}{\sqrt{2}\left[\left(V+h\right)^{2}+\left(N+\eta\right)^{2}\right]}\sin^{2}\frac{\sqrt{\left(V+h\right)^{2}+\left(N+\eta\right)^{2}}}{f}\overline{t}t.\\
\end{split}
\end{equation}
When the $H$ and $\zeta$ fields take their VEV we have
\begin{equation}\label{masatopnm}
\begin{split}
\mathcal{L}_{\textup{mass}}&=-c^{t}\frac{\lambda_{t_{L}}\lambda_{t_{R1}}}{g^{2}_{\ast}}m_{\ast}\frac{1}{2}\sqrt{\frac{V^{2}}{2\left(V^{2}+N^{2}\right)}}\sin\frac{2\sqrt{V^{2}+N^{2}}}{f}\overline{t}t\\
&-c^{t}\frac{\lambda_{t_{L}}\lambda_{t_{R2}}}{g^{2}_{\ast}}m_{\ast}\frac{NV}{\sqrt{2}\left(V^{2}+N^{2}\right)}\sin^{2}\frac{\sqrt{V^{2}+N^{2}}}{f}\overline{t}t\\
&=-c^{t}\frac{\lambda_{t_{L}}\lambda_{t_{R1}}}{g^{2}_{\ast}}m_{\ast}\sqrt{\frac{V^{2}\xi\left(1-\xi\right)}{2\left(V^{2}+N^{2}\right)}}\overline{t}t-c^{t}\frac{\lambda_{t_{L}}\lambda_{t_{R2}}}{g^{2}_{\ast}}m_{\ast}\frac{NV\xi}{\sqrt{2}\left(V^{2}+N^{2}\right)}\overline{t}t\\
&=-m_{t1}\overline{t}t-m_{t2}\overline{t}t\equiv-m_{t}\overline{t}t,
\end{split}
\end{equation}
where we have used expressions (\ref{xi}) and (\ref{pro}) from Appendix \ref{series}. Explicitly the top mass term is
\begin{equation}
m_{t}=c^{t}\frac{\lambda_{t_{L}}}{g^{2}_{\ast}}m_{\ast}\sqrt{\frac{V^{2}}{2\left(V^{2}+N^{2}\right)}}\left[\lambda_{t_{R1}}\sqrt{\xi\left(1-\xi\right)}+\lambda_{t_{R2}}\frac{N\xi}{\sqrt{V^{2}+N^{2}}}\right].
\end{equation}
With the above expressions we can write the top Yukawa Lagrangian as
\begin{equation}
\begin{split} 
\mathcal{L}^{t}_{\textup{Yuk}}=&-\frac{m_{t1}}{2}\sqrt{\frac{\left(V^{2}+N^{2}\right)\left(V+h\right)^{2}}{V^{2}\xi\left(1-\xi\right)\left[\left(V+h\right)^{2}+\left(N+\eta\right)^{2}\right]}}\sin\frac{2\sqrt{\left(V+h\right)^{2}+\left(N+\eta\right)^{2}}}{f}\overline{t}t\\
&-\frac{ m_{t2}}{NV\xi}\frac{\left(V^{2}+N^{2}\right)\left(V+h\right)\left(N+\eta\right)}{\left[\left(V+h\right)^{2}+\left(N+\eta\right)^{2}\right]}\sin^{2}\frac{\sqrt{\left(V+h\right)^{2}+\left(N+\eta\right)^{2}}}{f}\overline{t}t.
\end{split}
\end{equation}
We now use the expansions (\ref{thexp}) and (\ref{etamexp}) of Appendix \ref{series} in terms of physical fields, and we obtain
\begin{equation}
\begin{split} 
\mathcal{L}^{t}_{\textup{Yuk}}=&-m_{t1}\Bigg\{1+\frac{V}{\left(V^{2}+N^{2}\right)^{1/2}}\frac{\left(1-2\xi\right)}{\sqrt{1-\xi}}\frac{h_{p}}{v}-\frac{2V^{2}}{\left(V^{2}+N^{2}\right)}\xi\frac{h_{p}^{2}}{v^{2}}\\
&-\frac{N}{\left(V^{2}+N^{2}\right)^{1/2}}\frac{\eta_{p}}{v}-\frac{V^{2}}{2\left(V^{2}+N^{2}\right)}\frac{\eta_{p}^{2}}{v^{2}}+\frac{V^{3}}{2\left(V^{2}+N^{2}\right)}\frac{\left(1-2\xi\right)}{\sqrt{1-\xi}}\frac{\eta_{p}^{2}}{v^{3}}\\
&+\frac{N}{\left(V^{2}+N^{2}\right)}\frac{h_{p}\eta_{p}}{v}-\frac{NV}{\left(V^{2}+N^{2}\right)}\frac{\left(1-2\xi\right)}{\sqrt{1-\xi}}\frac{h_{p}\eta_{p}}{v^{2}}+\cdots\Bigg\}\overline{t}t,\\
&-m_{t2}\Bigg\{1+2V\sqrt{\frac{1-\xi}{V^{2}+N^{2}}}\frac{h_{p}}{v}+\frac{V^{2}}{\left(V^{2}+N^{2}\right)}\left(1-2\xi\right)\frac{h_{p}^{2}}{v^{2}}\\
&+\frac{\left(V^{2}-N^{2}\right)}{N\left(V^{2}+N^{2}\right)^{1/2}}\frac{\eta_{p}}{v}-\frac{2V^{2}}{\left(V^{2}+N^{2}\right)}\frac{\eta_{p}^{2}}{v^{2}}+\frac{V^{3}}{\left(V^{2}+N^{2}\right)}\sqrt{1-\xi}\frac{\eta_{p}^{2}}{v^{3}}\\
&+\frac{\left(N^{2}-V^{2}\right)}{N\left(V^{2}+N^{2}\right)}\frac{h_{p}\eta_{p}}{v}+\frac{2V\left(V^{2}-N^{2}\right)}{N\left(V^{2}+N^{2}\right)}\sqrt{1-\xi}\frac{h_{p}\eta_{p}}{v^{2}}+\cdots\Bigg\}\overline{t}t,
\end{split}
\end{equation}
which we write in a more compact way as 
\begin{align}
\notag
\mathcal{L}^{t}_{\textup{Yuk}}=&-m_{t}\overline{t}t-\frac{\left(k_{t1}^{\textbf{6}}m_{t1}+k_{t2}^{\textbf{6}}m_{t2}\right)}{v}h_{p}\overline{t}t-\frac{\left(C_{2t1}^{\textbf{6}}m_{t1}+C_{2t2}^{\textbf{6}}m_{t2}\right)}{v^{2}}h_{p}^{2}\overline{t}t\\\notag
&-\frac{\left(J_{t1}^{\textbf{6}}m_{t1}+J_{t2}^{\textbf{6}}m_{t2}\right)}{v}\eta_{p}\overline{t}t-\frac{\left(D_{2t1}^{\textbf{6}}m_{t1}+D_{2t2}^{\textbf{6}}m_{t2}\right)}{v^{2}}\eta_{p}^{2}\overline{t}t-\frac{\left(E_{2t1}^{\textbf{6}}m_{t1}+E_{2t2}^{\textbf{6}}m_{t2}\right)}{v^{3}}\eta_{p}^{2}\overline{t}t\\
&-\frac{\left(M_{t1}^{\textbf{6}}m_{t1}+M_{t2}^{\textbf{6}}m_{t2}\right)}{v}h_{p}\eta_{p}\overline{t}t-\frac{\left(R_{2t1}^{\textbf{6}}m_{t1}+R_{2t2}^{\textbf{6}}m_{t2}\right)}{v^{2}}h_{p}\eta_{p}\overline{t}t-\cdots
\end{align}
Using the above procedure we can also obtain the interactions in the bottom quark sector. Choosing $X=-1/3$, we can embed the operators in the 
fundamental representation $\textbf{6}_{-1/3}$, which decomposes under $\textup{SO}\left(4\right)\times\textup{SO}\left(2\right)_{\zeta}$ 
and $\textup{SU}\left(2\right)_{L}\times\textup{SU}\left(2\right)_{R}\times\textup{U}\left(1\right)_{\zeta}$ as
\begin{equation}
\textbf{6}_{-1/3}\rightarrow\textbf{4}_{-1/3}\oplus\textbf{1}_{-1/3}\oplus\textbf{1}_{-1/3}\rightarrow\textbf{2}_{1/6}\oplus\textbf{2}_{-5/6}\oplus\textbf{1}_{-1/3}\oplus\textbf{1}_{-1/3}.
\end{equation}
The interaction Lagrangian of the bottom quark sector is given by
\begin{equation}
\mathcal{L}_{\textup{Int}}^{b}=\lambda_{b_{L}}\left(\overline{Q}_{b_{L}}\right)^{I}\left(\mathcal{O}^{b_{L}}_{\hspace{0.1 cm}F}\right)_{I}+\lambda_{b_{R1}}\left(\overline{B}_{R1}\right)^{I}\left(\mathcal{O}^{b_{R1}}_{\hspace{0.1 cm}F}\right)_{I}+\lambda_{b_{R2}}\left(\overline{B}_{R2}\right)^{I}\left(\mathcal{O}^{b_{R2}}_{\hspace{0.1 cm}F}\right)_{I},
\end{equation}
where the sources are given by
\begin{equation}
\begin{split}
&B_{R1}=\{0,0,0,0,0,b_{R}\}^{T},\\ 
&B_{R2}=\{0,0,0,0,b_{R},0\}^{T},\\ 
&Q_{b_{L}}=\frac{1}{\sqrt{2}}\{-it_{L},t_{L},ib_{L},b_{L},0,0\}^{T}=\{\boldsymbol{q}_{L_{+}},0,0\}^{T}.
\end{split}
\end{equation}
The $Q_{b_{L}}$ source is obtained when we project on the $T^{3}_{R}=1/2$ doublet in Eq. (\ref{cuadrupleto}).\\
Again, the dressing procedure converts the sources into the following multiples
\begin{equation}
\begin{split} 
&B_{R1}^{\textbf{5}}=-\frac{\boldsymbol{\Pi}}{\Pi}b_{R}\sin\frac{\Pi}{f},\hspace{4.1 cm}B^{\textbf{1}}_{R1}=b_{R}\cos\frac{\Pi}{f},\\
&B^{\textbf{5}}_{R2}=-\frac{b_{R}}{\Pi^{2}}\begin{pmatrix}\zeta\boldsymbol{G}\left(1-\cos\frac{\Pi}{f}\right)\\G^{2}+\zeta^{2}\cos\frac{\Pi}{f}
\end{pmatrix},\hspace{1.76 cm}B^{\textbf{1}}_{R2}=b_{R}\frac{\zeta}{\Pi}\sin\frac{\Pi}{f},\\
&Q_{t_{L}}^{\textbf{5}}=\left[\mathds{1}-\left(1-\cos\frac{\Pi}{f}\right)\frac{\boldsymbol{\Pi}\cdot\boldsymbol{\Pi}^{T}}{\Pi^{2}}\right]\cdot\boldsymbol{q}_{L_{+}},\hspace{0.45 cm}Q_{t_{L}}^{\textbf{1}}=\left(\frac{\boldsymbol{\Pi}^{T}}{\Pi}\sin\frac{\Pi}{f}\right)\cdot\boldsymbol{q}_{L_{+}}.
\end{split}
\end{equation}
From these multiplets we can form invariants and we obtain the bottom Yukawa Lagrangian
\begin{equation}
\begin{split} 
\mathcal{L}^{b}_{\textup{Yuk}}=&-c^{b}\frac{\lambda_{b_{L}}\lambda_{b_{R1}}}{g^{2}_{\ast}}m_{\ast}\frac{1}{2\sqrt{2|H|^{2}+\zeta^{2}}}\sin\frac{2\sqrt{2|H|^{2}+\zeta^{2}}}{f}\overline{q}_{L}Hb_{R}\\
&-c^{b}\frac{\lambda_{b_{L}}\lambda_{b_{R2}}}{g^{2}_{\ast}}m_{\ast}\frac{\zeta}{2|H|^{2}+\zeta^{2}}\sin^{2}\frac{\sqrt{2|H|^{2}+\zeta^{2}}}{f}\overline{q}_{L}Hb_{R}+h.c.
\end{split}
\end{equation}
In the unitary and around the VEV of $\zeta$ we obtain
\begin{equation}
\begin{split} 
\mathcal{L}^{b}_{\textup{Yuk}}=&-c^{b}\frac{\lambda_{b_{L}}\lambda_{b_{R1}}}{g^{2}_{\ast}}\frac{m_{\ast}\left(V+h\right)}{2\sqrt{2\left[\left(V+h\right)^{2}+\left(N+\eta\right)^{2}\right]}}\sin\frac{2\sqrt{\left(V+h\right)^{2}+\left(N+\eta\right)^{2}}}{f}\overline{b}b\\
&-c^{b}\frac{\lambda_{b_{L}}\lambda_{b_{R2}}}{g^{2}_{\ast}}\frac{m_{\ast}\left(N+\eta\right)\left(V+h\right)}{\sqrt{2}\left[\left(V+h\right)^{2}+\left(N+\eta\right)^{2}\right]}\sin^{2}\frac{\sqrt{\left(V+h\right)^{2}+\left(N+\eta\right)^{2}}}{f}\overline{b}b,\\
\end{split}
\end{equation}
in this way we have the bottom mass term (when $H$ and $\zeta$ are set to their VEV) with
\begin{equation}
\begin{split}
m_{b}&=c^{b}\frac{\lambda_{b_{L}}\lambda_{b_{R1}}}{g^{2}_{\ast}}m_{\ast}\sqrt{\frac{V^{2}\xi\left(1-\xi\right)}{2\left(V^{2}+N^{2}\right)}}+c^{b}\frac{\lambda_{b_{L}}\lambda_{b_{R2}}}{g^{2}_{\ast}}m_{\ast}\frac{NV\xi}{\sqrt{2}\left(V^{2}+N^{2}\right)}\\
&=m_{b1}+m_{b2}.
\end{split}
\end{equation}
We use the Taylor expansion in terms of physical fields, in such a way that the bottom Yukawa Lagrangian becomes 
\begin{equation}
\begin{split} 
\mathcal{L}^{b}_{\textup{Yuk}}=&-m_{b}\overline{b}b-\frac{\left(k_{b1}^{\textbf{6}}m_{b1}+k_{b2}^{\textbf{6}}m_{b2}\right)}{v}h_{p}\overline{b}b-\frac{\left(J_{b1}^{\textbf{6}}m_{b1}+J_{b2}^{\textbf{6}}m_{b2}\right)}{v}\eta_{p}\overline{b}b\\
&-\frac{\left(M_{b1}^{\textbf{6}}m_{b1}+M_{b2}^{\textbf{6}}m_{b2}\right)}{v}h_{p}\eta_{p}\overline{b}b-\cdots
\end{split}
\end{equation}
The modifications to the couplings in the fermionic sector of the $(\textup{NMCHM})_{\textbf{6}}$ model are combinations of the contributions 
given by the singlets $T^{\textbf{1}}_{R1}$, $T^{\textbf{1}}_{R2}$, $B^{\textbf{1}}_{R1}$, $B^{\textbf{1}}_{R2}$, formed by the dressing. This is a
consequence of the possibility of embedding the quarks singlets in two different ways.
\chapter{Estimation of the composite Higgs potential}
In this last chapter we estimate the composite Higgs potential for the two cosets studied. The generation of the composite Higgs potential comes
from the explicit breaking of the Goldstone symmetry, that is to say from the elementary/composite interactions. The dominant contributions to the 
potential come from the interactions between the top quark sector and the composite sector, because their coupling constant must be much greater 
than the others in order to reproduce the large top Yukawa.\\
To estimate the potential we use the method of spurions \cite{Panico}. A spurion is a fictitious field which allows us to promote the parameters 
that break the symmetry to fields that transform under the $\mathcal{G}$ symmetry group. We can use the spurions fields to construct invariant 
operators under $\mathcal{G}$. As we are going to see the spurions will allow us to study the implications of the $\textup{SO}\left(N\right)$ 
explicit symmetry breaking, assigning them to transform in an $N$ of $\textup{SO}\left(N\right)$.
\section{Case MCHM\texorpdfstring{$_{\textbf{5}}$}{5}}
Our starting point will be the fermion Lagrangian presented in Chapter \ref{ChapMCHM}. We rewrite the expressions (\ref{tR}) and (\ref{qlint}) by 
introducing the spurion fields $\Lambda_{L}$ and $\Lambda_{R}$
\begin{equation}
\begin{split}
\mathcal{L}^{\slashed{\mathcal{G}}}_{\lambda_{t}}&=\lambda_{t_{L}}\left(\overline{Q}_{t_{L}}\right)^{I}\left(\mathcal{O}^{L}_{F}\right)_{I}+\lambda_{t_{R}}\left(\overline{T}_{R}\right)^{I}\left(\mathcal{O}^{R}_{F}\right)_{I}+h.c=\overline{q}_{L}\Lambda_{L}\mathcal{O}^{L}_{F}+\overline{t}_{R}\Lambda_{R}\mathcal{O}^{R}_{F}+h.c\\
&=\left(\overline{t}_{L}\Lambda_{L}^{t_{L}}+\overline{b}_{L}\Lambda_{L}^{b_{L}}\right)\mathcal{O}^{L}_{F}+\overline{t}_{R}\Lambda_{R}\mathcal{O}^{R}_{F}+h.c,
\end{split}
\end{equation}
where the spurion fields are given by
\begin{align}
&\Lambda_{L}^{t_{L}}=\frac{\lambda_{t_{L}}}{\sqrt{2}}\begin{pmatrix}0&0&i&1&0\end{pmatrix},\label{spurionLt}\\
&\Lambda_{L}^{b_{L}}=\frac{\lambda_{t_{L}}}{\sqrt{2}}\begin{pmatrix}i&-1&0&0&0\end{pmatrix},\\
&\Lambda_{R}^{t_{R}}=\lambda_{t_{R}}\begin{pmatrix}0&0&0&0&1\end{pmatrix}\label{spurionRt}.
\end{align}
Notice that since in Section \ref{FerMCHM} we have decided to assign $\mathcal{O}^{R,L}_{F}\sim\textbf{5}_{2/3}$ of 
$\textup{SO}\left(5\right)\times\textup{U}\left(1\right)_{X}$, to form an invariant we must have 
$\Lambda_{L}^{i},\Lambda_{R}^{t_{R}}\sim\textbf{5}_{2/3}$ of $\textup{SO}\left(5\right)\times\textup{U}\left(1\right)_{X}$. This explains why the
spurious in Eqs. (\ref{spurionLt})-(\ref{spurionRt}) are written as quintuplets. We can now form multiplets in the representations \textbf{4}
and \textbf{1} of $\textup{SO}\left(4\right)$ ``dressing'' the spurions, that is to say
\begin{equation}
\begin{pmatrix}\Lambda^{\textbf{4}}_D\\\Lambda^{\textbf{1}}_D\end{pmatrix}=U^{\dagger}\Lambda.
\end{equation}
Using the fact that the Goldstone matrix in the unitary gauge is given by 
\begin{equation}\label{golUGMCH}
U\underset{\textup{UG}}{=}\begin{pmatrix}\mathds{1}_{3}&0&0\\0&\cos\frac{\sqrt{2}H}{f}&\sin\frac{\sqrt{2}H}{f}\\0&-\sin\frac{\sqrt{2}H}{f}&\cos\frac{\sqrt{2}H}{f}\end{pmatrix},
\end{equation}
we have
\begin{align}
&\Lambda_{LD}^{t_{L}}=U^{-1}\Lambda_{L}^{t_{L}}\Rightarrow\left(\Lambda_{LD}^{\textbf{4}t_{L}}\right)^{T}=\frac{\lambda_{t_{L}}}{\sqrt{2}}\begin{pmatrix}0&0&i&\cos\frac{\sqrt{2}H}{f}\end{pmatrix};\hspace{0.585 cm}\Lambda_{LD}^{\textbf{1}t_{L}}=\frac{\lambda_{t_{L}}}{\sqrt{2}}\sin\frac{\sqrt{2}H}{f},\\
&\Lambda_{LD}^{b_{L}}=U^{-1}\Lambda_{L}^{b_{L}}\Rightarrow\left(\Lambda_{LD}^{\textbf{4}b_{L}}\right)^{T}=\frac{\lambda_{t_{L}}}{\sqrt{2}}\begin{pmatrix}i&-1&0&0\end{pmatrix};\hspace{1.27 cm}\Lambda_{LD}^{\textbf{1}b_{L}}=0,\\
&\Lambda_{RD}^{t_{R}}=U^{-1}\Lambda_{R}^{t_{R}}\Rightarrow\left(\Lambda_{RD}^{\textbf{4}t_{R}}\right)^{T}=\lambda_{t_{R}}\begin{pmatrix}0&0&0&-\sin\frac{\sqrt{2}H}{f}\end{pmatrix};\hspace{0.18 cm}\Lambda_{RD}^{\textbf{1}t_{R}}=\lambda_{t_{R}}\cos\frac{\sqrt{2}H}{f}.
\end{align}
Notice that the spurion associated with $b_{L}$ is a constant (it does not contain the Higgs field) and can thus be ignored. Moreover, the 
remaining spurions multiplets are not independent, since we have 
\begin{equation}
\begin{split}
&\left(\Lambda_{LD}^{t_{L}}\right)^{\dagger}\Lambda_{LD}^{t_{L}}+\left(\Lambda_{LD}^{b_{L}}\right)^{\dagger}\Lambda_{LD}^{b_{L}}=\lambda_{t_{L}}^{2},\\
&\left(\Lambda_{RD}^{t_{R}}\right)^{\dagger}\Lambda_{RD}^{t_{R}}=\lambda_{t_{R}}^{2}.
\end{split}
\end{equation}
This implies that we can always trade $\Lambda_{LD}^{\textbf{4}t_{L}}$ for $\Lambda_{LD}^{\textbf{1}t_{L}}$ ($\Lambda_{RD}^{\textbf{4}t_{R}}$ for
$\Lambda_{RD}^{\textbf{1}t_{L}}$), and we have only 2 invariant combinations of spurions at $\mathcal{O}\left(\lambda_{t_{L},t_{R}}^{2}\right)$
\begin{equation}
\begin{split} 
&\mathcal{O}_{\lambda_{L}^{2}}=\left(\Lambda_{LD}^{\textbf{1}t_{L}}\right)^{\dagger}\Lambda_{LD}^{\textbf{1}t_{L}}=\frac{\lambda_{t_{L}}^{2}}{2}\sin^{2}\frac{\sqrt{2}H}{f},\\
&\mathcal{O}_{\lambda_{R}^{2}}=\left(\Lambda_{RD}^{\textbf{1}t_{R}}\right)^{\dagger}\Lambda_{RD}^{\textbf{1}t_{R}}=\lambda_{t_{R}}^{2}\left(1-\sin^{2}\frac{\sqrt{2}H}{f}\right).
\end{split}
\end{equation}
This mean that, to $\mathcal{O}\left(\lambda_{t_{L},t_{R}}^{2}\right)$ the contribution to the potential are
\begin{equation}
V_{\lambda^{2}}\propto\left(\frac{c_{L}}{2}\lambda_{t_{L}}^{2}-c_{R}\lambda_{t_{R}}^{2}\right)\sin^{2}\frac{\sqrt{2}H}{f}+\textup{const}.,
\end{equation}
where $c_{L}$ and $c_{R}$ are unknown constants. We see that this contribution to the potential does not provide 
$\xi=\sin^{2}\sqrt{2}\langle H\rangle/f\ll 1$ since it has a minimum in $\langle H\rangle=\pi f/2\sqrt{2}$ or $\langle H\rangle=0$. Let us then 
add operators of order $\mathcal{O}\left(\lambda^{4}_{L}\right)$, $\mathcal{O}\left(\lambda^{4}_{R}\right)$ and 
$\mathcal{O}\left(\lambda^{2}_{L}\lambda^{2}_{R}\right)$. The structure of such operators is given by the following terms
\begin{equation}
\begin{split} 
&\mathcal{O}_{\lambda_{L,1}^{4}}=\left(\mathcal{O}_{\lambda_{L}^{2}}\right)^{2}=\frac{\lambda_{t_{L}}^{4}}{4}\sin^{4}\frac{\sqrt{2}H}{f},\\ 
&\mathcal{O}_{\lambda_{L,2}^{4}}=\mathcal{O}_{\lambda_{L}^{2}}\left[\left(\Lambda_{LD}^{\textbf{4}t_{L}}\right)^{\dagger}\Lambda_{LD}^{\textbf{4}t_{L}}+\left(\Lambda_{LD}^{\textbf{4}b_{L}}\right)^{\dagger}\Lambda_{LD}^{\textbf{4}b_{L}}\right]=\frac{\lambda_{t_{L}}^{4}}{4}\left(4\sin^{2}\frac{\sqrt{2}H}{f}-\sin^{4}\frac{\sqrt{2}H}{f}\right),\\
&\mathcal{O}_{\lambda_{R,1}^{4}}=\left(\mathcal{O}_{\lambda_{R}^{2}}\right)^{2}=\lambda_{t_{R}}^{4}\left(1-2\sin^{2}\frac{\sqrt{2}H}{f}+\sin^{4}\frac{\sqrt{2}H}{f}\right),\\
&\mathcal{O}_{\lambda_{R,2}^{4}}=\mathcal{O}_{\lambda_{R}^{2}}\left[\left(\Lambda_{RD}^{\textbf{4}t_{R}}\right)^{\dagger}\Lambda_{RD}^{\textbf{4}t_{R}}\right]=\lambda_{t_{R}}^{4}\left(\sin^{2}\frac{\sqrt{2}H}{f}-\sin^{4}\frac{\sqrt{2}H}{f}\right),\\ 
&\mathcal{O}_{\lambda_{L}^{2}\lambda_{R}^{2},1}=\mathcal{O}_{\lambda_{L}^{2}}\mathcal{O}_{\lambda_{R}^{2}}=\frac{\lambda_{t_{L}}^{2}\lambda_{t_{R}}^{2}}{2}\left(\sin^{2}\frac{\sqrt{2}H}{f}-\sin^{4}\frac{\sqrt{2}H}{f}\right),\\
&\mathcal{O}_{\lambda_{L}^{2}\lambda_{R}^{2},2}=\mathcal{O}_{\lambda_{L}^{2}}\left[\left(\Lambda_{RD}^{\textbf{4}t_{R}}\right)^{\dagger}\Lambda_{RD}^{\textbf{4}t_{R}}\right]=\frac{\lambda_{t_{L}}^{2}\lambda_{t_{R}}^{2}}{2}\sin^{4}\frac{\sqrt{2}H}{f}.
\end{split}
\end{equation}
We note that the invariants constructed for $\lambda^{4}$ order are linear combinations of $\sin^{2}\theta$ and $\sin^{4}\theta$, so we can write 
the $\mathcal{O}\left(\lambda^{4}\right)$ potential as
\begin{equation}
\begin{split}
V_{\lambda^{4}}&\propto\left(c_{LL}\lambda_{t_{L}}^{4}+c_{RR}\lambda_{t_{R}}^{4}+c_{LR}\lambda_{t_{L}}^{2}\lambda_{t_{R}}^{2}\right)\sin^{2}\frac{\sqrt{2}H}{f}\\
&+\left(c'_{LL}\lambda_{t_{L}}^{4}+c'_{RR}\lambda_{t_{R}}^{4}+c'_{LR}\lambda_{t_{L}}^{2}\lambda_{t_{R}}^{2}\right)\sin^{4}\frac{\sqrt{2}H}{f}.
\end{split}
\end{equation}
Terms with the same trigonometric structure are also obtained in the gauge sector to order $g^{2}$ and $g'^{2}$ \cite{Panico}. Finally we can write 
the potential in the generic form
\begin{equation}\label{PotencialMCHM}
V\left(H\right)=-\alpha f^{2}\sin^{2}\frac{\sqrt{2}H}{f}+\beta f^{2}\sin^{4}\frac{\sqrt{2}H}{f},
\end{equation}
where the contributions coming from the gauge sectors and the fermion sectors are encoded in the parameters $\alpha$ and $\beta$. We have chosen 
the $f^{2}$ normalization and the signs for convenience. Using the minimum condition, we have
\begin{equation}
\begin{split}
\left.\frac{dV}{dH}\right|_{min.}=0=\left(2\beta\sin^{2}\frac{\sqrt{2}\langle H\rangle}{f}-\alpha\right)\sin\frac{\sqrt{2}\langle H\rangle}{f}\cos\frac{\sqrt{2}\langle H\rangle}{f},
\end{split}
\end{equation} 
from which we see that either $\langle H\rangle=0$ or
\begin{equation}
\sin^{2}\frac{\sqrt{2}\langle H\rangle}{f}=\frac{\alpha}{2\beta},
\end{equation}
using Eq. (\ref{gap}) and (\ref{masa}), we rewrite this equation as 
\begin{equation}\label{conalpha}
\xi=\frac{\alpha}{2\beta}. 
\end{equation}
Given that the mass of a scalar particle can be read from the potential, we have
\begin{equation}\label{conbeta}
m_{H}^{2}=\left.\frac{d^{2}V}{d H^{2}}\right|_{min.,\alpha=2\beta\xi}=8\xi\left(1-\xi\right)\beta.
\end{equation}
With Eqs. (\ref{conalpha}) and (\ref{conbeta}) we can write the potential in terms of the $\xi$ parameter, finding
\begin{equation}
V\left(H\right)=\frac{m_{H}^{2}f^{2}}{8\xi\left(1-\xi\right)}\left(\sin^{2}\frac{\sqrt{2}H}{f}-\xi\right)^{2}-\frac{m_{H}^{2}f^{2}\xi}{8\left(1-\xi\right)}.
\end{equation}
Let us stress that the minimum condition (\ref{conalpha}) implies
\begin{equation}
\xi=\frac{v^{2}}{f^{2}}=\frac{\alpha}{2\beta}\ll 1,
\end{equation}
to be phenomenologically admittable, so that some tuning between the parameters is needed to ensure $\alpha\ll\beta$.
\section{Case NMCHM\texorpdfstring{$_{\textbf{6}}$}{5}}
As in the case of MCHM$_{\textbf{5}}$ we can write the interactions that cause the explicit breaking of the symmetry and that give the largest contributions to 
the potential as
\begin{equation}
\begin{split}
\mathcal{L}^{\slashed{\mathcal{G}}}_{\lambda_{t}}&=\lambda_{t_{L}}\left(\overline{Q}_{t_{L}}\right)^{I}\left(\mathcal{O}^{L}_{F}\right)_{I}+\lambda_{t_{R1}}\left(\overline{T}_{R1}\right)^{I}\left(\mathcal{O}^{t_{R1}}_{F}\right)_{I}+\lambda_{t_{R2}}\left(\overline{T}_{R2}\right)^{I}\left(\mathcal{O}^{t_{R2}}_{F}\right)_{I}+h.c\\
&=\left(\overline{t}_{L}\Lambda_{L}^{t_{L}}+\overline{b}_{L}\Lambda_{L}^{b_{L}}\right)\mathcal{O}^{L}_{F}+\overline{t}_{R1}\Lambda_{R1}\mathcal{O}^{t_{R1}}_{F}+\overline{t}_{R2}\Lambda_{R2}\mathcal{O}^{t_{R2}}_{F},
\end{split}
\end{equation}
with spurion fields belonging to the \textbf{6} of $\textup{SO}\left(6\right)$ written as
\begin{align}
&\Lambda_{L}^{t_{L}}=\frac{\lambda_{t_{L}}}{\sqrt{2}}\begin{pmatrix}0&0&i&1&0&0\end{pmatrix},\\
&\Lambda_{L}^{b_{L}}=\frac{\lambda_{t_{L}}}{\sqrt{2}}\begin{pmatrix}i&-1&0&0&0&0\end{pmatrix},\\
&\Lambda_{R1}=\lambda_{t_{R1}}\begin{pmatrix}0&0&0&0&0&1\end{pmatrix},\\
&\Lambda_{R2}=\lambda_{t_{R2}}\begin{pmatrix}0&0&0&0&1&0\end{pmatrix}.
\end{align}
In this case the Goldstone matrix in the unitary gauge is
\begin{equation}
U\underset{\textup{UG}}{=}\begin{pmatrix}\mathds{1}_{3}&0&0&0\\0&E_{1}&E_{2}&E_{3}\\0&E_{2}&E_{4}&E_{5}\\0&-E_{3}&-E_{5}&E_{6}\end{pmatrix},
\end{equation}
where the $E_{k}$ elements are
\begin{equation}
\begin{split}
&E_{1}=1-\frac{H^{2}}{H^{2}+\zeta^{2}}\left(1-\cos\frac{\sqrt{H^{2}+\zeta^{2}}}{f}\right),\\
&E_{2}=-\frac{H\zeta}{H^{2}+\zeta^{2}}\left(1-\cos\frac{\sqrt{H^{2}+\zeta^{2}}}{f}\right),\\
&E_{3}=\frac{H}{\sqrt{H^{2}+\zeta^{2}}}\sin\frac{\sqrt{H^{2}+\zeta^{2}}}{f},\hspace{0.5cm}E_{4}=1-\frac{\zeta^{2}}{H^{2}+\zeta^{2}}\left(1-\cos\frac{\sqrt{H^{2}+\zeta^{2}}}{f}\right),\\
&E_{5}=\frac{\zeta}{\sqrt{H^{2}+\zeta^{2}}}\sin\frac{\sqrt{H^{2}+\zeta^{2}}}{f},\hspace{0.5 cm}E_{6}=\cos\frac{\sqrt{H^{2}+\zeta^{2}}}{f},
\end{split}
\end{equation}
where we use $H$ as the neutral component of the Higgs field times $\sqrt{2}$. By dressing procedure we obtain the multiplets 
\begin{align}
&\Lambda_{LD}^{t_{L}}=U^{-1}\Lambda_{L}^{t_{L}}\Rightarrow\left(\Lambda_{LD}^{\textbf{5}t_{L}}\right)^{T}=\frac{\lambda_{t_{L}}}{\sqrt{2}}\begin{pmatrix}0&0&i&E_{1}&E_{2}\end{pmatrix};\hspace{1.03 cm}\Lambda_{LD}^{\textbf{1}t_{L}}=\frac{\lambda_{t_{L}}}{\sqrt{2}}E_{3},\\
&\Lambda_{LD}^{b_{L}}=U^{-1}\Lambda_{L}^{b_{L}}\Rightarrow\left(\Lambda_{LD}^{\textbf{5}b_{L}}\right)^{T}=\frac{\lambda_{t_{L}}}{\sqrt{2}}\begin{pmatrix}i&-1&0&0&0\end{pmatrix};\hspace{1.08 cm}\Lambda_{LD}^{\textbf{1}b_{L}}=0,\\
&\Lambda^{t_{R1}}_{D}=U^{-1}\Lambda_{R1}\Rightarrow\left(\Lambda_{D}^{\textbf{5}t_{R1}}\right)^{T}=\lambda_{t_{R1}}\begin{pmatrix}0&0&0&-E_{3}&-E_{5}\end{pmatrix};\hspace{0.19 cm}\Lambda_{D}^{\textbf{1}t_{R1}}=\lambda_{t_{R1}}E_{6},\\
&\Lambda^{t_{R2}}_{D}=U^{-1}\Lambda_{R2}\Rightarrow\left(\Lambda_{D}^{\textbf{5}t_{R2}}\right)^{T}=\lambda_{t_{R2}}\begin{pmatrix}0&0&0&E_{2}&E_{4}\end{pmatrix};\hspace{0.85 cm}\Lambda_{D}^{\textbf{1}t_{R2}}=\lambda_{t_{R2}}E_{5}.
\end{align}
Again the multiplets are not independent because of the following relations
\begin{equation}
\begin{split}
&\left(\Lambda_{LD}\right)^{\dagger}\Lambda_{LD}=\frac{\lambda_{t_{L}}^{2}}{2}\left(3+E_{1}^{2}+E_{2}^{2}+E_{3}^{2}\right)=2\lambda_{t_{L}}^{2},\\
&\left(\Lambda^{t_{R1}}_{D}\right)^{\dagger}\Lambda^{t_{R1}}_{D}=\lambda_{t_{R1}}^{2}\left(E_{3}^{2}+E_{5}^{2}+E_{6}^{2}\right)=\lambda_{t_{R1}}^{2},\\
&\left(\Lambda^{t_{R2}}_{D}\right)^{\dagger}\Lambda^{t_{R2}}_{D}=\lambda_{t_{R2}}^{2}\left(E_{2}^{2}+E_{4}^{2}+E_{5}^{2}\right)=\lambda_{t_{R2}}^{2},
\end{split}
\end{equation}
from which we have the following combinations of spurions
\begin{equation}
\begin{split} 
&\mathcal{O}_{\lambda_{L}^{2}}=\left(\Lambda_{LD}^{\textbf{1}t_{L}}\right)^{\dagger}\Lambda_{LD}^{\textbf{1}t_{L}}=\lambda_{t_{L}}^{2}\frac{H^{2}}{H^{2}+\zeta^{2}}\sin^{2}\frac{\sqrt{H^{2}+\zeta^{2}}}{f},\\
&\mathcal{O}_{\lambda_{R1}^{2}}=\left(\Lambda_{D}^{\textbf{1}t_{R1}}\right)^{\dagger}\Lambda_{D}^{\textbf{1}t_{R1}}=\lambda_{t_{R1}}^{2}\left[1-\sin^{2}\frac{\sqrt{H^{2}+\zeta^{2}}}{f}\right],\\
&\mathcal{O}_{\lambda_{R2}^{2}}=\left(\Lambda_{D}^{\textbf{1}t_{R2}}\right)^{\dagger}\Lambda_{D}^{\textbf{1}t_{R2}}=\lambda_{t_{R2}}^{2}\frac{\zeta^{2}}{H^{2}+\zeta^{2}}\sin^{2}\frac{\sqrt{H^{2}+\zeta^{2}}}{f}.
\end{split}
\end{equation}
Then in the NMCHM at $\lambda^{2}$, the form of the potential is
\begin{equation}
\begin{split}
V_{\lambda^{2}}\left(H,\zeta\right)\propto&\Bigg\{c_{L}\frac{\lambda_{t_{L}}^{2}H^{2}}{H^{2}+\zeta^{2}}+c_{R2}\frac{\lambda_{t_{R2}}^{2}\zeta^{2}}{H^{2}+\zeta^{2}}-c_{R1}\lambda_{t_{R1}}^{2}\Bigg\}\sin^{2}\frac{\sqrt{H^{2}+\zeta^{2}}}{f}+\textup{const}.
\end{split}
\end{equation}
This expression has critical points corresponding to $\left(\langle H\rangle,\langle\zeta\rangle\right)=\left(0,\left(k+\frac{1}{2}\right)f\pi\right)$,\\
$\left(\langle H\rangle,\langle\zeta\rangle\right)=\left(\left(k+\frac{1}{2}\right)f\pi,0\right)$,
$\left(\langle H\rangle,\langle\zeta\rangle\right)=\left(0,0\right)$ and the family $\langle H\rangle^{2}+\langle\zeta\rangle^{2}=\left(kf\pi\right)^{2}$.
Independently on the nature of the critical point, Eq. (\ref{VEVNM}) gives either $v=0$ or $v=f$, both phenomenologically unacceptable. Let us 
see the contribution of operators to next order, where the structure of the potential is provided by the following terms:
\begin{align}
\notag
\mathcal{O}_{\lambda_{L,1}^{4}}&=\left(\mathcal{O}_{\lambda_{L}^{2}}\right)^{2}=\lambda_{t_{L}}^{4}\frac{H^{4}}{\left(H^{2}+\zeta^{2}\right)^{2}}\sin^{4}\frac{\sqrt{H^{2}+\zeta^{2}}}{f},\\\notag
\mathcal{O}_{\lambda_{L,2}^{4}}&=\mathcal{O}_{\lambda_{L}^{2}}\left[\left(\Lambda_{LD}^{\textbf{5}t_{L}}\right)^{\dagger}\Lambda_{LD}^{\textbf{5}t_{L}}+\left(\Lambda_{LD}^{\textbf{5}b_{L}}\right)^{\dagger}\Lambda_{LD}^{\textbf{5}b_{L}}\right]\\\notag
&=\lambda_{t_{L}}^{4}\frac{H^{2}}{H^{2}+\zeta^{2}}\left(4\sin^{2}\frac{\sqrt{H^{2}+\zeta^{2}}}{f}-\frac{H^{2}}{H^{2}+\zeta^{2}}\sin^{4}\frac{\sqrt{H^{2}+\zeta^{2}}}{f}\right),\\\notag
\mathcal{O}_{\lambda_{R1,1}^{4}}&=\left(\mathcal{O}_{\lambda_{R1}^{2}}\right)^{2}=\lambda_{t_{R1}}^{4}\left(1-2\sin^{2}\frac{\sqrt{H^{2}+\zeta^{2}}}{f}+\sin^{4}\frac{\sqrt{H^{2}+\zeta^{2}}}{f}\right),\\\notag
\mathcal{O}_{\lambda_{R1,2}^{4}}&=\mathcal{O}_{\lambda_{R1}}^{2}\left[\left(\Lambda_{D}^{\textbf{5}R1}\right)^{\dagger}\left(\Lambda_{D}^{\textbf{5}R1}\right)^{\dagger}\right]=\lambda_{t_{R1}}^{4}\left(\sin^{2}\frac{\sqrt{H^{2}+\zeta^{2}}}{f}-\sin^{4}\frac{\sqrt{H^{2}+\zeta^{2}}}{f}\right),\\\notag
\mathcal{O}_{\lambda_{R2,1}^{4}}&=\left(\mathcal{O}_{\lambda_{R2}^{2}}\right)^{2}=\lambda_{t_{R2}}^{4}\frac{\zeta^{4}}{\left(H^{2}+\zeta^{2}\right)^{2}}\sin^{4}\frac{\sqrt{H^{2}+\zeta^{2}}}{f},\\\notag
\mathcal{O}_{\lambda_{R2,2}^{4}}&=\mathcal{O}_{\lambda_{R2}}^{2}\left[\left(\Lambda_{D}^{\textbf{5}R2}\right)^{\dagger}\left(\Lambda_{D}^{\textbf{5}R1}\right)^{\dagger}\right]\\
&=\lambda_{t_{R2}}^{4}\frac{\zeta^{2}}{H^{2}+\zeta^{2}}\left(\sin^{2}\frac{\sqrt{H^{2}+\zeta^{2}}}{f}-\frac{\zeta^{2}}{H^{2}+\zeta^{2}}\sin^{4}\frac{\sqrt{H^{2}+\zeta^{2}}}{f}\right),\\\notag
\mathcal{O}_{\lambda_{L}^{2}\lambda_{R1,1}^{2}}&=\mathcal{O}_{\lambda_{L}^{2}}\mathcal{O}_{\lambda_{R1}^{2}}=\lambda_{t_{L}}^{2}\lambda_{t_{R1}}^{2}\frac{H^{2}}{H^{2}+\zeta^{2}}\left(\sin^{2}\frac{\sqrt{H^{2}+\zeta^{2}}}{f}-\sin^{4}\frac{\sqrt{H^{2}+\zeta^{2}}}{f}\right),\\\notag
\mathcal{O}_{\lambda_{L}^{2}\lambda_{R1,2}^{2}}&=\mathcal{O}_{\lambda_{L}^{2}}\left(\Lambda_{D}^{\textbf{5}R1}\right)^{\dagger}\Lambda_{D}^{\textbf{5}R1}=\lambda_{t_{L}^{2}}\lambda_{t_{R1}^{2}}\frac{H^{2}}{H^{2}+\zeta^{2}}\sin^{4}\frac{\sqrt{H^{2}+\zeta^{2}}}{f},\\\notag
\mathcal{O}_{\lambda_{L}^{2}\lambda_{R2,1}^{2}}&=\mathcal{O}_{\lambda_{L}^{2}}\mathcal{O}_{\lambda_{R2}^{2}}=\lambda_{t_{L}}^{2}\lambda_{t_{R2}}^{2}\frac{H^{2}\zeta^{2}}{\left(H^{2}+\zeta^{2}\right)^{2}}\sin^{4}\frac{\sqrt{H^{2}+\zeta^{2}}}{f},\\\notag
\mathcal{O}_{\lambda_{L}^{2}\lambda_{R2,2}^{2}}&=\mathcal{O}_{\lambda_{L}^{2}}\left(\Lambda_{D}^{\textbf{5}R2}\right)^{\dagger}\Lambda_{D}^{\textbf{5}R2}\\\notag
&=\lambda_{t_{L}^{2}}\lambda_{t_{R2}^{2}}\frac{H^{2}}{H^{2}+\zeta^{2}}\left(\sin^{2}\frac{\sqrt{H^{2}+\zeta^{2}}}{f}-\frac{\zeta^{2}}{H^{2}+\zeta^{2}}\sin^{4}\frac{\sqrt{H^{2}+\zeta^{2}}}{f}\right),\\\notag
\mathcal{O}_{\lambda_{R1}^{2}\lambda_{R2}^{2}}&=\lambda_{t_{R1}}^{2}\lambda_{t_{R2}}^{2}\frac{\zeta^{2}}{H^{2}+\zeta^{2}}\left(\sin^{2}\frac{\sqrt{H^{2}+\zeta^{2}}}{f}-\sin^{4}\frac{\sqrt{H^{2}+\zeta^{2}}}{f}\right).\notag
\end{align}
From the above invariants, and factoring common terms, we found the $\mathcal{O}\left(\lambda^{4}\right)$ potential as
\begin{equation}
\begin{split} 
V_{\lambda^{4}}&\propto\Bigg\{\left(c'_{LL}\lambda^{4}_{t_{L}}+c'_{R1R1}\lambda^{4}_{t_{R1}}+c'_{LR1}\lambda^{2}_{t_{L}}\lambda^{2}_{t_{R1}}+c'_{LR2}\lambda^{2}_{t_{L}}\lambda^{2}_{t_{R2}}\right)H^{2}\\
&+\left(c'_{R1R1}\lambda^{4}_{t_{R1}}+c'_{R2R2}\lambda^{4}_{t_{R2}}+c'_{R1R2}\lambda_{t_{R1}}^{2}\lambda_{t_{R2}}^{2}\right)\zeta^{2}\Bigg\}\frac{1}{H^{2}+\zeta^{2}}\sin^{2}\frac{\sqrt{H^{2}+\zeta^{2}}}{f}\\
&+\Bigg\{\left(c_{LL}\lambda^{4}_{t_{L}}+c_{LR1}\lambda^{2}_{t_{L}}\lambda^{2}_{t_{R1}}+c_{R1R1}\lambda^{4}_{t_{R1}}\right)H^{4}\\
&+\left(c_{LR1}\lambda^{2}_{t_{L}}\lambda^{2}_{t_{R1}}+c_{LR2}\lambda^{2}_{t_{L}}\lambda^{2}_{t_{R2}}+c_{R1R2}\lambda_{t_{R1}}^{2}\lambda_{t_{R2}}^{2}+2c_{R1R1}\lambda^{4}_{t_{R1}}\right)H^{2}\zeta^{2}\\
&+\left(c_{R1R1}\lambda^{4}_{t_{R1}}+c_{R2R2}\lambda^{4}_{t_{R2}}+c_{R1R2}\lambda_{t_{R1}}^{2}\lambda_{t_{R2}}^{2}\right)\zeta^{4}\Bigg\}\frac{1}{\left(H^{2}+\zeta^{2}\right)^{2}}\sin^{4}\frac{\sqrt{H^{2}+\zeta^{2}}}{f}.
\end{split}
\end{equation}
Finally, the final form of the potential is
\begin{equation}
\begin{split}
V\left(H,\zeta\right)=&-f^{2}\frac{\left(\alpha_{1}H^{2}+\alpha_{2}\zeta^{2}\right)}{H^{2}+\zeta^{2}}\sin^{2}\frac{\sqrt{H^{2}+\zeta^{2}}}{f}\\
&+f^{2}\frac{\left(\beta_{1}H^{4}+\beta_{2}\zeta^{4}+\beta_{3}H^{2}\zeta^{2}\right)}{\left(H^{2}+\zeta^{2}\right)^{2}}\sin^{4}\frac{\sqrt{H^{2}+\zeta^{2}}}{f}.
\end{split}
\end{equation}
Again, $\alpha_{i}$ and $\beta_{i}$ encoded the contributions from gauge and the fermion sectors. The minimum conditions of the potential are
\begin{align}
\notag
\left.\frac{\partial V}{\partial H}\right|_{min.}=&\frac{H}{\left(H^{2}+\zeta^{2}\right)^{2}}\sin\theta\Bigg\{\frac{f\zeta^{2}}{H^{2}+\zeta^{2}}\left[\zeta^{2}\left(\beta_{3}-2\beta_{2}\right)+H^{2}\left(2\beta_{1}-\beta_{3}\right)\right]\sin^{3}\theta\\\label{nomHiggs}
&+\frac{1}{\left(H^{2}+\zeta^{2}\right)^{1/2}}\bigg\{H^{4}\left(2\beta_{1}\sin^{2}\theta-\alpha_{1}\right)+\zeta^{4}\left(2\beta_{2}\sin^{2}\theta-\alpha_{2}\right)\\&+H^{2}\zeta^{2}\left(2\beta_{3}\sin^{2}\theta-\alpha_{1}-\alpha_{2}\right)\bigg\}\cos\theta+f\zeta^{2}\left(\alpha_{2}-\alpha_{1}\right)\sin\theta\left.\Bigg\}\right|_{min.}=0,\notag\\[0.5cm]
\notag
\left.\frac{\partial V}{\partial\zeta}\right|_{min.}=&\frac{\zeta}{\left(H^{2}+\zeta^{2}\right)^{2}}\sin\theta\Bigg\{\frac{fH^{2}}{H^{2}+\zeta^{2}}\left[H^{2}\left(\beta_{3}-2\beta_{1}\right)+\zeta^{2}\left(2\beta_{2}-\beta_{3}\right)\right]\sin^{3}\theta\\
&+\frac{1}{\left(H^{2}+\zeta^{2}\right)^{1/2}}\bigg\{H^{4}\left(2\beta_{1}\sin^{2}\theta-\alpha_{1}\right)+\zeta^{4}\left(2\beta_{2}\sin^{2}\theta-\alpha_{2}\right)\\&+H^{2}\zeta^{2}\left(2\beta_{3}\sin^{2}\theta-\alpha_{1}-\alpha_{2}\right)\bigg\}\cos\theta+fH^{2}\left(\alpha_{1}-\alpha_{2}\right)\sin\theta\left.\Bigg\}\right|_{min.}=0,\notag
\end{align}
where
\begin{equation}
\theta\left(H,\zeta\right)=\frac{\sqrt{H^{2}+\zeta^{2}}}{f}.
\end{equation}
Considering $\langle H\rangle=0$ or the family $\langle H\rangle^{2}+\langle\zeta\rangle^{2}=\left(kf\pi\right)^{2}$, we have an $v=0$,
phenomenologically unacceptable. But if $\langle\zeta\rangle=0$, then from (\ref{nomHiggs}) we obtain the additional possibility
\begin{equation}
\left(2\beta_{1}\sin^{2}\langle\theta\rangle-\alpha_{1}\right)\sin\langle\theta\rangle\cos\langle\theta\rangle=0.
\end{equation}
Again we see that either $\langle H\rangle=0$ (phenomenologically unacceptable) or
\begin{equation}\label{NMcon}
\sin^{2}\langle\theta\rangle=\xi=\frac{\alpha_{1}}{2\beta_{1}}.
\end{equation}
So when $\langle\zeta\rangle=0$ and $\langle H\rangle\neq 0$ we have a critical point that under certain conditions produces a minimum of potential.
From the potential and the above considerations, we obtain the mass of $H$ and $\zeta$
\begin{align}
&m^{2}_{H}=\left.\frac{\partial^{2}V}{\partial H^{2}}\right|_{min.,\alpha_{1}=2\beta_{1}\xi}=8\xi\left(1-\xi\right)\beta_{1},\\
&m^{2}_{\zeta}=\left.\frac{\partial^{2}V}{\partial\zeta^{2}}\right|_{min.,\alpha_{1}=2\beta_{1}\xi}=\frac{2f^{2}\xi}{\langle H\rangle^{2}}\left(\xi\beta_{3}-\alpha_{2}\right).
\end{align}
The condition (\ref{NMcon}) implies that to ensure $\alpha_{1}\ll\beta_{2}$, some tuning between the parameters is necessary. 
Also under these conditions the (NMCHM)$_{\textbf{6}}$ results are
\begin{align}
\notag
\mathcal{L}=&-\frac{1}{4}W^{a}_{\mu\nu}W_{a}^{\mu\nu}-\frac{1}{4}B_{\mu\nu}B^{\mu\nu}+\frac{1}{2}\partial_{\mu}h_{p}\partial^{\mu}h_{p}+\frac{1}{2}\partial_{\mu}\eta_{p}\partial^{\mu}\eta_{p}+\frac{g^{2}v^{2}}{4}\left(|W_{\mu}|^{2}+\frac{1}{2c^{2}_{w}}Z_{\mu}^{2}\right)\\\notag
&+\frac{g^{2}v^{2}}{4}\left(|W_{\mu}|^{2}+\frac{1}{2c^{2}_{w}}Z_{\mu}^{2}\right)
\Bigg\{2\sqrt{1-\xi}\frac{h_{p}}{v}+\left(1-2\xi\right)\frac{h_{p}^{2}}{v^{2}}\\\notag
&-\frac{\eta_{p}^{2}}{v^{2}}+V\sqrt{1-\xi}\frac{\eta_{p}^{2}}{v^{3}}+\cdots\Bigg\}-m_{t}\overline{t}t-k_{t}^{\textbf{6}}\frac{m_{t}}{v}h_{p}\overline{t}t-C_{2t}^{\textbf{6}}\frac{m_{t}}{v^{2}}h_{p}^{2}\overline{t}t\\
&-D_{2t}^{\textbf{6}}\frac{m_{t}}{v^{2}}\eta_{p}^{2}\overline{t}t-E_{2t}^{\textbf{6}}\frac{m_{t}}{v^{3}}\eta_{p}^{2}\overline{t}t-m_{b}\overline{b}b-k_{b}^{\textbf{6}}\frac{m_{b}}{v}h_{p}\overline{b}b-\cdots,
\end{align}
where $D_{2t}^{\textbf{6}}=-\frac{1}{2}$ and $E_{2t}^{\textbf{6}}=\frac{V\left(1-2\xi\right)}{2\sqrt{1-\xi}}$ and $k_{t}^{\textbf{6}}$, 
$C_{2t}^{\textbf{6}}$, $k_{b}^{\textbf{6}}$ are given by (\ref{coeff}). In the Lagrangian we have that the $\eta_{p}$ state has parity symmetry 
$\eta_{p}\rightarrow-\eta_{p}$, that makes $\eta_{p}$ to be stable and as good Dark Matter candidate (see \cite{Frigerio,Gripaios}).
\chapter{Conclusions}
In this work we have studied and presented the basic characteristics of the composite Higgs scenario, which have been applied to the cosets 
$\textup{SO}\left(6\right)/\textup{SO}\left(5\right)$ and $\textup{SO}\left(5\right)/\textup{SO}\left(4\right)$. In both cases we obtained 
a doublet for the Higgs, as well as modifications with respect to the Standard Model of the couplings between the physical Higgs and 
the Fermionic and Gauge sectors.\\[0.5 cm]
In the minimal model the modifications to the couplings depend on the $\xi$ parameter, unlike in the non-minimal model where the modifications
depend not only on $\xi$, but on the Vacuum Expected Values of the $H$ and $\zeta$ fields as well.\\[0.5 cm]
We have presented in a systematic way the fermions embedding process, which we did in the fundamental representation of the symmetry group,
but such a procedure is similar for any representation that we take. The phenomenology of the models depends on the representation of the fermions 
embedding process, since this determines the shape of the Invariants to form.\\[0.5 cm]
In the two studied cosets the Higgs-Gauge interactions are established by symmetry breaking pattern, however in the fermionic sector the 
Higgs-quark interactions depend on the representation used.\\[0.5 cm]
The estimated composite Higgs potential for the minimal case has a functional dependence on $H$ through a linear combination of
$\sin^{2}$ and $\sin^{4}$, while the non-minimal case is a linear combination of functions on $H$ and $\zeta$.
\appendix
\chapter{The Standard Model}\label{SM}
The Standard Model (SM) of particle physics based on the works of Weinberg \cite{Weinberg}, Glashow \cite{Glashow} and Salam \cite{Salam}
is a theory which describes the electromagnetic, weak, and strong interactions. A complete discussion of the SM is given in \cite{Schwartz}.
It is a non-abelian quantum field theory, invariant under the Lie symmetry group of local transformations
$G_{SM}=\textup{SU}(3)_{C}\times\textup{SU}(2)_{L}\times\textup{U}(1)_{Y}$, where the indices $C$, $L$ and $Y$ denote respectively  color, 
left hand isospin and hypercharge. To be phenomenologically viable, the SM gauge group must be spontaneously broken to $\textup{U}(1)_{em}$. 
The gauge group determines the interactions and the number of vector bosons that correspond 
to the generators of the group, for which 
\begin{itemize}
 \item $\textup{SU}(3)_{C}$ has eight non-massive gauge bosons called gluons ($G_{\mu}$), which are the mediators of strong interaction;
 \item ${SU}(2)_{L}\times\textup{U}(1)_{Y}$ has four gauge bosons, three of which are massive ($W^{\pm}_{\mu}$ and $Z_{\mu}$) and one 
 massless (the photon, $A_{\mu}$). These are responsible for weak and electromagnetic interactions.
\end{itemize}
The matter content  (quarks and leptons) in the theory with their gauge transformation properties under 
$\textup{SU}(3)_{C}\times\textup{SU}(2)_{L}\times\textup{U}(1)_{Y}$ are presented in Table \ref{tableMatter}.\\
All the matter content presented in Table \ref{tableMatter} comes in three generations or flavors, see Table \ref{leptons} 
for leptons\footnote{Neutrino masses are exactly zero in the Standard Model which is not correct according to the neutrino oscillation 
experiments \cite{neutrinos,neutrinos2}.} and quarks.
\begin{table}[t]
\begin{center}
\begin{tabular}{|c|c|c|c|}
\hline
$\textup{Matter}$ & $\textup{SU}(3)_{C}$ & $\textup{SU}(2)_{L}$ & $\textup{U}(1)_{Y}$\\\hline
$l_{L}$ & 1 & 2 & -1/2 \\  \hline
$e_{R}$ & 1 & 1 & -1 \\  \hline
$q_{L}$ & 3 & 2 & 1/6  \\  \hline
$u_{R}$ & 3 & 1 & 2/3 \\ \hline
$d_{R}$ & 3 & 1 & -1/3 \\ \hline
\end{tabular}\caption{The matter content in the SM. The subscripts $L$ and $R$ distinguish between between left and right
handed fields.}\label{tableMatter}
\end{center}
\end{table}
\begin{table}[t]
\begin{center}
\begin{tabular}{|c|c|c|c|c|c|c|}\hline\multicolumn{7}{|c|}{\textbf{Leptons}} \\
  \hline
$Q$ & Flavor & Mass (MeV) & Flavor & Mass (MeV) & Flavor & Mass (MeV)\\\hline
-1 & Electron $e$ & 0.511 & Muon $\mu$ & 105.7 & Tau $\tau$ & 1777\\ \hline
0& $\nu_{e}$ & 0 & $\nu_{\mu}$ & 0 & $\nu_{\tau}$ & 0\\\hline
\multicolumn{7}{|c|}{\textbf{Quarks}}\\\hline
+2/3 & Up $u$ & 2.2 & Charm $c$ & 1.28$\times 10^{3}$ & Top $t$ & 173.1$\times 10^{3}$\\ \hline
-1/3 & Down $d$ & 4.7 & Strange $s$ & 96 & Bottom $b$ & 4.18$\times 10^{3}$\\\hline  
\end{tabular}\caption{Leptons $l$ (Antileptons $\overline{l}$), with spin=1/2. The $\nu_{j}$ ($j=e,\mu,\tau$) are the corresponding neutrino 
flavors. Quarks $q$ (Antiquarks $\overline{q}$), with spin=1/2. Taken from \cite{PDG}}\label{leptons}
\end{center}
\end{table}
\newpage
\justify
We need a scalar sector to drive the spontaneous symmetry breaking (in this case is called ElectroWeak Symmetry Breaking, EWSB), which must 
be charged under $\textup{SU}(2)_{L}$, because otherwise it cannot break the symmetry. This complex scalar field $H$, known as the Higgs field, 
transforms in the representation $(1,2,1/2)$ of $\textup{SU}(3)_{C}\times\textup{SU}(2)_{L}\times\textup{U}(1)_{Y}$.
The Higgs field acquires a nonzero VEV that spontaneously breaks $\textup{SU}(2)_{L}\times\textup{U}(1)_{Y}$ into $\textup{U}(1)_{em}$.\\
The dynamics is described by the following Lagrangian invariant under the $G_{SM}$ gauge group
\begin{equation}
\mathcal{L}=\mathcal{L}_{Gauge}+\mathcal{L}_{Matter}+\mathcal{L}_{Higgs}+\mathcal{L}_{Yuk}.
\end{equation}
The Lagrangian $\mathcal{L}_{Gauge}$ consists in terms of Yang- Mills type that explain the kinetic terms of the gauge bosons that correspond to
each group of the SM, that is to say
\begin{equation}
\mathcal{L}_{Gauge}=-\frac{1}{4}G^{\mu\nu}_{a}G_{\mu\nu}^{a}-\frac{1}{4}W^{\mu\nu}_{a}W_{\mu\nu}^{a}-\frac{1}{4}B^{\mu\nu}B_{\mu\nu},
\end{equation}
where
\begin{equation}
\begin{split} 
&G_{\mu\nu}^{a}=\partial_{\mu}G^{a}_{\nu}-\partial_{\nu}G^{a}_{\mu}+g_{s}f^{bca}G_{\mu}^{b}G_{\nu}^{c},\\
&W_{\mu\nu}^{a}=\partial_{\mu}W^{a}_{\nu}-\partial_{\nu}W^{a}_{\mu}+gf^{bca}W_{\mu}^{b}W_{\nu}^{c},\\
&B_{\mu\nu}=\partial_{\mu}B_{\nu}-\partial_{\nu}B_{\mu},
\end{split}
\end{equation}
with $g_{s}$ and $g$ the couplings of the $\textup{SU}(3)_{c}$, and $\textup{SU}(2)_{L}$ groups, respectively. The terms 
$g_{s}f^{bca}G_{\mu}^{b}G_{\nu}^{c}$ and $gf^{bca}W_{\mu}^{b}W_{\nu}^{c}$, typical of non-Abelian gauge theories, give rise to the three and 
four-gauge bosons interactions. The structure constants $f^{bca}$ ($a,b,c=1,\cdots,N^{2}-1$) are defined by
\begin{equation}
\left[T^{a},T^{b}\right]=if^{abc}T^{c}.
\end{equation}
The matter content of the SM is described in $\mathcal{L}_{Matter}$. The coupling of the matter fields (described in Table \ref{leptons}) 
with the gauge fields is done through the covariant derivative, namely
\begin{equation}
\mathcal{L}_{Matter}=i\sum_{k=1}^{3}\left(\overline{q}_{L}^{k}\slashed{D}^{k}q_{L}^{k}+\overline{l}_{L}^{k}\slashed{D}^{k}l_{L}^{k}+
\overline{e}_{R}^{k}\slashed{D}^{k}e_{R}^{k}+\overline{u}_{R}^{k}\slashed{D}^{k}u_{R}^{k}+\overline{d}_{R}^{k}\slashed{D}^{k}d_{R}^{k}\right),
\end{equation}
where $\slashed{D}=\gamma^{\mu}D_{\mu}$, and the index $k$ runs over each of the flavors of the fermions. For example, considering leptons (which 
transforms as $\sim(1,2,-1/2)$), we have
\begin{equation}
\overline{l}_{L}^{k}\slashed{D}^{k}l_{L}^{k}=\overline{l}_{L}^{k}\gamma^{\mu}\left(\partial_{\mu}-i\frac{g}{2}W_{\mu}^{a}\sigma_{a}
+i\frac{g'}{2}B_{\mu}\right)l_{L}^{k}.
\end{equation}
The Higgs sector is responsible for the EWSB (as we will see in section \ref{EWSB}) and the Higgs Lagrangian is given by
\begin{equation}
\mathcal{L}_{Higgs}=\left(D_{\mu}H\right)^{\dagger}\left(D^{\mu}H\right)+\mu^{2}H^{\dagger}H-\lambda\left(H^{\dagger}H\right)^{2}, 
\end{equation}
where $\lambda$ and $\mu^{2}$ are constants $>0$. The Higgs field is a doublet of $\textup{SU}\left(2\right)_{L}$ and can be written as  
\begin{equation}\label{Higgs}
H=\begin{pmatrix}H^{+}\\H_{0}\end{pmatrix}=\frac{1}{\sqrt{2}}\begin{pmatrix}H_{2}+iH_{1}\\H_{4}+iH_{3}\end{pmatrix},
\end{equation}
with covariant derivative given by
\begin{equation}\label{DcovHiggs}
D_{\mu}H=\left(\partial_{\mu}-igW^{\alpha}_{\mu}\frac{\sigma_{\alpha}}{2}-ig'B_{\mu}\frac{1}{2}\right)H. 
\end{equation}
Finally there are interactions between two SM fermions and the Higgs boson given by Yukawa type terms, namely
\begin{equation}
\mathcal{L}_{Yuk}=-\sum_{k,j=1}^{3}\left(Y^{kj}_{e}\overline{l}^{k}_{L}He^{j}_{R}+Y^{kj}_{d}\overline{q}^{k}_{L}Hd^{j}_{R}+Y^{kj}_{u}\overline{q}^{k}_{L}H^{c}u^{j}_{R}\right)+ h.c.,
\end{equation}
where $Y^{kj}_{e}$, $Y^{kj}_{d}$ and $Y^{kj}_{u}$ are the coupling constants between the quarks and the Higgs field and remembering that $H^{c}$ is
the conjugate of the doublet $H$.
\section{Electroweak Symmetry Breaking in the Standard Model}\label{EWSB}
In order to provide mass to the $W^{\pm}_{\mu}$, $Z_{\mu}$ bosons and to the SM fermions, the gauge group must be spontaneously broken through the
Higgs mechanism, for which the $\mathcal{L}_{Higgs}$ sector is used. The Higgs field acquires a VEV such that the symmetry 
$\textup{SU}\left(2\right)_{L}\times\textup{U}\left(1\right)_{Y}$ is spontaneously broken. This VEV is induced by the potential $V(H)=-\mu^{2}|H|^{2}+\lambda|H|^{4}$, namely
\begin{equation}
\left.\frac{dV}{d|H|^{2}}\right|_{min}=0=-\mu^{2}+2\lambda|H|^{2}\Rightarrow|\left\langle H\right\rangle|^{2}=v^{2}=\frac{\mu^{2}}{2\lambda}.
\end{equation}
There are infinite degenerate vacuums which are equivalent. Conventionally, we choose 
\begin{equation}
\left\langle H_{1}\right\rangle=\left\langle H_{2}\right\rangle=\left\langle H_{3}\right\rangle=0,\hspace{0.5 cm}\left\langle H_{4}\right\rangle=v=\sqrt{\frac{\mu^{2}}{2\lambda}},
\end{equation}
from which
\begin{equation}\label{vacio}
\left\langle H\right\rangle=\frac{1}{\sqrt{2}}\begin{pmatrix}0\\v\end{pmatrix}.
\end{equation}
Once the Higgs acquires the VEV, we obtain
\begin{equation}\label{parmasa}
D_{\mu}\left\langle H\right\rangle=-\frac{iv}{2\sqrt{2}}\begin{pmatrix}g\left(W^{1}_{\mu}-iW^{2}_{\mu}\right)\\-gW_{\mu}^{3}+g'B_{\mu}\end{pmatrix},
\end{equation}
with which we have
\begin{equation}
\mathcal{L}_{mass}=\frac{1}{4}g^{2}v^{2}W_{\mu}^{+}W^{\mu}_{-}+\frac{1}{8}v^{2}\left(g^{2}+g'^{2}\right)Z_{\mu}Z^{\mu},
\end{equation}
where we have defined the fields of the charged weak interactions as
\begin{equation}
W^{\pm}_{\mu}=\frac{1}{\sqrt{2}}\left(W^{1}_{\mu}\mp iW^{2}_{\mu}\right),
\end{equation}
and the boson $Z_{\mu}$ that mediates the neutral weak interactions as a combination of bosons $W^{3}_{\mu}$ and $B_{\mu}$. We see that after EWSB 
the bosons $W^{\pm}_{\mu}$ and $Z_{\mu}$ have acquired a mass
\begin{equation}\label{masabosones}
M_{W}=\frac{1}{2}gv,\hspace{0.5 cm}M_{Z}=\frac{1}{2}\left(g^{2}+g'^{2}\right)v,
\end{equation}
and the boson $A_{\mu}$ that is one combination orthogonal to $Z_{\mu}$, has not acquired mass. Such a boson is thus identified with the photon.\\
When the VEV causes the spontaneous symmetry breaking, $\textup{SU}\left(2\right)_{L}\times\textup{U}\left(1\right)_{Y}\rightarrow\textup{U}\left(1\right)_{em}$, the unbroken subgroup
$\textup{U}\left(1\right)_{em}$ that corresponds to the electromagnetic interaction has as generator the electric charge $Q$, which is a linear 
combination of generators that annihilates the vacuum (\ref{vacio}) and is identified as
\begin{equation}
Q=\frac{1}{2}\sigma_{3}+Y=T_{3}+Y. 
\end{equation}
After EWSB in the Yukawa sector, we have
\begin{equation}
\mathcal{L}_{Yuk}=-v\overline{e}_{L}Y_{e}e_{R}-v\overline{d}_{L}Y_{d}e_{R}-v\overline{u}_{L}Y_{u}e_{R}.
\end{equation}
We have obtained the Dirac mass terms for leptons and quarks.\\[0.5 cm]
Now, without loss of generality, the Higgs doublet (\ref{Higgs}) can be written in the direction of the broken generators $T_{1}$, $T_{2}$ and 
$T_{3}-Y$ (CCWZ) as
\begin{equation}
H=\frac{1}{\sqrt{2}}\exp\left[\frac{i}{2v}\left(\sigma^{a}H_{a}-\mathds{1}H_{3}\right)\right]\begin{pmatrix}0\\h\left(x\right)+v\end{pmatrix}.
\end{equation}
Due to the local symmetry of the theory, we can make a gauge transformation
\begin{equation}
H\rightarrow H'=UH=\exp\left[-\frac{i}{2v}\left(\sigma^{a}H_{a}-\mathds{1}H_{3}\right)\right]H,
\end{equation}
to eliminate the would-be Goldstone bosons from the theory (these Goldstone bosons are ``eaten" by the three gauge bosons that acquire mass 
in the EWSB), obtaining 
\begin{equation}
H=\frac{1}{\sqrt{2}}\begin{pmatrix}0\\h\left(x\right)+v\end{pmatrix}.
\end{equation}
This is the so-called unitary gauge, where $H\left(x\right)$ has one degree of freedom and the physical implications of the 
theory are clear to see.
\section{Custodial symmetry}\label{Custodial}
From the expression for the masses of the $W$ and $Z$ bosons eq. (\ref{masabosones}), at the tree level we have that
\begin{equation}\label{parametrorho}
\rho=\frac{M_{W}^{2}}{M_{Z}^{2}c_{w}^{2}}=1. 
\end{equation}
Experimentally the value of $\rho$ has been confirmed to a per mille accuracy \cite{PDG}. This result is a consequence of the spontaneous symmetry 
breaking by $\textup{SU}\left(2\right)$-doublets and guaranteed by a symmetry. To see this symmetry let us see that the potential $V(H)$ is 
invariant under a $\textup{SO}(4)$ symmetry. Given that $2H^{\dagger}H=H_{1}^{2}+H_{2}^{2}+H_{3}^{2}+H_{4}^{2}$, we can write
\begin{equation}
V\left(H\right)=-\frac{\mu^{2}}{2}\boldsymbol{H}^{T}\boldsymbol{H}+\frac{\lambda}{4}\left(\boldsymbol{H}^{T}\boldsymbol{H}\right)^{2},
\end{equation}
where $\boldsymbol{H}^{T}=\left(H_{1},H_{2},H_{3},H_{4}\right)$. This means that $V\left(H\right)$ is invariant under the transformation
\begin{equation}
\begin{split}
&\boldsymbol{H}\rightarrow\boldsymbol{H}'=R\boldsymbol{H},\\
&\boldsymbol{H}\rightarrow\boldsymbol{H}'^{T}=\boldsymbol{H}^{T}R^{T},
\end{split}
\end{equation} 
such that $R^{T}R=\mathds{1}$ which means that $V\left(H\right)$ is invariant under  $\textup{SO}(4)$. When $H$ acquires a VEV
($\left\langle H_{4}\right\rangle=v$, and $\left\langle H_{1}\right\rangle=\left\langle H_{2}\right\rangle=\left\langle H_{3}\right\rangle=0$), 
it causes the symmetry breaking pattern $\textup{SO}(4)\rightarrow\textup{SO}(3)$. So in the Higgs sector there remain three directions of 
unbroken symmetry, that is there is a residual symmetry, $\textup{SO}(3)\simeq\textup{SU}(2)$, after the EWSB, which is known as custodial symmetry
\footnote{The original literature is given in \cite{cust}.}. From Eq. (\ref{parmasa}) we can write the $\mathcal{L}_{mass}$ as
\begin{equation}
\mathcal{L}_{mass}=\boldsymbol{W}^{T}\mathcal{M}^{2}\boldsymbol{W},
\end{equation}
where $\mathcal{M}^{2}$ is the mass-squared matrix and $\boldsymbol{W}$ the vector of gauge fields, given by
\begin{equation}
\mathcal{M}^{2}=\frac{v^{2}}{8}\begin{pmatrix}g^{2}&0&0&0\\0&g^{2}&0&0\\0&0&g^{2}&-g'g\\0&0&-gg'&g'^{2}\end{pmatrix},\hspace{0.5 cm}
\boldsymbol{W}=
\begin{pmatrix}W_{\mu}^{1}\\W_{\mu}^{2}\\W_{\mu}^{3}\\B_{\mu}\end{pmatrix}.
\end{equation}
In the limit $g'\rightarrow 0$, we have $c_{w}=g/\sqrt{g^{2}+g'^{2}}\rightarrow 1$, and under $\textup{SO}(3)$ the $W_{\mu}^{a}$ gauge bosons transform 
as a triplet, and are thus degenerate. We notice that the ratio
\begin{equation}
\frac{M_{W}^{2}}{M_{Z}^{2}}=1,
\end{equation}
is required by the custodial symmetry. When $g'\neq 0$, we obtain the relation (\ref{parametrorho}).\\
We see that this symmetry is responsible for keeping the ratio between the mass of $Z$ and $W$ ensuring $\rho=1$, and also ensures small 
corrections to $\rho$. Theories without  custodial symmetry will give large corrections to the Peskin-Takeuchi $T$ parameter, and in order to avoid
this, a viable composite Higgs models should include such symmetry.
\chapter{The Goldstone matrix}\label{CalGoldstone}
In this dissertation we saw that the Goldstone matrix is the main object of the CCWZ construction, since it allows us to see the properties of the
Goldstone bosons under the action of the $\mathcal{G}$ group, it also allows us to build the symbols $d_{\mu}$ and $e_{\mu}$, it is present in 
the construction of invariants, etc. So its explicit form for the cosets studied is necessary.
\section{Explicit computation of the Goldstone matrix}
The Goldstone matrix of Eq. (\ref{MGoldstone}) can be computed easily for the $\textup{SO}\left(N\right)\rightarrow\textup{SO}(N-1)$ breaking. Since the structure is the same for every N \cite{Panico}, we will explicitly compute the matrix for the $\textup{SO}(3)\rightarrow\textup{SO}(2)$ case. The broken generators for this pattern are
\begin{equation}
\hat{T}_{1}=\frac{1}{\sqrt{2}}\begin{pmatrix}
0 & 0 & 0 \\ 0 & 0 & -i \\ 0 & i & 0
\end{pmatrix}, \hspace{0.5 cm}\hat{T}_{2}=\frac{1}{\sqrt{2}}\begin{pmatrix}
0 & 0 & -i \\ 0 & 0 & 0 \\ i & 0 & 0
\end{pmatrix},
\end{equation}
from which
\begin{equation}
\begin{split} 
i\frac{\sqrt{2}}{f}\Pi_{\hat{a}}\left(x\right)\hat{T}^{\hat{a}}&=i\frac{\sqrt{2}}{f}\left\lbrace\frac{\Pi_{1}}{\sqrt{2}}\begin{pmatrix}
0 & 0 & 0\\ 0 & 0 & -i\\0 & i & 0\end{pmatrix}+\frac{\Pi_{2}}{\sqrt{2}}
\begin{pmatrix}0 & 0 & -i\\ 0 & 0 & 0\\ i & 0 & 0\end{pmatrix}\right\rbrace\\
&=-\frac{1}{f}\begin{pmatrix}0 & 0 & -\Pi_{2}\\0 & 0 & -\Pi_{1}\\\Pi_{2} &\Pi_{1} & 0\end{pmatrix}=\frac{\boldsymbol{X}}{f},
\end{split}
\end{equation}
the Goldstone matrix can be written as
\begin{equation}\label{expGoldstone}
U\left[\Pi\right]=e^{\frac{\boldsymbol{X}}{f}}=\sum_{n=0}^{\infty}\left(\frac{\boldsymbol{X}}{f}\right)^{n}\frac{1}{n!}=\boldsymbol{1}+\frac{\boldsymbol{X}}{f}+\frac{\boldsymbol{X}^{2}}{2!f^{2}}+\frac{\boldsymbol{X}^{3}}{3!f^{3}}+\cdots
\end{equation}
Now let's calculate the terms $\boldsymbol{X}^{n}$ for $n\geq 2$ of the expansion:
\begin{equation}\label{x2}
\begin{split}
\boldsymbol{X}^{2}&=\begin{pmatrix}-\Pi^{2}_{2} & -\Pi_{1}\Pi_{2} & 0\\-\Pi_{1}\Pi_{2} & -\Pi^{2}_{1} & 0\\0 & 0 & -\left(\Pi^{2}_{1}+\Pi^{2}_{2}\right)\end{pmatrix}\\
&=-\begin{pmatrix}\Pi^{2}_{2} & \Pi_{1}\Pi_{2} & 0\\
\Pi_{1}\Pi_{2} & \Pi^{2}_{1} & 0\\0 & 0 & \Pi^{2}\end{pmatrix};\hspace{2 cm}\Pi^{2}=\Pi^{2}_{1}+\Pi^{2}_{1}.
\end{split}
\end{equation}
\begin{equation}\label{x3}
\begin{split}
\boldsymbol{X}^{3}=&\boldsymbol{X}^{2}\boldsymbol{X}=-\begin{pmatrix}0 & 0 & \Pi_{2}\Pi^{2}\\
0 & 0 & \Pi_{1}\Pi^{2}\\-\Pi^{2}\Pi_{2}^{2} & -\Pi^{2}\Pi_{1}^{2} & 0\end{pmatrix}=-\Pi^{2}\begin{pmatrix}0 & 0 & \Pi_{2}\\
0 & 0 & \Pi_{1}\\-\Pi_{2}^{2} & \Pi_{1}^{2} & 0\end{pmatrix}\\=&-\Pi^{2}\boldsymbol{X}.
\end{split}
\end{equation}
Eqs. (\ref{x2}) and (\ref{x3}) are the starting point to obtain recursion relations for the following terms 
\begin{equation}
\begin{split}
&\boldsymbol{X}^{4}=\boldsymbol{X}^{3}\boldsymbol{X}=-\Pi^{2}\boldsymbol{X}^{2},\hspace{0.5 cm}\boldsymbol{X}^{5}=\boldsymbol{X}^{4}\boldsymbol{X}=-\Pi^{2}\boldsymbol{X}^{3}=\Pi^{4}\boldsymbol{X},\\
&\boldsymbol{X}^{6}=\boldsymbol{X}^{5}\boldsymbol{X}=\Pi^{4}\boldsymbol{X}^{2},\hspace{0.5 cm}\boldsymbol{X}^{7}=\Pi^{4}\boldsymbol{X}^{3}=-\Pi^{6}\boldsymbol{X},
\end{split}
\end{equation}
giving the following recursion formulas:
\begin{equation}
\begin{split}
&\boldsymbol{X}^{2n+1}=\left(-1\right)^{n}\Pi^{2n}\boldsymbol{X},\hspace{0.90 cm}n=0,1,2,3,\cdots\\
&\boldsymbol{X}^{2\left(n+1\right)}=\left(-1\right)^{n}\Pi^{2n}\boldsymbol{X}^{2},\hspace{0.5 cm}n=0,1,2,3,\cdots
\end{split}
\end{equation}
Eq. (\ref{expGoldstone}) can thus be written as
\begin{equation}
\begin{split}
e^{\frac{\boldsymbol{X}}{f}}&=\mathds{1}+\sum_{n=0}^{\infty}\frac{\left(-1\right)^{n}\Pi^{2n}}{\left(2n+1\right)!f^{2n+1}}\boldsymbol{X}+\sum_{n=0}^{\infty}\frac{\left(-1\right)^{n}\Pi^{2n}}{\left[2\left(n+1\right)\right]!f^{2\left(n+1\right)}}\boldsymbol{X}^{2}\\
&=\mathds{1}+\sum_{n=0}^{\infty}\frac{\left(-1\right)^{n}}{\left(2n+1\right)!}\left(\frac{\Pi}{f}\right)^{2n+1}\frac{\boldsymbol{X}}{\Pi}+\sum_{n=0}^{\infty}\frac{\left(-1\right)^{n}}{\left[2\left(n+1\right)\right]!}\left(\frac{\Pi}{f}\right)^{2\left(n+1\right)}\frac{\boldsymbol{X}^{2}}{\Pi^{2}}.
\end{split}
\end{equation}
Remembering that
\begin{equation}
\sin x=\sum_{n=0}^{\infty}\frac{\left(-1\right)^{n}}{\left(2n+1\right)!}x^{2n+1},
\end{equation}
and
\begin{equation}
\begin{split} 
\sum_{n=0}^{\infty}\frac{\left(-1\right)^{n}}{\left[2\left(n+1\right)\right]!}\left(\frac{\Pi}{f}\right)^{2\left(n+1\right)}&=\sum_{n=1}^{\infty}\frac{\left(-1\right)^{n-1}}{\left(2n\right)!}\left(\frac{\Pi}{f}\right)^{2n}=-\sum_{n=1}^{\infty}\frac{\left(-1\right)^{n}}{\left(2n\right)!}\left(\frac{\Pi}{f}\right)^{2n}\\
&=1-\left(1+\sum_{n=1}^{\infty}\frac{\left(-1\right)^{n}}{\left(2n\right)!}\left(\frac{\Pi}{f}\right)^{2n}\right)\\
&=1-\sum_{n=0}^{\infty}\frac{\left(-1\right)^{n}}{\left(2n\right)!}\left(\frac{\Pi}{f}\right)^{2n}=1-\cos\frac{\Pi}{f},
\end{split}
\end{equation}
the Goldstone matrix is
\begin{equation}\label{GoldstoneSerie}
\begin{split}
U\left[\Pi\right]&=\mathds{1}+\left(\sin\frac{\Pi}{f}\right)\frac{\boldsymbol{X}}{\Pi}+\left(1-\cos\frac{\Pi}{f}\right)\frac{\boldsymbol{X}^{2}}{\Pi^{2}}.
\end{split}
\end{equation}
Using the explicit expression for $\boldsymbol{X}$ and Eq.(\ref{x2}), the Goldstone matrix $U\left[\Pi\right]$ can be written explicitly as
\begin{equation}
\begin{split}
U\left[\Pi\right]=\begin{pmatrix}1-\frac{\Pi^{2}_{2}}{\Pi^{2}}\left(1-\cos\frac{\Pi}{f}\right) &-\frac{\Pi_{1}\Pi_{2}}{\Pi^{2}}\left(1-\cos\frac{\Pi}{f}\right) & \frac{\Pi_{2}}{\Pi}\sin\frac{\Pi}{f}\\
-\frac{\Pi_{1}\Pi_{2}}{\Pi^{2}}\left(1-\cos\frac{\Pi}{f}\right) & 1-\frac{\Pi^{2}_{1}}{\Pi^{2}}\left(1-\cos\frac{\Pi}{f}\right) & \frac{\Pi_{1}}{\Pi}\sin\frac{\Pi}{f}\\
-\frac{\Pi_{2}}{\Pi}\sin\frac{\Pi}{f} & -\frac{\Pi_{1}}{\Pi}\sin\frac{\Pi}{f} & \cos\frac{\Pi}{f}\end{pmatrix}. 
\end{split}
\end{equation}
Now, making the following definitions
\begin{equation}
\boldsymbol{\Pi}=\begin{pmatrix}\Pi_{2}\\\Pi_{1}\end{pmatrix},\hspace{0.5 cm}\boldsymbol{\Pi}^{T}=\begin{pmatrix}\Pi_{2} & \Pi_{1}\end{pmatrix},
\hspace{0.5 cm}\Pi=\sqrt{\boldsymbol{\Pi}^{T}\cdot\boldsymbol{\Pi}},
\end{equation}
\begin{equation}
\boldsymbol{\Pi}\cdot\boldsymbol{\Pi}^{T}=\begin{pmatrix}\Pi_{2}\\\Pi_{1}\end{pmatrix}\cdot\begin{pmatrix}\Pi_{2} & \Pi_{1}\end{pmatrix}=\begin{pmatrix}\Pi^{2}_{2} & \Pi_{1}\Pi_{2}\\
\Pi_{1}\Pi_{2} & \Pi_{1}^{2}\end{pmatrix},
\end{equation}
we see that the Goldstone matrix can be written as
\begin{equation}
U\left[\Pi\right]=\begin{pmatrix}\mathds{1}-\left(1-\cos\frac{\Pi}{f}\right)\frac{\boldsymbol{\Pi}\cdot\boldsymbol{\Pi}^{T}}{\Pi^{2}} & \frac{\boldsymbol{\Pi}}{\Pi}\sin\frac{\Pi}{f}\\
-\frac{\boldsymbol{\Pi}^{T}}{\Pi}\sin\frac{\Pi}{f} & \cos\frac{\Pi}{f}\end{pmatrix}.
\end{equation}
This expression calculated for the $\textup{SO}(3)\rightarrow \textup{SO}(2)$ breaking pattern is general and valid for any $\textup{SO}(N)\rightarrow \textup{SO}(N-1)$ breaking pattern, provided that the 
broken generators have non-zero entries in their last row and column.
\section{Inverse and derived of the Goldstone matrix}
The Goldstone matrix is orthogonal, so its inverse is equal to the transpose given by
\begin{equation}\label{InvGoldstone}
U^{-1}\left[\Pi\right]=\begin{pmatrix}\mathds{1}-\left(1-\cos\frac{\Pi}{f}\right)\frac{\boldsymbol{\Pi}\cdot\boldsymbol{\Pi}^{T}}{\Pi^{2}} & -\frac{\boldsymbol{\Pi}}{\Pi}\sin\frac{\Pi}{f}\\
\frac{\boldsymbol{\Pi}^{T}}{\Pi}\sin\frac{\Pi}{f} & \cos\frac{\Pi}{f}\end{pmatrix},
\end{equation}
and the derivative of the Goldstone matrix is
\footnotesize
\begin{equation}
\partial_{\mu}U\left[\Pi\right]=\begin{pmatrix}-\frac{\boldsymbol{\Pi}\cdot\boldsymbol{\Pi}^{T}}{f\Pi^{2}}\partial_{\mu}\Pi\sin\frac{\Pi}{f}-\left(1-\cos\frac{\Pi}{f}\right)\partial_{\mu}\left(\frac{\boldsymbol{\Pi}\cdot\boldsymbol{\Pi}^{T}}{\Pi^{2}}\right)
&\frac{\boldsymbol{\Pi}}{f\Pi}\partial_{\mu}\Pi\cos\frac{\Pi}{f}+\partial_{\mu}\left(\frac{\boldsymbol{\Pi}}{\Pi}\right)\sin\frac{\Pi}{f}\\
-\frac{\boldsymbol{\Pi}^{T}}{f\Pi}\partial_{\mu}\Pi\cos\frac{\Pi}{f}-\partial_{\mu}\left(\frac{\boldsymbol{\Pi}^{T}}{\Pi}\right)\sin\frac{\Pi}{f} & -\frac{1}{f}\partial_{\mu}\Pi\sin\frac{\Pi}{f}\end{pmatrix},
\end{equation}
\normalsize 
where
\begin{equation}
\begin{split} 
&\partial_{\mu}\left(\frac{\boldsymbol{\Pi}\cdot\boldsymbol{\Pi}^{T}}{\Pi^{2}}\right)=\frac{1}{\Pi^{2}}\partial_{\mu}\left(\boldsymbol{\Pi}\cdot\boldsymbol{\Pi}^{T}\right)-\frac{2}{\Pi^{3}}\boldsymbol{\Pi}\cdot\boldsymbol{\Pi}^{T}\partial_{\mu}\Pi,\\
&\partial_{\mu}\left(\frac{\boldsymbol{\Pi}}{\Pi}\right)=\frac{1}{\Pi}\partial_{\mu}\boldsymbol{\Pi}-\frac{\boldsymbol{\Pi}}{\Pi^{2}}\partial_{\mu}\Pi.
\end{split}
\end{equation}
The above results are used in the main text.
\chapter{Useful Taylor series.}\label{series}
In this appendix we present the Taylor series used to obtain the results of Chapter \ref{CNMCHM} in the main text.\\
The first terms of the Taylor series of a $B\left(h,\eta\right)$ function that depends on the $h$ and $\eta$ fields, such that 
$B\left(h,\eta\right)$ is infinitely differentiable, is given by
\begin{equation}
\begin{split}
B\left(h,\eta\right)\simeq\left.B\left(h,\eta\right)\right|_{a}&+h\left.\frac{\partial B}{\partial h}\right|_{a}+\eta\left.\frac{\partial B}{\partial\eta}\right|_{a}\\
&+\frac{1}{2}\left(h^{2}\left.\frac{\partial^{2}B}{\partial h^{2}}\right|_{a}+\eta^{2}\left.\frac{\partial^{2}B}{\partial\eta^{2}}\right|_{a}+2h\eta\left.\frac{\partial^{2}B}{\partial\eta\partial h}\right|_{a}\right)+\cdots
\end{split}
\end{equation}
Where the series is evaluated at any point $a$ of the space-time.\\
Let be a function
\begin{equation}\label{gaugint}
B\left(h,\eta\right)=\frac{f^{2}\left(V+h\right)^{2}}{\left(V+h\right)^{2}+\left(\eta+N\right)^{2}}\sin^{2}\frac{\sqrt{\left(V+h\right)^{2}+\left(\eta+N\right)^{2}}}{f},
\end{equation}
by Taylor-expanding around $a=\left(0,0\right)$ we have
\begin{equation}
\begin{split} 
B\left(h,\eta\right)&=\frac{V^{2}}{V^{2}+N^{2}}f^{2}\sin^{2}\theta+\frac{2V^{3}}{\left(V^{2}+N^{2}\right)^{3/2}}fh\cos\theta\sin\theta\\
&+\frac{2N^{2}V}{\left(V^{2}+N^{2}\right)^{2}}f^{2}h\sin^{2}\theta+\frac{2NV^{2}}{\left(V^{2}+N^{2}\right)^{3/2}}f\eta\cos\theta\sin\theta\\
&-\frac{2NV^{2}}{\left(V^{2}+N^{2}\right)^{2}}f^{2}\eta\sin^{2}\theta+\frac{V^{4}}{\left(V^{2}+N^{2}\right)^{2}}\left(1-2\sin^{2}\theta\right)h^{2}\\
&+\frac{5N^{2}V^{2}}{\left(V^{2}+N^{2}\right)^{5/2}}fh^{2}\cos\theta\sin\theta+\frac{N^{2}\left(N^{2}-3V^{2}\right)}{\left(V^{2}+N^{2}\right)^{3}}f^{2}h^{2}\sin^{2}\theta\\
&+\frac{N^{2}V^{2}}{\left(V^{2}+N^{2}\right)^{2}}\left(1-2\sin^{2}\theta\right)\eta^{2}+\frac{V^{2}\left(V^{2}-4N^{2}\right)}{\left(V^{2}+N^{2}\right)^{5/2}}f\eta^{2}\cos\theta\sin\theta\\
&+\frac{V^{2}\left(3N^{2}-V^{2}\right)}{\left(V^{2}+N^{2}\right)^{3}}f^{2}\eta^{2}\sin^{2}\theta+\frac{2NV^{3}}{\left(V^{2}+N^{2}\right)^{2}}\left(1-2\sin^{2}\theta\right)h\eta\\
&+\frac{2NV\left(2N^{2}-3V^{2}\right)}{\left(V^{2}+N^{2}\right)^{5/2}}fh\eta\cos\theta\sin\theta+\frac{4NV\left(V^{2}-N^{2}\right)}{\left(V^{2}+N^{2}\right)^{3}}f^{2}h\eta\sin^{2}\theta+\cdots,
\end{split}
\end{equation}
where
\begin{equation}
\theta=\frac{\sqrt{V^{2}+N^{2}}}{f}. 
\end{equation}
From ElectroWeak VEV Eq. (\ref{VEVNM}) we can define the next relations
\begin{align}
&\xi=\frac{v^{2}}{f^{2}}\frac{\left(V^{2}+N^{2}\right)}{V^{2}}=\sin^{2}\theta,\label{xi}\\
&\cos\theta\sin\theta=\frac{v}{f}\frac{\sqrt{\left(1-\xi\right)\left(V^{2}+N^{2}\right)}}{V},\label{pro}
\end{align}
from which we have
\begin{equation}
\begin{split} 
B\left(h,\eta\right)&=v^{2}\Bigg\{1+\frac{2V^{2}}{\left(V^{2}+N^{2}\right)}\sqrt{1-\xi}\frac{h}{v}+\frac{2N^{2}}{\left(V^{2}+N^{2}\right)}\frac{h}{V}\\
&+\frac{V^{4}}{\left(V^{2}+N^{2}\right)^{2}}\left(1-2\xi\right)\frac{h^{2}}{v^{2}}+\frac{5N^{2}V}{\left(V^{2}+N^{2}\right)^{2}}\sqrt{1-\xi}\frac{h^{2}}{v}+N^{2}\frac{\left(N^{2}-3V^{2}\right)}{\left(V^{2}+N^{2}\right)^{2}}\frac{h^{2}}{V^{2}}\\
&+\frac{2NV}{\left(V^{2}+N^{2}\right)}\sqrt{1-\xi}\frac{\eta}{v}-\frac{2N}{\left(V^{2}+N^{2}\right)}\eta+\frac{N^{2}V^{2}}{\left(V^{2}+N^{2}\right)^{2}}\left(1-2\xi\right)\frac{\eta^{2}}{v^{2}}\\
&+V\frac{\left(V^{2}-4N^{2}\right)}{\left(V^{2}+N^{2}\right)^{2}}\sqrt{1-\xi}\frac{\eta^{2}}{v}+\frac{\left(3N^{2}-V^{2}\right)}{\left(V^{2}+N^{2}\right)^{2}}\eta^{2}+\frac{2NV^{3}}{\left(V^{2}+N^{2}\right)^{2}}\left(1-2\xi\right)\frac{h\eta}{v^{2}}\\
&+\frac{2N\left(2N^{2}-3V^{2}\right)}{\left(V^{2}+N^{2}\right)^{2}}\sqrt{1-\xi}\frac{h\eta}{v}+4\frac{N\left(V^{2}-N^{2}\right)}{\left(V^{2}+N^{2}\right)^{2}}\frac{h\eta}{V}+\cdots
\Bigg\}
\end{split}
\end{equation}
Similarly for the following expressions
\begin{align}
B\left(h,\eta\right)=&\frac{\left(V+h\right)}{\sqrt{\left(V+h\right)^{2}+\left(N+\eta\right)^{2}}}\sin\frac{2\sqrt{\left(V+h\right)^{2}+\left(N+\eta\right)^{2}}}{f},\\[0.5 cm]\notag
B\left(h,\eta\right)=&2V\sqrt{\frac{\xi}{V^{2}+N^{2}}}\Bigg\{\sqrt{1-\xi}+\frac{V^{2}}{\left(V^{2}+N^{2}\right)}\left(1-2\xi\right)\frac{h}{v}\\\notag\label{thexp}
&+\frac{N^{2}}{\left(V^{2}+N^{2}\right)}\sqrt{1-\xi}\frac{h}{V}+\frac{3N^{2}V}{2\left(V^{2}+N^{2}\right)^{2}}\left(1-2\xi\right)\frac{h^{2}}{v}\\\notag
&-\frac{3N^{2}}{2\left(V^{2}+N^{2}\right)^{2}}\sqrt{1-\xi}h^{2}-\frac{2V^{4}}{\left(V^{2}+N^{2}\right)^{2}}\xi\sqrt{1-\xi}\frac{h^{2}}{v^{2}}\\
&+\frac{NV}{\left(V^{2}+N^{2}\right)}\left(1-2\xi\right)\frac{\eta}{v}-\frac{N}{\left(V^{2}+N^{2}\right)}\sqrt{1-\xi}\eta\\\notag
&+\frac{\left(2N^{2}-V^{2}\right)}{2\left(V^{2}+N^{2}\right)^{2}}\sqrt{1-\xi}\eta^{2}+\frac{V\left(V^{2}-2N^{2}\right)}{2\left(V^{2}+N^{2}\right)^{2}}\left(1-2\xi\right)\frac{\eta^{2}}{v}\\\notag
&-\frac{2N^{2}V^{2}}{\left(V^{2}+N^{2}\right)^{2}}\xi\sqrt{1-\xi}\frac{\eta^{2}}{v^{2}}+\frac{N\left(N^{2}-2V^{2}\right)}{\left(V^{2}+N^{2}\right)^{2}}\left(1-2\xi\right)\frac{h\eta}{v}\\\notag
&+\frac{N\left(2V^{2}-N^{2}\right)}{\left(V^{2}+N^{2}\right)^{2}}\sqrt{1-\xi}\frac{h\eta}{V}-\frac{4NV^{3}}{\left(V^{2}+N^{2}\right)^{2}}\xi\sqrt{1-\xi}\frac{h\eta}{v^{2}}+\cdots\Bigg\}.
\end{align}
\begin{align}
B\left(h,\eta\right)=&\frac{\left(V+h\right)\left(N+\eta\right)}{\left(V+h\right)^{2}+\left(N+\eta\right)^{2}}\sin^{2}\frac{\sqrt{\left(V+h\right)^{2}+\left(N+\eta\right)^{2}}}{f},\\[0.5 cm]\notag
B\left(h,\eta\right)=&\frac{V\xi}{V^{2}+N^{2}}\Bigg\{N+\frac{2V^{2}N}{\left(V^{2}+N^{2}\right)}\sqrt{1-\xi}\frac{h}{v}+\frac{N\left(N^{2}-V^{2}\right)}{\left(V^{2}+N^{2}\right)}\frac{h}{V}\\\notag\label{etamexp}
&+\frac{NV^{4}}{\left(V^{2}+N^{2}\right)^{2}}\left(1-2\xi\right)\frac{h^{2}}{v^{2}}+\frac{NV\left(3N^{2}-2V^{2}\right)}{\left(V^{2}+N^{2}\right)^{2}}\sqrt{1-\xi}\frac{h^{2}}{v}\\\notag
&+\frac{N\left(V^{2}-3N^{2}\right)}{\left(V^{2}+N^{2}\right)^{2}}h^{2}+\frac{2N^{2}V}{\left(V^{2}+N^{2}\right)}\sqrt{1-\xi}\frac{\eta}{v}+\frac{\left(V^{2}-N^{2}\right)}{\left(V^{2}+N^{2}\right)}\eta\\
&+\frac{V^{2}N^{3}}{\left(V^{2}+N^{2}\right)^{2}}\left(1-2\xi\right)\frac{\eta^{2}}{v^{2}}+\frac{NV\left(3V^{2}-2N^{2}\right)}{\left(V^{2}+N^{2}\right)^{2}}\sqrt{1-\xi}\frac{\eta^{2}}{v}\\\notag
&+\frac{N\left(N^{2}-3V^{2}\right)}{\left(V^{2}+N^{2}\right)^{2}}\eta^{2}+\frac{2N^{2}V^{3}}{\left(V^{2}+N^{2}\right)^{2}}\left(1-2\xi\right)\frac{h\eta}{v^{2}}\\\notag
&+\frac{2\left(N^{4}-3N^{2}V^{2}+V^{4}\right)}{\left(V^{2}+N^{2}\right)^{2}}\sqrt{1-\xi}\frac{h\eta}{v}+\frac{\left(6V^{2}N^{2}-N^{4}-V^{4}\right)}{\left(V^{2}+N^{2}\right)^{2}}\frac{h\eta}{V}+\cdots\Bigg\}.
\end{align}
\singlespace

\end{document}